\documentclass[12pt]{article}
\usepackage{comment}
\usepackage{textcomp}
\usepackage{chetnew}
\usepackage{amssymb,amsmath,amsfonts,mathrsfs,braket}
\usepackage[utf8]{inputenc}
\usepackage{graphicx}

\usepackage{bm}

\def\one{{\,\hbox{1\kern-.8mm l}}}

\newcommand{\nn}{\nonumber}
\newcommand{\bz}{\bar{z}}

\allowdisplaybreaks

\def\makeatletter{\catcode`\@=11}
\makeatletter
\def\mathbox#1{\hbox{$\m@th#1$}}%
\def\math@ccstyles#1#2#3#4#5#6#7{{\leavevmode
      \setbox0\mathbox{#6#7}%
      \setbox2\mathbox{#4#5}%
      \dimen@ #3%
      \baselineskip\z@\lineskiplimit#1\lineskip\z@
      \vbox{\ialign{##\crcr
             \hfil \kern #2\box2 \hfil\crcr
             \noalign{\kern\dimen@}%
             \hfil\box0\hfil\crcr}}}}
\def\mathaccstyles{\math@ccstyles\maxdimen}
\def\maththroughstyles{\math@ccstyles{-\maxdimen}}
\def\unity%
 {\maththroughstyles{.45\ht0}\z@\displaystyle {\mathchar"006C}\displaystyle 1}

\def\beq{\begin{equation}}
\def\eeq{\end{equation}}
\newcommand{\bea}{\begin{eqnarray}}
\newcommand{\eea}{\end{eqnarray}}
\def\bal{\begin{align}}
\def\eal{\end{align}}

\preprint{CCTP-2022-4 \hfill QMUL-PH-22-25 \\ ITCP-IPP-2022/2}

\title{\vspace{-1.cm} 6D (2,0) Bootstrap with soft-Actor-Critic\\ 
}

\author{
Gergely K\'antor$^{a,\clubsuit}$,
Vasilis Niarchos\;$^{b,\diamondsuit}$, Constantinos Papageorgakis\;$^{a,\spadesuit}$, Paul Richmond\;$^{a,\heartsuit}$}

\affiliation{
$^a$ Centre for Theoretical Physics, Department of Physics and Astronomy\\ Queen Mary University of London, London E1 4NS, UK \vspace{0.3cm} $ $ \\
$^b$ ITCP \& CCTP, Department of Physics,\\
University of Crete, 71003 Heraklion, Greece
\vspace{0.3cm} $ $\\

\vspace{0.3cm}
{\tt \small$^\clubsuit$g.kantor@qmul.ac.uk,
$^\diamondsuit$niarchos@physics.uoc.gr,  
$^\spadesuit$c.papageorgakis@qmul.ac.uk, 

$^{\heartsuit}$p.richmond@qmul.ac.uk}}

\abstract{We study numerically the 6D (2,0) superconformal bootstrap using the soft-Actor-Critic (SAC) algorithm as a stochastic optimizer. We focus on the four-point functions of scalar superconformal primaries in the energy-momentum multiplet. Starting from the supergravity limit, we perform searches for adiabatically varied central charges and derive two curves for a collection of 80 CFT data (70 of these data correspond to unprotected long multiplets and 10 to protected short multiplets). We conjecture that the two curves capture the $A$- and $D$-series (2,0) theories. Our results are competitive when compared to the existing bounds coming from standard numerical bootstrap methods, and data obtained using the OPE inversion formula. With this paper we are also releasing our Python implementation of the SAC algorithm, BootSTOP. The paper discusses the main functionality features of this package.}

\date{\today}

\begin{document}

\maketitle

\hypersetup{pageanchor=true}

\setcounter{tocdepth}{2}

\toc

\section{Introduction and Summary}
\label{summary}

In a recent paper, \cite{Kantor:2021kbx,Kantor:2021jpz}, we proposed that stochastic optimization methods could be a useful new tool in the arsenal of the conformal bootstrap program. More specifically, in the context of truncation methods (where one looks for approximate solutions to the crossing equations with a truncated spectrum; see e.g. \cite{Gliozzi:2013ysa, Gliozzi:2014jsa, Gliozzi:2015qsa, Gliozzi:2016cmg, Li:2017ukc}) one can view the conformal bootstrap as a Large-Scale, non-convex, continuous optimization problem. Such complex non-linear programming problems are ubiquitous in many scientific areas and applications, and are commonly treated by stochastic algorithms. 

In \cite{Kantor:2021kbx,Kantor:2021jpz} we used an optimization method based on a popular soft-Actor-Critic (SAC) Reinforcement-Learning (RL) algorithm, first developed and employed in the context of robotics, \cite{DBLP:journals/corr/abs-1801-01290}, where it was primarily geared towards continuous control tasks. The use of RL in optimization problems is not typical.  In fact, there is a vast number of stochastic algorithms and metaheuristics in the market designed to attack Large Scale Optimization problems that do not employ Machine Learning (ML) techniques. Depending on the problem, some algorithms may outperform others, but there is no single general-purpose algorithm that dominates. The development of new (hybrid) optimization algorithms and their improvement with the use of ML methods is an active research direction. In our opinion, the use of RL in this general context has promising features and deserves further study.

In this paper we present the results of a very specific exercise: We study the crossing equation of the 4-point function for the superconformal primary in the energy-momentum multiplet of interacting 6D (2,0) superconformal field theories (SCFTs). These theories, which do not have a known Lagrangian description, play a central role in M theory (they capture the low-energy dynamics of multiple M5 branes) and provide canonical examples of the AdS/CFT correspondence in string theory. Even though we study a single-correlator conformal bootstrap problem, the superconformal algebra combined with known facts from chiral--algebra techniques \cite{Beem:2014kka,Beem:2015aoa} leads to a crossing equation, which is expected to admit a constrained set of solutions describing known 6D (2,0) SCFTs. 

Throughout the paper we employ a fixed truncation on the spectrum of superconformal primaries that can appear in the conformal block decomposition of the above crossing equation with 45 operators (10 protected up to spin 17, and 35 unprotected up to spin 12). This yields a continuous, non-convex optimization problem in an 80-dimensional configuration space, where the algorithm aims to maximize a reward. We have chosen the reward as the inverse of a cost function that measures the violation of the truncated crossing equation on a uniform grid of 180 points in cross-ratio space. The coordinates in the configuration search space include 35 scaling dimensions of non-protected operators (in leading and subleading Regge trajectories),\footnote{The scaling dimensions of the remaining 10 protected operators in short superconformal multiplets are fixed by superconformal symmetry, and therefore remain constant throughout our search.} and 45 unknown OPE-squared coefficients. We do not solve for the OPE-squared coefficients at each step of the computation to reduce the dimensionality of the search space for reasons that are explained in Sec.\ \ref{implementation}.  

Our 80-dimensional search signifies a factor of 2 increase in search-space dimensionality compared to our previously published study cases, in examples of 2D CFTs \cite{Kantor:2021kbx,Kantor:2021jpz}. It is also, as far as we know, one of the largest searches in truncated crossing equations to-date. However, it is still very far from our ultimate goal of truncations with thousands of operators. The latter would constitute a truly Large Scale Optimization problem, powerful enough to start attacking with good accuracy a wide range of situations, including the bootstrap of multiple correlators. Nevertheless, the fixed truncation used in this paper allows us to demonstrate clearly a particularly interesting implementation of our approach that produces very promising results in a demanding context, where traditional Lagrangian methods are inapplicable and other conformal bootstrap methods have produced relatively few results.

\begin{figure}[t]
\centering
\includegraphics[width=12cm, height=6cm]{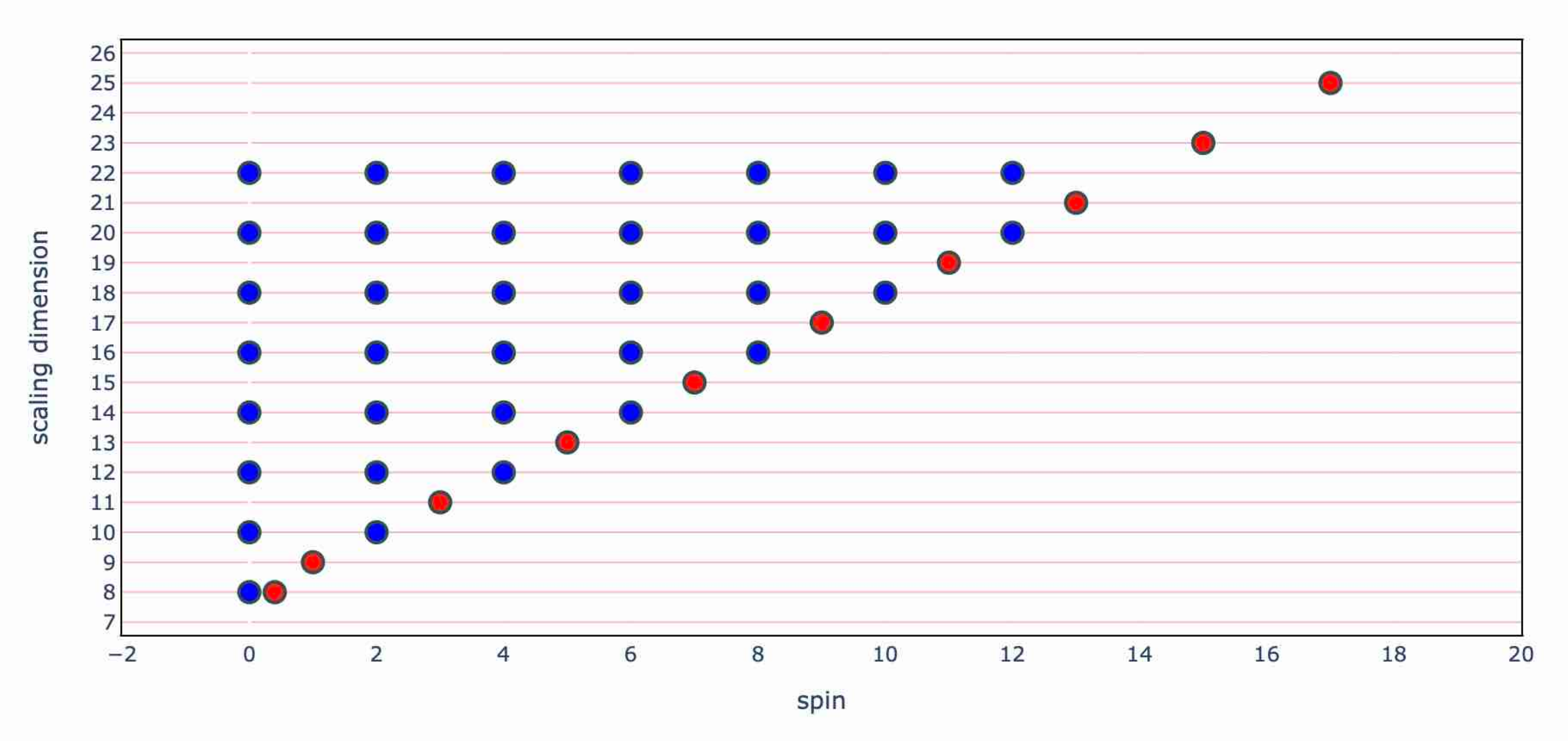}
\caption{This figure depicts the chosen spectrum of 45 operators in the supergravity limit, where the operators are generalized-free with known scaling dimensions and OPE-squared coefficients. The blue dots represent non-protected operators in $\mathcal L[0,0]$ superconformal multiplets---see Sec.\ \ref{main1}. The red dots along the lowest diagonal represent protected operators in the $\mathcal D[0,4]$ (at spin 0 and scaling dimension 8) and $\mathcal B[0,2]$ multiplets (at odd spin).}
\label{spin_partition}
\end{figure}

For the purposes of our application, we first fixed the structure of the unknown\footnote{We emphasize that this truncation refers specifically to the crossing equation after the use of superconformal algebra relations and the known analytic results from chiral--algebra techniques \cite{Beem:2014kka,Beem:2015aoa}. The full crossing equation involves additional operators whose CFT data are known and have been already incorporated---see Sec.\ \ref{main1} for details.} truncated spectrum (namely, the number of operators at each spin) in the supergravity limit (at $c=\infty$) where the spectrum is generalized-free (see Fig.\ \ref{spin_partition}). We then checked that optimizing around the analytic generalized-free configuration at $c=\infty$, our algorithm recovers a slightly modified configuration with high reward.\footnote{The slight deviation of the obtained configuration from the exact generalized-free result is the expected effect of the truncation.} That is a reassuring sign that this particular truncation is an acceptable approximation in the supergravity regime.\footnote{This does not guarantee, however, that the truncation remains a good approximation for all values of $c$. One of our goals is to probe this aspect.} Then, by adiabatically changing the value of $c$ while keeping the number of operators fixed, we tracked the evolution of the supergravity spectrum of scaling dimensions and OPE-squared coefficients and produced curves for the 80 CFT data as a function of the central charge $c$. At the level of the single crossing equation that we are studying it is impossible to detect the discrete nature of $c$, so throughout our computation the latter is treated as a continuous parameter.\footnote{A recent, very interesting, study of the Ising CFT using a different adiabatic deformation in spacetime dimension appeared in \cite{Henriksson:2022gpa}. This work combined the standard numerical conformal bootstrap technology with the navigator method \cite{Reehorst:2021ykw} and the extremal functional method \cite{El-Showk:2012vjm,El-Showk:2016mxr}.}    

The most prominent and encouraging features of the results obtained in this manner are the following:

\begin{itemize}
    \item[(1)] Adopting a process that is outlined in Sec.\ \ref{implementation}, we managed to produce two distinct sets of curves of high reward. We conjecture that they correspond to the $A$- and $D$-series of the 6D (2,0) theories. In previous applications of the numerical conformal bootstrap to 6D (2,0) SCFTs, \cite{Beem:2015aoa,Lemos:2021azv}, it had been impossible to make a distinction between the two series. This distinction was unclear even in \cite{Lemos:2021azv}, where a very interesting attempt was made to go beyond the rigorous bounds by employing an iterative procedure based on the OPE-inversion formula \cite{Caron-Huot:2017vep}.
    \item[(2)] The overall comparison of the CFT data that we produced against the known results from the literature is very promising. Most notably, we demonstrate that the data of the lowest-lying protected and unprotected scalar operators in Figs\ \ref{lowest_protected}-\ref{lowest_unprotected},  are within the known numerical conformal bootstrap bounds, and compare well with the results obtained using the OPE-inversion formula in \cite{Lemos:2021azv}. For CFT data at higher spin, some data behave  worse compared to the results of \cite{Beem:2015aoa,Lemos:2021azv}, but the overall emerging picture supports the message that our approach produces competitive results for a wide range of central charge values (even with a fixed truncation of only 45 operators).
    \item[(3)] In total, we obtained two sets of curves as a function of $c$ for 80 CFT data. Most of these data refer to operators of unprotected long multiplets in subleading Regge trajectories. It is the first time that predictions have been made for these data. However, the obvious word of caution applies here: As one goes higher in scaling dimension some of these predictions are less reliable. This reflects clearly in, e.g., the statistical standard deviations of our searches. 
\end{itemize}

In our opinion, these results are very encouraging and motivate further work in this direction. With improved algorithms and better implementation one should be able to achieve more reliable, higher-quality numerics. These can then be fed back to the standard numerical bootstrap methods to produce even better results. Along the same lines, for the 6D (2,0) numerical bootstrap, the results presented in this paper are merely a first step towards a more complete investigation and should not be viewed as our final best answer to this problem. As an immediate next step, an enlargement of the truncation by tens of operators is already possible with the current algorithm. We have made our Python code, BootSTOP, public\footnote{BootSTOP can  be publicly accessed on GitHub at \href{https://github.com/vniarchos/BootSTOP}{this link}.} and encourage the interested reader to explore the many possibilities. The broader question of scalability for the present algorithm, or the development of a different more efficient one is currently under investigation.

The rest of this article is organized as follows. In Sec.\ \ref{main1} we review the key features of the 6D (2,0) crossing equation that we analyze in our computations. In Sec.\ \ref{implementation} we outline the details of the BootSTOP package, detail the improvements made compared to \cite{Kantor:2021kbx,Kantor:2021jpz}, and highlight some of the potential alternatives that we have not yet explored. Sec.\ \ref{adiabatic} contains the main results of the paper on the adiabatic evolution of the truncated spectrum as a function of the central charge $c$. We conclude with a summary and brief comments on future directions in Sec.\ \ref{outlook}. In App.\ \ref{blind} we present a blind run of the SAC algorithm with 10k agents, to demonstrate how it performs in a very wide search (starting with search windows of size 10 in scaling dimensions, and 20 in OPE-squared coefficients).

\section{6D (2,0) Crossing Equations}\label{main1}
Our main purpose in this section is to set up the language needed to follow our numerical approach. For a more complete review of the pertinent details of 6D (2,0) SCFTs in the context of the conformal bootstrap, we refer the reader to the excellent presentation of Ref.\ \cite{Beem:2015aoa}. 

In the conformal bootstrap analysis of 4-point functions, in any dimension, one needs to solve crossing equations of the general form
\beq
\label{impleaa}
\sum_{I} {\mathfrak C}_I F_{\Delta_I}(z,\bar z) = 0
~,
\eeq
where $\mathfrak C_I$ are OPE-squared coefficients for the $I$-th operator that runs in the conformal block decomposition and $F_{\Delta_I}(z,\bar z)$ a conformal block (combination) for the corresponding operator, the scaling dimension of which is  denoted as $\Delta_I$.\footnote{In superconformal theories, like the 6D (2,0) theory in this paper, it is convenient to work with the full superconformal blocks, which are then used to express the functions $F_{\Delta_I}(z,\bar z)$.} The unknowns of the problem are the CFT data $\{\Delta_I, {\mathfrak C}_I\}$. Eq.\ \eqref{impleaa} gets contributions from two channels and depends on the complex conjugate parameters $(z,\bar z)$, which express the conformal cross ratios of the four operator insertions. It has to be satisfied at all values of the $z$ parameters where the conformal block expansion (in both channels) is convergent.

To set up the exact crossing equations of the 6D (2,0) theory, we follow closely the Refs.\ \cite{Beem:2015aoa,Lemos:2021azv}. We focus on the crossing equations arising from the 4-point function of superconformal primaries in the energy-momentum multiplet. We remind the reader that unitary representations of the 6D (2,0) superconformal algebra are characterized by linear relations between the quantum numbers of the superconformal primaries. They can be classified as \cite{Minwalla:1997ka, Dobrev:2002dt}:\footnote{See also \cite{Buican:2016hpb, Cordova:2016emh} for the explicit  multiplet construction.}
\begin{equation}\label{shortening}
\begin{array}{lll}
\mathcal{L}: & \Delta>h_{1}+h_{2}-h_{3}+2\left(d_{1}+d_{2}\right)+6, \qquad & h_{1} \geqslant h_{2} \geqslant h_{3} \\
\mathcal{A}: & \Delta=h_{1}+h_{2}-h_{3}+2\left(d_{1}+d_{2}\right)+6, \qquad & h_{1} \geqslant h_{2} \geqslant h_{3} \\
\mathcal{B}: & \Delta=h_{1}+2\left(d_{1}+d_{2}\right)+4, \qquad & h_{1} \geqslant h_{2}=h_{3} \\
\mathcal{C}: & \Delta=h_{1}+2\left(d_{1}+d_{2}\right)+2, \qquad & h_{1}=h_{2}=h_{3} \\
\mathcal{D}: & \Delta=2\left(d_{1}+d_{2}\right), \qquad & h_{1}=h_{2}=h_{3}=0 \, ,
\end{array}
\end{equation}
where the quantum numbers of a state are given by $[h_1,h_2,h_3;d_1,d_2;\Delta]$. These correspond to eigenvalues of the $so(6)$ generators of rotations in three orthogonal planes in $\mathbb{R}^6$, generators of a Cartan subalgebra of the $R$-symmetry $so(5)_R$, and the dilatation generator respectively. The conformal dimensions of the short (BPS) multiplets are fixed in terms of the other quantum numbers, while the spin of a state is related to the $so(6)$ quantum numbers through $\ell = h_1 + h_2 - h_3$. The energy-momentum multiplet is denoted $\mathcal{D}[2,0] \equiv \mathcal{D}[0,0,0;2,0;4]$, the superconformal primary of which is a spacetime scalar $\Phi^{IJ}$, $I,J=1,\ldots,5$, of conformal dimension $\Delta=4$, transforming in the symmetric-traceless representation of $so(5)_R$.

The OPE selection rules for two energy-momentum multiplets in an interacting 6D (2,0) theory include \cite{Heslop:2004du, Eden:2001wg}
\begin{align}
	\mathcal{D}[2,0] \times \mathcal{D}[2,0] \ =& \ \mathbf{1} + \mathcal{D}[4,0] + \mathcal{D}[2,0] + \mathcal{D}[0,4] \nn \\
	&+ \sum_{\ell=0,2,\ldots} ( \mathcal{B}[2,0]_\ell  + \mathcal{B}[0,2]_{\ell +1} ) + \sum_{\substack{\ell=0,2,\ldots \\ \Delta>6+\ell}} \, \mathcal{L}[0,0]_{\Delta,\ell} \, . \label{SelfOPE}
\end{align}

With this information, the corresponding exact crossing equations boil down to an expression of the form \cite{Beem:2015aoa}
\begin{align}\label{crossingeqs}
	c(z,\bz) - c(1-z,1-\bz) = 0\, ,
\end{align}
where $(z,\bar z)$ are complex coordinates on the cross-ratio plane, and 
\begin{align}\label{cees}
	c(z,\bz)  =  z \bz ( a^u(z,\bz) + a^\chi(z,\bz) ) + \mathcal{C}_{h}(z, \bz) \, .
\end{align}
The function $\mathcal{C}_{h}$ is known exactly:
\begin{align}
	\mathcal{C}_{h}(z, \bar{z}) =  \frac{1}{(z-\bar{z})^{3}} \frac{h(z)-h(\bar{z})}{z \bar{z}}
\end{align}
with 
\begin{align}
	h(z) & =  -\left(\frac{z^{3}}{3}-\frac{1}{z-1}-\frac{1}{(z-1)^{2}}-\frac{1}{3(z-1)^{3}}-\frac{1}{z}\right)-\frac{8}{c}\left(z-\frac{1}{z-1}+\log (1-z)\right)\cr & \qquad \qquad- \frac{1}{6} + \frac{8}{c} \, 
\end{align}
and $c$ is the central charge of the 6D (2,0) theory, whose value for the $A_{N-1}/D_N$-series is respectively 
\begin{align}
	c_{A_{N-1}} =  4 N^3 - 3 N - 1 \, , \qquad c_{D_N} =  16 N^3 - 24 N^2 + 9 N    \; .
\end{align}
To complete the definition of \eqref{crossingeqs}, consider the ``atomic" function
\begin{align}
	a_{\Delta, \ell}^{\mathrm{at}}(z, \bar{z})=\frac{4}{z^{6} \bar{z}^{6}(\Delta-\ell-2)(\Delta+\ell+2)} \mathcal{G}_{\Delta+4}^{(\ell)}(0,-2 ; z, \bar{z}) \, , \label{aatomic}
\end{align}
where $\mathcal{G}_{\Delta}^{(\ell)}(\Delta_1 - \Delta_2, \Delta_3 - \Delta_4 ; z, \bar{z})$ are the 6D non-supersymmetric conformal blocks \cite{Dolan:2003hv, Dolan:2011dv}.\footnote{These can also be explicitly found in App.~B of \cite{Beem:2015aoa}.} Then in \eqref{cees} the piece
\begin{align}\label{achi}
a^\chi(z,\bz)  =  \sum_{\substack{\ell=0,2, \ldots \\ \ell \text{ even}}} 2^\ell b_\ell \, a_{\ell+4, \ell}^{\mathrm{at}}(z, \bar{z})
\end{align}
represents the contributions from the identity, $\mathcal{D}[2,0]$, $\mathcal{D}[4,0]$ and $\mathcal{B}[2,0]_{\ell-2}$ superconformal multiplets appearing in the self-OPE of $\Phi^{IJ}$ from \eqref{SelfOPE}. The OPE coefficients $b_\ell$ can be fixed through the associated 2D chiral algebra \cite{Beem:2014kka,Beem:2015aoa} and read
\begin{align}
b_{\ell} & =  \frac{(\ell+1)(\ell+3)(\ell+2)^{2} \frac{\ell}{2} !\left(\frac{\ell}{2}+2\right) ! !\left(\frac{\ell}{2}+3\right) ! !(\ell+5) ! !}{18(\ell+2) ! !(2 \ell+5) ! !} \cr 
&\qquad \qquad +\frac{8}{c} \frac{\left(2^{-\frac{\ell}{2}-1}(\ell(\ell+7)+11)(\ell+3) ! ! \Gamma\left(\frac{\ell}{2}+2\right)\right)}{(2 \ell+5) ! !} \, .
\end{align}
The remaining piece in \eqref{cees}, $a^u(z,\bz)$, contains the unknown data that one hopes to constrain through the crossing equations. These include the conformal dimension and OPE-squared coefficients corresponding to the long multiplets $\mathcal{L}[0,0]_{\Delta,\ell}$, and the OPE-squared coefficients corresponding to the short $\mathcal{D}[0,4]$ and $\mathcal{B}[0,2]_{\ell-1}$ multiplets. In full,
\begin{align}\label{au}
	a^u(z,\bz) \ =& \ \sum_{\substack{\Delta \geq \ell+6 \\ \ell \geq 0 , \, \ell \text{ even}}} \lambda^2_{\Delta,\ell} a^{at}_{\Delta,\ell}(z,\bz) \, .
\end{align}
The conformal dimensions of the $\mathcal{D}[0,4]$ and $\mathcal{B}[0,2]_{\ell-1}$ (with $\ell>0$) multiplets are fixed via \eqref{shortening} and given by $8$ and $\ell+7$ respectively ($\ell$ is always an even integer).

For the purposes of truncating the 6D (2,0) crossing equations to the spin partition displayed in Fig.\ \ref{spin_partition}, we specifically used in place of Eqs\ \eqref{achi}, \eqref{au}, the truncated versions
\begin{align}
a_{\rm tr}^\chi(z,\bz)  =  \sum_{\substack{\ell=0}}^{30} 2^\ell b_\ell \, a_{\ell+4, \ell}^{\mathrm{at}}(z, \bar{z})
\end{align}
and
\begin{align}
	a_{\rm tr}^u(z,\bz) \ =& \ (\lambda^{\mathcal{D}[0,4]}_{8,0})^2 a^{at}_{6,0}(z,\bz) + \sum_{\ell=2}^{18} (\lambda^{\mathcal{B}[0,2]_{\ell-1}}_{\ell+7,\ell})^2 a^{at}_{\ell+6,\ell}(z,\bz) + \sum_{\substack{\ell=0\\ \Delta > \ell+6}}^{12} (\lambda^{\mathcal{L}[0,0]_{\Delta,\ell}}_{\Delta,\ell})^2 a^{at}_{\Delta,\ell}(z,\bz)\;,
\end{align}
where $a_{\Delta, \ell}^{\mathrm{at}}(z, \bar{z})$ is given in \eqref{aatomic}.

\section{BootSTOP: Remarks on SAC Implementation}
\label{implementation}

Throughout this paper we make use of the soft-Actor-Critic (SAC) Reinforcement Learning algorithm as a continuous optimizer, as already outlined in the context of the conformal bootstrap in \cite{Kantor:2021kbx,Kantor:2021jpz}. We will not review the details of the SAC algorithm here, but instead refer the reader to \cite{Kantor:2021kbx,Kantor:2021jpz} and \cite{Kantor:thesis} for further details and useful references. We are also making our Python implementation, BootSTOP, available on GitHub at \href{https://github.com/vniarchos/BootSTOP}{this link}. The README file in that repository contains additional information. In this section we summarize the main components of the package, its functionalities, and highlight the key improvements included in BootSTOP compared to our earlier work \cite{Kantor:2021kbx,Kantor:2021jpz}. These improvements extend beyond the current application of BootSTOP to the 6D (2,0) theory, and we plan on upgrading its functionality in the near future to incorporate CFTs in diverse dimensions, as well as a choice of different continuous optimizers, in addition to SAC.

When the sum over the intermediate operators $I$ in \eqref{impleaa} is truncated, the crossing equations cannot be satisfied exactly. Instead, one attempts to minimize an associated semi-positive-definite cost function $\boldsymbol{C}$. The choice of this function is not unique. In this paper, we have chosen to discretize the $z$-plane using a uniform grid of 180 points (following the multipoint scheme of \cite{CastedoEcheverri:2016fxt}); these are detailed in a file called $\displaystyle{\bf data\_ z\_ sample.py}$.\footnote{The specific values that we used in our adiabatic runs can be found at \href{https://github.com/vniarchos/BootSTOP/blob/main/Applications/6D/2022_08_results/data_z_sample.py}{this link}.} We have also chosen to define $\boldsymbol{C}$ as the Euclidean quadratic norm of a vector, the components of which are given by the LHS of \eqref{impleaa}, 
\begin{align}
\sum_{I}^{\rm cutoff} {\mathfrak C}_I F_{\Delta_I}(z_i,\bar z_i)
~,
\end{align}
evaluated at each point $z_i$ on the grid. It is straightforward to implement other possibilities, e.g.\ use Taylor-expansion coefficients around the $z=\frac{1}{2}$ point (which is common practice in the numerical conformal bootstrap), or define a different cost function. Once the cost function is specified, we define the reward function as $\boldsymbol{R} = {\boldsymbol{C}}^{-1}$, which the algorithm aims to maximize.

\subsection{Python Libraries, Numerical Accuracy and Speed-up Techniques}

BootSTOP has been implemented using the PyTorch package for Python 3.7. One of the main factors affecting the speed of the code is the evaluation of the conformal blocks, which takes place at every cycle of the algorithm. The conformal blocks contain several hypergeometric functions, the numerical evaluation of which is subtle and costs time. Two commonly used mathematical libraries in Python are ``mpmath"  and ``SciPy". The former is slower, but offers greater accuracy (including the possibility of arbitrary numerical precision). The latter is considerably faster, but exhibits known inaccuracies in the evaluation of the hypergeometric functions at specific arguments. We have chosen to employ the SciPy library in BootSTOP and have avoided the evaluation inaccuracies by carefully selecting the points on the $z$-grid. 

An additional and considerable speed-up (typically by a factor of 10) can be obtained by computing the conformal blocks only once outside the main loop. This was carried out by performing a discretization in the space of the unknown scaling dimensions with 60k lattice points, evaluating the conformal blocks and storing them in a separate set of CSV files.\footnote{For the runs presented in this paper this amounts to a lattice separation of $5 \times 10^{-4}$ in the scaling dimensions that enter the hypergeometric functions in the conformal blocks.} When BootSTOP starts running, it pre-loads all the conformal blocks before the initialization of the main loop. Once the main loop starts, the agent explores continuously the space of scaling dimensions, but the code rounds them to the closest value on the pre-evaluated scaling-dimension lattice to obtain information about the reward. This procedure limits numerical accuracy but offers a significant speed-up compared to the original implementation in \cite{Kantor:2021kbx,Kantor:2021jpz}. 

Another possibility in the implementation of the SAC algorithm is the following. When using a quadratic Euclidean norm cost function it is in principle straightforward to take advantage of the fact that the OPE-squared coefficients appear quadratically: one could solve the corresponding extremization equations to obtain the OPE-squared coefficients at each step as a function of the unknown scaling dimensions. Inserting the answer back into the cost function, results into a new cost function that depends only on the scaling dimensions. The obvious benefit of this method is that it allows one to reduce the dimensionality of the search space significantly. However, there are also considerable disadvantages. 

First, by employing this intermediate extremization, one changes the optimization problem and finds different solutions at finite truncation. As the size of the truncation grows and one approximates the exact solutions better and better, one expects the global minima of the cost functions in both approaches to converge to each other. It is not, however, obvious how the search is affected by the intermediate extremization at finite truncation (especially, if there are multiple extrema of interest, which are not global minima). Second, the solution of the extremization equations for the OPE-squared coefficients involves the inversion of a matrix, the size of which is set by the number of unknown scaling dimensions. This inversion becomes increasingly computationally expensive as the dimension of the search space grows. Moreover, it can become quite unstable if some scaling dimensions are almost degenerate. Third, this approach is harder to implement with non-quadratic cost functions, which can be useful for some problems. For all these reasons, we have chosen to treat the OPE-squared coefficients as independent unknowns in BootSTOP, over which we optimize. Nevertheless, the option of intermediate extremization can be incorporated into our code with the obvious modifications.

\subsection{Summary of Parameters and BootSTOP Functionalities}\label{functionality}

We now dive a little deeper into the options built into BootSTOP. The code has several sets of parameters which we broadly describe as: neural network hyperparameters, learning loop, automation and environment parameters. In this section we briefly summarize the functionality of each set, and highlight them in {\bf bold} for easy reference. The reader can find the values that were used in our searches on \href{https://github.com/vniarchos/BootSTOP}{GitHub}.
\vspace{-.3cm}
\paragraph{\it Neural network hyperparameters:} 
These control the behavior of the neural networks employed in the SAC algorithm. The relevant details on the architecture of the SAC algorithm can be found in \cite{Kantor:2021kbx,Kantor:2021jpz}, \cite{Kantor:thesis} and references therein.
\vspace{-.3cm}
\paragraph{\it Learning loop and automation parameters:} These control how the code handles the quenching of the size of the search windows and the re-initialization of the SAC algorithm during a run. When a run is initialized, the agent moves through the environment and continuously updates the memory buffer, optimizing the SAC neural networks and the stochastic policies of the associated Markov Decision Process. After a number of iterations the reward stops improving and the result saturates. Before quenching the size of the search windows, it is useful to repeat the search anew, with the same search windows but retaining  knowledge of the previous highest reward configuration. During such a re-initialization, the memory buffer is flushed and the agent starts learning from scratch. Typically, this leads to immediate improvement. The parameter controlling  the maximum time spent without improving before re-initialization is called $\displaystyle{\bf faff\_ max}$. Another parameter, $\displaystyle{\bf pc\_ max}$, controls how many re-initializations without improvement in the reward are made before the sizes of the search windows are quenched. Once $\displaystyle{\bf pc\_ max}$ is saturated, the search windows are quenched by some percentage, which is controlled by the $\displaystyle{\bf window\_ rate}$ parameter. The parameter $\displaystyle{\bf max\_ window\_ exp}$ specifies how many window quenches are carried out before the run ends. We note that this automation was missing in \cite{Kantor:2021kbx,Kantor:2021jpz}, where all reported results were the outcome of single runs on a laptop computer, and with the user manually performing the re-initializations and search-window quenches. BootSTOP on the other hand, does not require user supervision and can be run in parallel on a computing cluster.
\vspace{-.3cm}
\paragraph{\it Environment parameters:} This set comprises parameters that specify the environment and how the agent interacts with it. In the former class are parameters like the central charge, which enters explicitly the 6D crossing equations that we are analyzing in this paper. The latter class gives the user control over several functionalities. There are two arrays of Booleans, called $\displaystyle{\bf guessing\_run\_list\_deltas}$ and $\displaystyle{\bf guessing\_run\_list\_opes}$, which specify for each CFT datum if the run starts in ``guessing mode". In this mode, every subsequent value in the search is generated stochastically within a fixed range of values specified by the search windows. In non-guessing mode, instead, every next value is generated stochastically around the previous best result. The latter gives the algorithm the capacity to move dynamically over an arbitrarily large, in principle, region even if the search windows are small. The guessing mode is ideal for blind searches with little prior knowledge of where to look. This mode was applied in the blind-search example of App.~\ref{blind}. The non-guessing mode is suitable for searches focused in a particular region of configuration space. This mode was employed in the adiabatic searches of Sec.\ \ref{adiabatic}. In BootSTOP a search that starts in guessing mode reverts automatically to non-guessing mode after the first $\displaystyle{\bf pc\_ max}$ is reached.    

The initial sizes of the search windows are specified by the arrays $\displaystyle{\bf guess\_sizes\_deltas}$ and $\displaystyle{\bf guess\_sizes\_opes}$. The user can also specify the lower bounds of each CFT datum using two arrays: $\displaystyle{\bf shifts\_ deltas}$ and $\displaystyle{\bf shifts\_ opecoeffs}$. This is useful as an option that enforces the unitarity bounds in unitary CFTs. Additionally, BootSTOP can impose a minimum separation of scaling dimension between operators of the same spin through use of the flag $\displaystyle{\bf same\_ spin\_ hierarchy}$. In this context, the uniform gap is set with the parameter $\displaystyle{\bf dyn\_ shift}$.

\subsection{Statistics, Parallel Runs and the Search for Basins of Attraction}\label{stochastic}

The SAC algorithm is in principle able to navigate continuous search spaces and escape local minima through a carefully balanced explore/exploit strategy. However, for high-dimensional, continuous search spaces, where there can be many local minima and the number of saddles increases exponentially \cite{NIPS2014_17e23e50}, this approach is not feasible in finite time and can be affected by the choice of parameters. For this reason, and in order to improve performance, we took advantage of the automation introduced in the previous subsection, and ran BootSTOP in parallel on a computing cluster.

In stochastic searches the collection of statistics with a large number of independent searches is very important, because every run produces a different result. BootSTOP can easily run in parallel on computing clusters (with each run being completely independent). For instance, the results in App.\ \ref{blind} were obtained using 10k parallel runs. For the adiabatic results in Sec. \ref{adiabatic} we had to perform a rather large number of sequential runs at different values of $c$, which imposed practical restrictions on our running times on the computing cluster. As a result, we ended up performing a parallel run of 200 agents for each value of $c$, which still produced useful statistics.\footnote{These searches were implemented on the Queen Mary University of London High Performance Compute cluster Apocrita \cite{king_thomas_2017_438045}. Each run utilized 1 core and 5GB of memory.}

It is well known that global-search algorithms can perform well at locating the basin of the optimal solution, but are less efficient at identifying the unique, most optimal solution within that basin. The problem becomes harder in high-dimensional searches even when stochastic algorithms are combined with local-search gradient-descent methods. For that reason, we would like to put emphasis on a strategy that focuses primarily on the identification of basins of attraction, and secondarily on the exploration of the small-scale structure within such basins, which can typically be very spiky. 

The results of App.~\ref{blind} illustrate how such a strategy would work for the 6D (2,0) SCFT at $c=25$ in a blind search without any additional, prior, theory-dependent information. In that search, which begins in guessing mode, large arbitrary search windows were chosen for the CFT data (10 for scaling dimensions and 20 for OPE-squared coefficients). The SAC algorithm proves quite effective in identifying basins of attraction in this manner, albeit with significant width. We refer the reader to App. \ref{blind} for a more detailed discussion of this example and how this run performed against independent expectations. We note that the initial search, and the corresponding identification of the basin, can improve by suitable choices of parameters. Subsequent runs in non-guessing modes around the initial statistical average will further improve the reward and decrease the statistical spread of the results. This, however, can be a delicate and time-consuming process without any guarantees that the global minimum will be identified. In general, the quality of the results arising from this process depend on the specifics of the problem and the corresponding complexity of the search. There are also situations, where the global minimum may not be the only minimum of physical interest. The 6D bootstrap problem that we analyze in this paper poses such a situation.

An alternative approach is based on a guided, {\it adiabatic} implementation of SAC starting from the known results for the CFT data in a convenient limit of parameters (in this paper, this is the supergravity limit) and gradually changing the parameters to explore the theory in a general parametric region. In our 6D context, this process involves the adiabatic decrease of the value of the central charge, which appears explicitly in the crossing equation. This procedure can be applied in many different situations. It avoids the pitfalls of having to navigate the full high-dimensional potential, leading to much narrower basins of attraction. We implemented this approach using BootSTOP in Sec.\ \ref{adiabatic} to obtain very promising results in the 6D (2,0) theory exhibiting two separate basins of attraction.

\subsection{Protocols for Adiabatic Searches}
\label{protocols}

\begin{figure}[t]
\centering
\includegraphics[width=13cm, height=6cm]{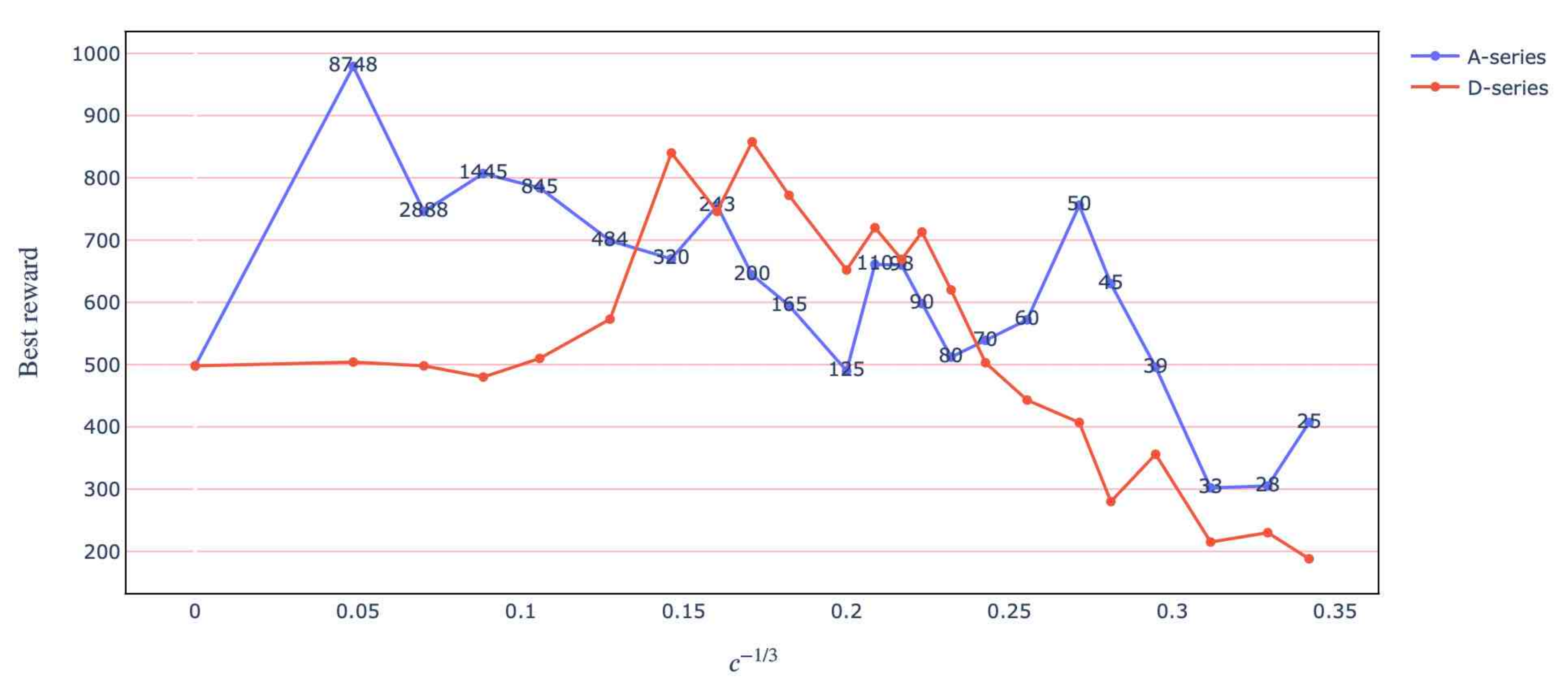}
\caption{A plot of the rewards obtained in the adiabatic searches of Sec.\ \ref{adiabatic} with the two distinct protocols described in Sec.\ \ref{protocols} The blue data are conjectured in Sec.\ \ref{adiabatic} to capture the $A$-series (2,0) SCFTs, while the red data the $D$-series (2,0) SCFTs.}
\label{reward_plot}
\end{figure}

As already noted, in Sec.\ \ref{adiabatic} we present results for adiabatic searches starting from the supergravity limit at large $c$ and gradually flowing towards smaller values up to $c=25$, where one expects to find the $A_1$ (2,0) theory. This theory is expected to be the (2,0) SCFT with the smallest possible value of central charge. Anticipating two distinct sets of solutions (one corresponding to the $A_{N-1}$-series and the other to the $D_N$-series 6D (2,0) SCFTs), we performed the adiabatic search using two distinct protocols. In both protocols, and for each value of $c$, we executed two sequential 8hr runs with 200 parallel agents in non-guessing mode with large $\displaystyle{\bf faff\_ max}$.\footnote{We used the parameters:$\;\displaystyle{\bf faff\_ max}=10000$, $\displaystyle{\bf pc\_ max}=5$ and $\displaystyle{\bf window\_ rate}=0.7$.} This process is especially subtle in the supergravity limit where the two series are expected to be very close. Nevertheless, both protocols worked surprisingly well producing two distinct curves of comparable reward. In Fig.\ \ref{reward_plot} we present the corresponding values of the reward function in each of the two curves. The details of each protocol are as follows. 

\vspace{-.3cm}
\paragraph{\it Protocol 1:} The first protocol is the most natural one in an adiabatic search: after changing the value of $c$ by a small amount, the next run is performed around the statistical average of the previous one.\footnote{The configuration around which a run is performed can be defined in BootSTOP by suitably setting the array ${\displaystyle{\bf global\_ best}}$.} During the first run for each new value of $c$, the $\displaystyle{\bf guess\_ sizes\_ deltas}$ parameter was set uniformly at the value 0.01, while the $\displaystyle{\bf guess\_ sizes\_ opes}$ was set uniformly at the value 0.001. In the second run at each value of $c$, these parameters were set at the 1$\sigma$ standard deviation of the statistics collected during the first run. We conjecture in Sec.\ \ref{adiabatic} that this protocol produces a curve for the $D_N$-series (2,0) SCFTs.
\vspace{-.3cm}
\paragraph{\it Protocol 2:} The second protocol, which is less obvious and more subtle to implement, is motivated by the following observations. The OPE-squared coefficient of the single scalar superconformal primary in the $\mathcal D[0,4]$ multiplet, $\lambda^2_{\mathcal D[0,4]}$, has the distinguishing characteristic that it vanishes exactly in the $A_1$ theory. Therefore, one might suspect that the value of $\lambda^2_{\mathcal D[0,4]}$ is smaller in the $A_{N-1}$-series compared to the $D_N$-series, and that the same happens in the corresponding basins of attraction that represent these solutions in our optimization problem.\footnote{Recall that for the purposes of the single-correlator conformal bootstrap exercise that we are performing, $c$ is viewed as a continuous parameter that can be extended to possibly unphysical values. For each $c$ one is, therefore, looking for both a potential $A$- and $D$-series extremum. In accordance with these expectations, in independent blind searches with the SAC algorithm at low values of $c$, we identified frequently two distinct types of basins---one at a higher value of $\lambda^2_{\mathcal D[0,4]}$ and another at a lower value.} This was, indeed, also the expectation in \cite{Alday:2020tgi}.\footnote{We note that in \cite{Alday:2022ldo} it was conjectured that the bootstrap bounds for the physical values of $c$ are saturated by AdS$_7$ maximal supergravity.} As a result, the second protocol was set up in a manner that would potentially allow us to identify the basin with the smaller value of $\lambda^2_{\mathcal D[0,4]}$. In Sec.\ \ref{adiabatic} we will provide further supporting evidence for the conjecture that the second protocol captures the $A_{N-1}$-series (2,0) SCFTs. The two runs of Protocol 2 involved the same guess sizes as in Protocol 1, but all runs were performed around a point with comparatively high reward and smaller value of $\lambda^2_{\mathcal D[0,4]}$. Prior to each run, this point was selected after the inspection of the full set of configurations obtained during the previous 200 independent runs. It was chosen as the highest-reward configuration in this set that fit the above criteria on $\lambda^2_{\mathcal D[0,4]}$. In many cases, a suitable configuration immediately stood out in the dataset.  
 
\section{Results: Adiabatic Runs from the Supergravity Limit}
\label{adiabatic}

We will now present and analyze the CFT data produced by the SAC algorithm when used to perform adiabatic deformations of the truncation given in Fig.\ \ref{spin_partition}. For this task we employed the two distinct protocols outlined in Sec.\ \ref{protocols} while slowly varying the central charge from the supergravity limit, $c= \infty$, down to $c=25$. This section contains the main results of the paper. 

We begin with a discussion of the CFT data analyzed in Refs\ \cite{Beem:2015aoa,Lemos:2021azv} using standard numerical conformal bootstrap techniques and the OPE inversion formula. The comparison with \cite{Beem:2015aoa,Lemos:2021azv} is useful as a performance check and towards building a consistent interpretation of our results. We conclude with results on CFT data that have not appeared previously in the literature and are new predictions.

\subsection{Lowest Protected and Unprotected Operators}

\begin{figure}[t]
\centering
\includegraphics[width=8.22cm, height=5cm]{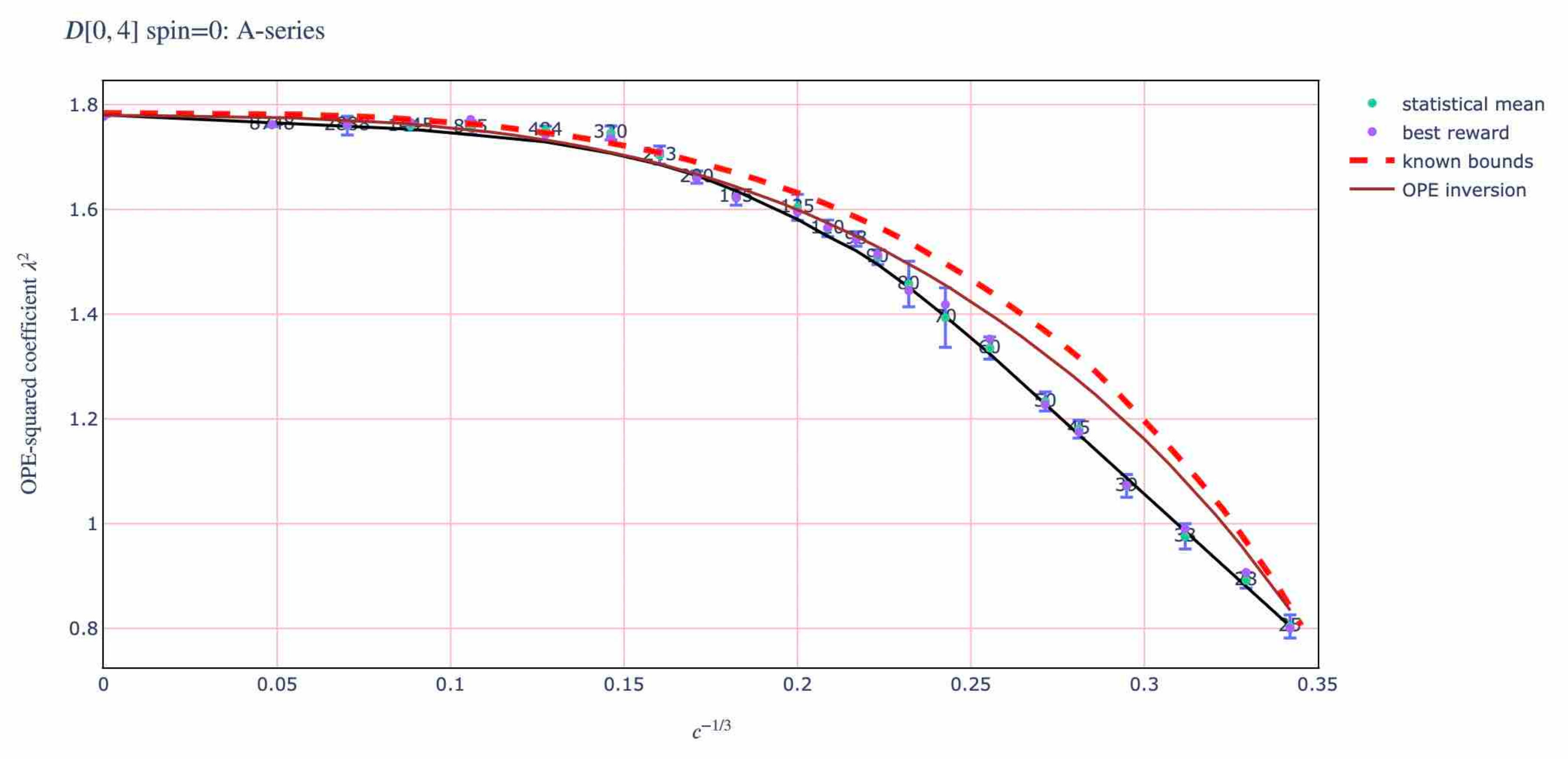}
\includegraphics[width=8.22cm, height=5cm]{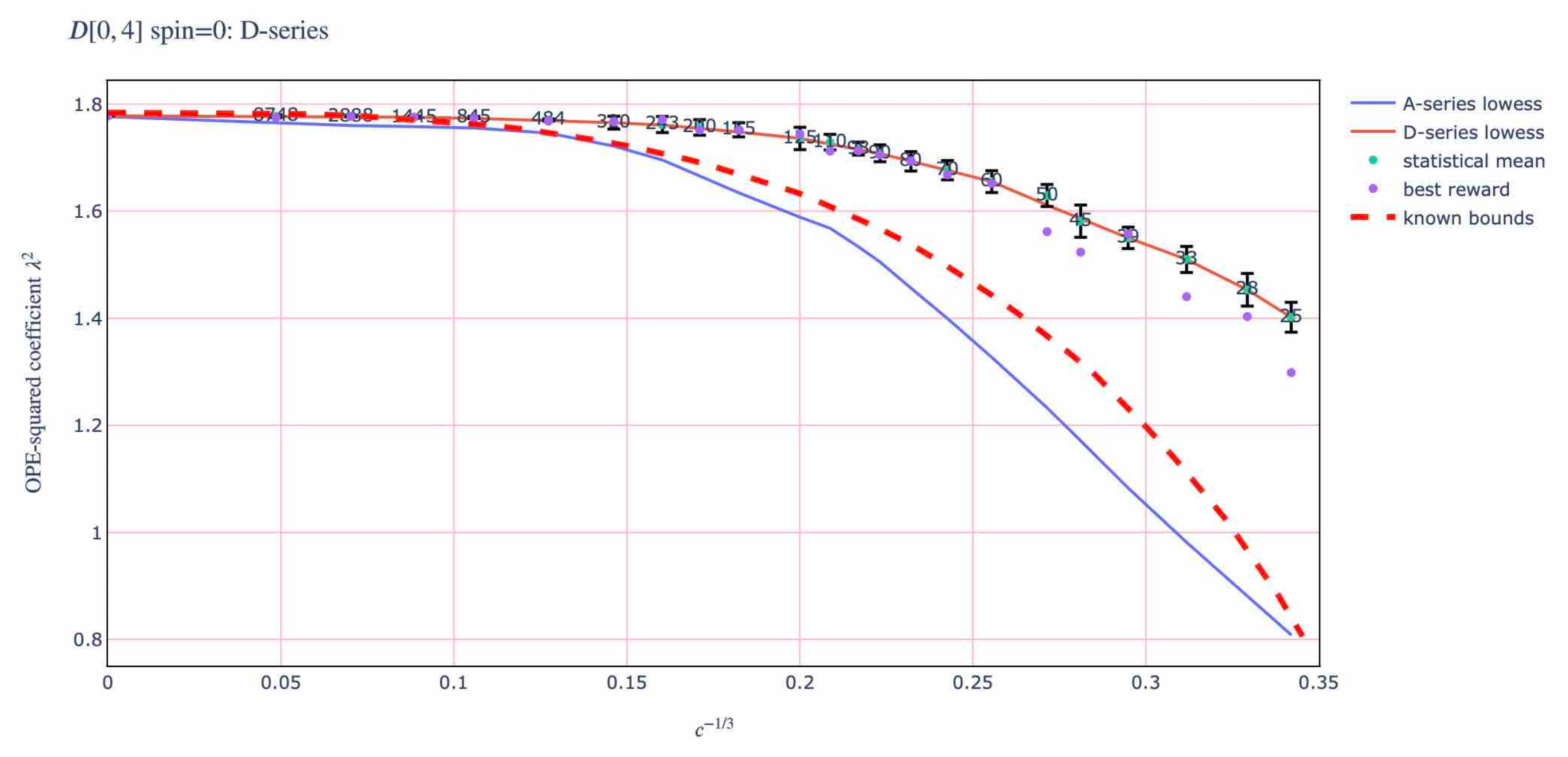}
\caption{Plots of the $A$- and $D$-series curves for the OPE-squared coefficient of the lowest protected operator in the $\mathcal D[0,4]$ multiplet with fixed scaling dimension $\Delta=8$. The dashed red curve represents the numerical conformal bootstrap upper bound obtained in \cite{Beem:2015aoa} and the brown curve on the left plot the long-inverted curve of \cite{Lemos:2021azv}. Regression fits to our data (blue and orange curves) are obtained with the LOWESS method.}
\label{lowest_protected}
\end{figure}

In this subsection we focus on the short $\mathcal D[0,4]$ multiplet, and the first scalar in the long $\mathcal L[0,0]$ multiplet. Let us first discuss the results on the $\mathcal D[0,4]$ multiplet outlined in Fig.\ \ref{lowest_protected}. 

The scaling dimension of the $\mathcal D[0,4]$ multiplet is fixed by superconformal symmetry at the value $\Delta=8$, but the OPE-squared coefficient between this multiplet and the energy-momentum multiplet depends non-trivially on the central charge $c$. The left plot in Fig.\ \ref{lowest_protected} depicts the results obtained with Protocol 2, while the right plot depicts the results obtained with Protocol 1. There are two sets of data points in each plot. The green points represent statistical averages of the 200 parallel runs, and the purple dots the result with the highest reward. The error bars are determined by the 1$\sigma$ standard deviation of the statistics obtained on the second 8hr run at each value of $c$. In most cases, the result of highest reward is inside the error bars. We observe that in both cases the errors are relatively small, which reflects the fact that the 200 parallel agents concentrated naturally inside a small region for the specific datum during the second run. The blue and orange curves are local regression curves produced using the Locally Weighted Scatterplot Smoothing (LOWESS) method.  The dashed red curve represents the numerical bootstrap upper bounds obtained in Ref.\ \cite{Beem:2015aoa}. Anything above this red curve is expected to be inconsistent in an exact (2,0) SCFT.

Comparing with the brown curve obtained using the OPE-inversion formula in \cite{Lemos:2021azv}, we notice a strong similarity with the Protocol-2 curve and the corresponding data points. Both stay closely below the numerical bootstrap bound and the long-inverted value at $c=25$ from \cite{Lemos:2021azv} is also very close to our result, slightly above 0.8.\footnote{The long-inverted-corrected result at $c=25$ quoted in \cite{Lemos:2021azv} is a bit lower, approximately at 0.65.} In summary, similar to \cite{Lemos:2021azv}, we have produced a curve below the numerical bootstrap bound, which has not achieved at this level of approximation the expected analytic result of $\lambda^2_{\mathcal D[0,4]}=0$ at $c=25$. The agreement with \cite{Lemos:2021azv} is our first strong indication that the SAC algorithm with Protocol 2 has identified a sensible approximation of a family of (2,0) SCFTs. Since the values of the blue curve are consistently below the ones of the orange curve (in the right plot of Fig.\ \ref{lowest_protected}) we are tempted to conjecture (using intuition from the known supergravity results \cite{Heslop:2004du, Alday:2020tgi} and the related discussion in \cite{Alday:2020tgi}) that the blue curve represents the $A$-series (2,0) theories and the orange curve the $D$-series (2,0) theories. But would such an interpretation be consistent with the existing expectations about the behavior of the $D$-series curve?

At first sight, the data-points on the right plot (fitted by the orange LOWESS curve) appear to behave rather badly compared to the numerical bootstrap bounds. They remain close to the bounds up to roughly $c^{-1/3} \sim 0.11$ and then gradually start diverging significantly from the bound-curve. The error bars are also increasing towards the small $c$ region and for the last few points close to $c=25$ the highest-reward values are significantly outside the 1$\sigma$ error bars. In other words, several features of these data demonstrate that the quality of these results deteriorates at lower values of $c$. The same qualitative feature is also present in the values of the best reward as a function of $c$, as depicted in Fig.\ \ref{reward_plot}. In that figure the Protocol 1 curve is the red curve. There is an apparent peak around $c^{-1/3}\sim 0.15$ beyond which the reward gradually falls. Interestingly, the smallest central charge for a non-trivial (2,0) SCFT in the $D$-series is $c=676$, the central charge of the $D_4$ theory. The $D_3$ theory (at $c=243$) is dual to the $A_3$ theory, while the $D_2$ theory (at $c=50$) is equivalent to the $A_1 \times A_1$ theory. The central charge of the $D_4$ theory has $c^{-1/3} \sim 0.11$, which is tantalizingly close to the region of 0.11 where we notice the deviation of the orange curve from the bootstrap bound curve. Hence, if we were to terminate the orange curve at $c=676$ (as an exact analysis of the $D$-series theories would suggest), the remaining set of data would be satisfyingly close to the numerical bootstrap bounds. Similar observations can be made in all the other CFT data that we have collected, reinforcing the conjecture that the orange curve is consistent with the $D$-series interpretation. 

\begin{figure}[t]
\centering
\includegraphics[width=8.22cm, height=5cm]{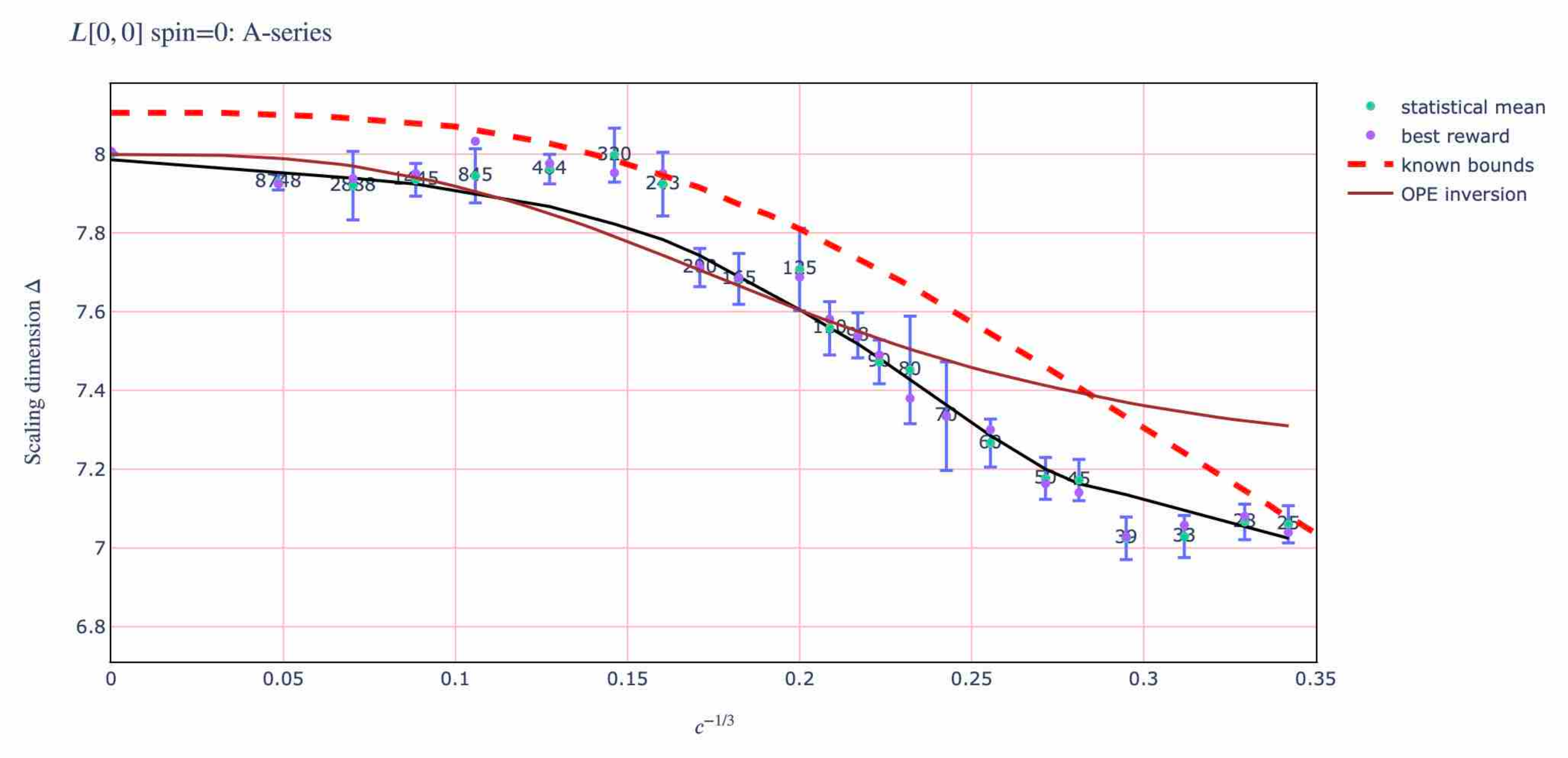}
\includegraphics[width=8.22cm, height=5cm]{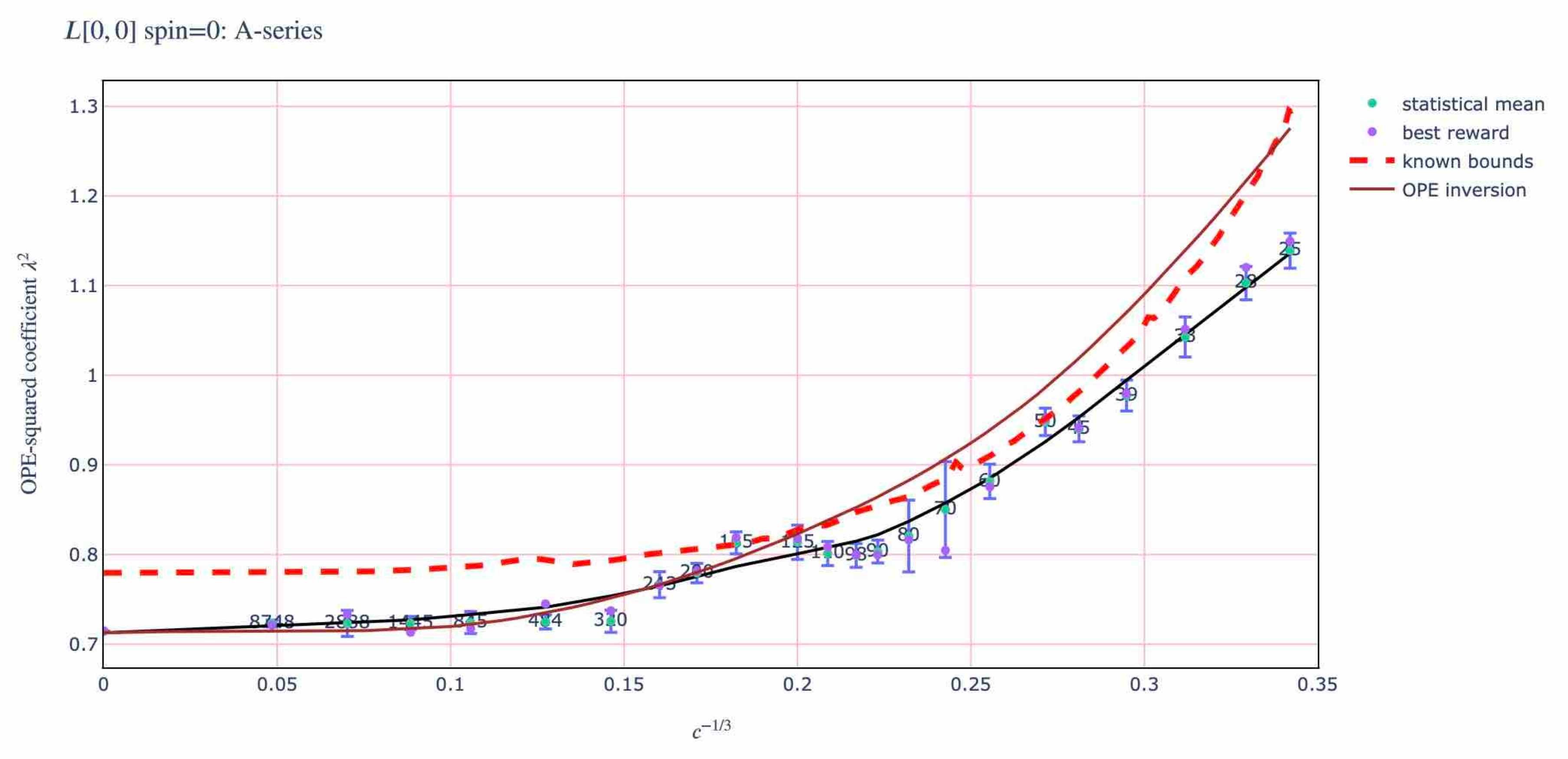} \\
\includegraphics[width=8.22cm, height=5cm]{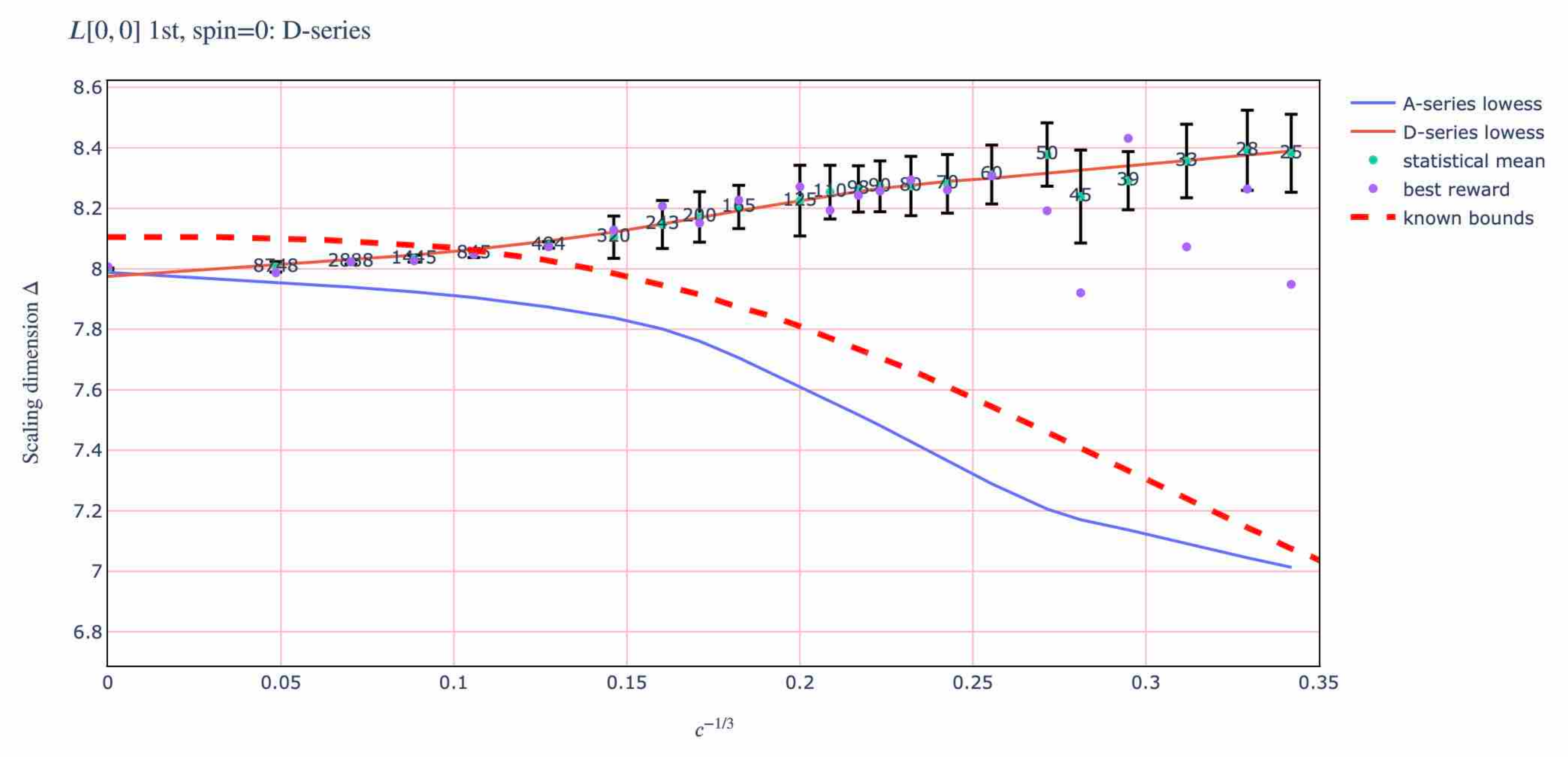}
\includegraphics[width=8.22cm, height=5cm]{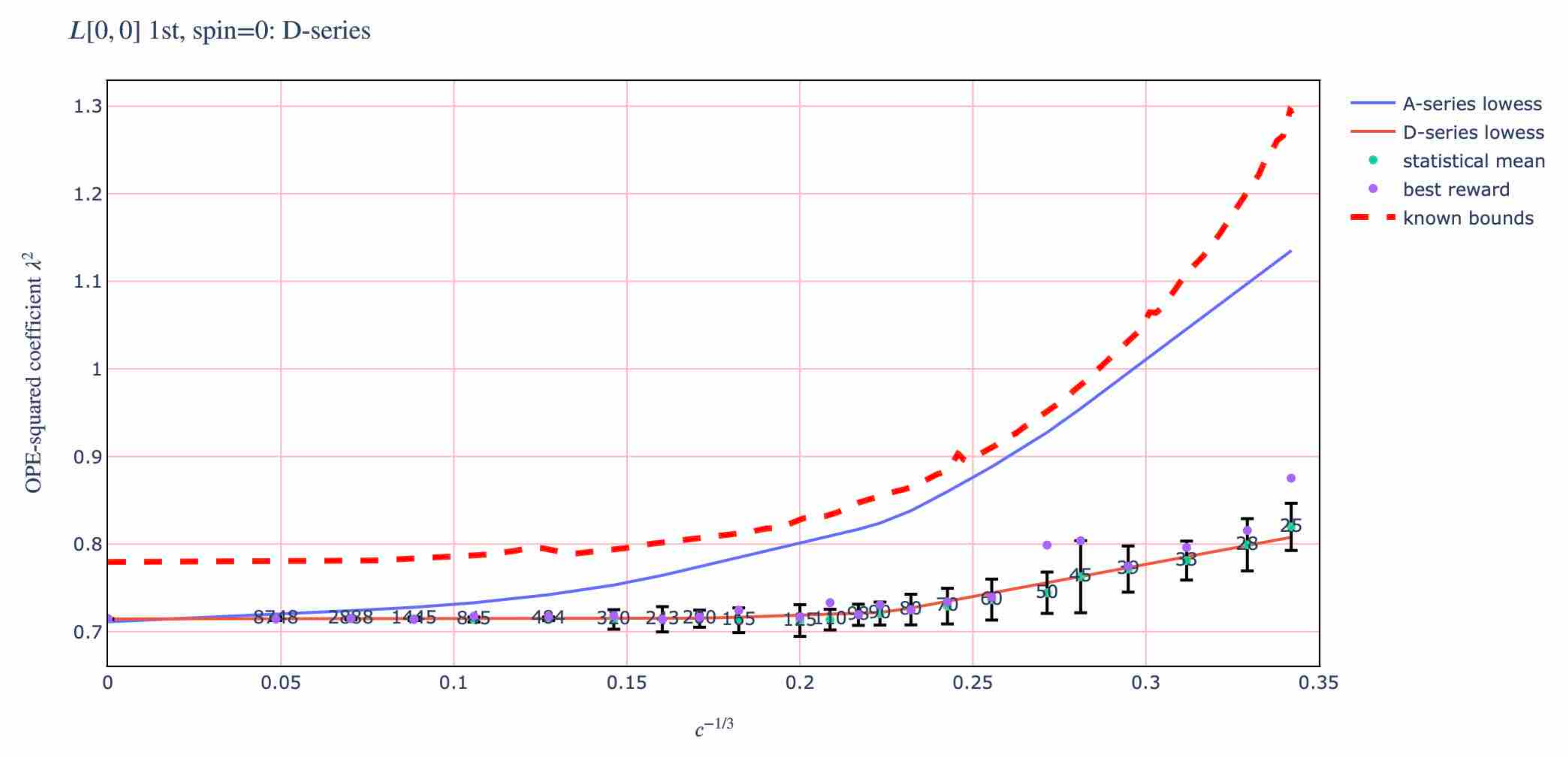}
\caption{Plots of the $A$- and $D$-series curves for the scaling dimension and OPE-squared coefficient of the lowest non-protected operator in the $\mathcal L[0,0]$ multiplet. The plots on the left depict the scaling-dimension results and the plots on the right the OPE-squared coefficient results. The brown curves on the top two plots exhibit the long-inverted curves of \cite{Lemos:2021azv}.}
\label{lowest_unprotected}
\end{figure}

Consider next the results on the CFT data for the lowest $\mathcal L[0,0]$ multiplet, which are displayed in Fig.\ \ref{lowest_unprotected}. The two top-row plots depict the Protocol 2 (conjectured $A$-series) results for the scaling dimension and OPE-squared coefficient, while the two bottom-row plots the corresponding Protocol 1 (conjectured $D$-series) results. Repeating the cautionary comment of Footnote 27 in Ref.\ \cite{Lemos:2021azv}, we point out that the dashed-red numerical bootstrap bounds for the OPE-squared coefficient on the right plots were obtained under the assumption that the corresponding conformal dimensions saturate their own bounds, which is not directly applicable here. As in \cite{Lemos:2021azv}, we depict these bounds under the expectation that they are a decent estimate to the actual bounds.

The plots in Fig.\ \ref{lowest_unprotected} reveal a picture consistent with the observations of the previous paragraphs. The $A$-series data are consistently below the numerical bootstrap bounds for all values of $c$ both for the scaling dimension and OPE-squared coefficient. In fact, when compared to the brown curve obtained using the OPE inversion formula  in \cite{Lemos:2021azv}, we see that our results are behaving slightly better in relation to the known numerical bootstrap bounds. For the $D$-series data we observe that they exhibit the same qualitative features that were noted in the context of the $\mathcal D[0,4]$ multiplet above. Their quality deteriorates beyond the region of $c^{-1/3}\sim 0.15$, and a violation of the numerical bootstrap bounds for the scaling dimension starts in the vicinity of $c^{-1/3}\sim 0.11$. Hence, if one were to terminate the curve at $c=676$, the agreement with the bounds would be very good both for the scaling dimension and OPE-squared coefficient.

We believe that all these results, collectively, are very suggestive for the efficiency of the SAC algorithm in this particular context, and the conjectured interpretation of the results.

\subsection{Comparison with the Remaining Previously Analyzed CFT Data}

We continue with the presentation of results for higher-spin operators for which certain predictions have already been made in \cite{Beem:2015aoa,Lemos:2021azv}. More specifically, we discuss scaling dimensions and OPE-squared coefficients for long $\mathcal L[0,0]$ multiplets at spin $\ell =2,4,6$ and OPE-squared coefficients for short $\mathcal B[0,2]$ multiplets at spin $\ell=1,3,5$. For the depicted numerical bootstrap bounds of the OPE-squared coefficients of the long multiplets the comment of Footnote 27 in \cite{Lemos:2021azv} still applies.

\subsubsection{$\mathcal L[0,0]$ for Spin $\ell=2,4,6$}

\begin{figure}[t]
\centering
\includegraphics[width=8.22cm, height=5cm]{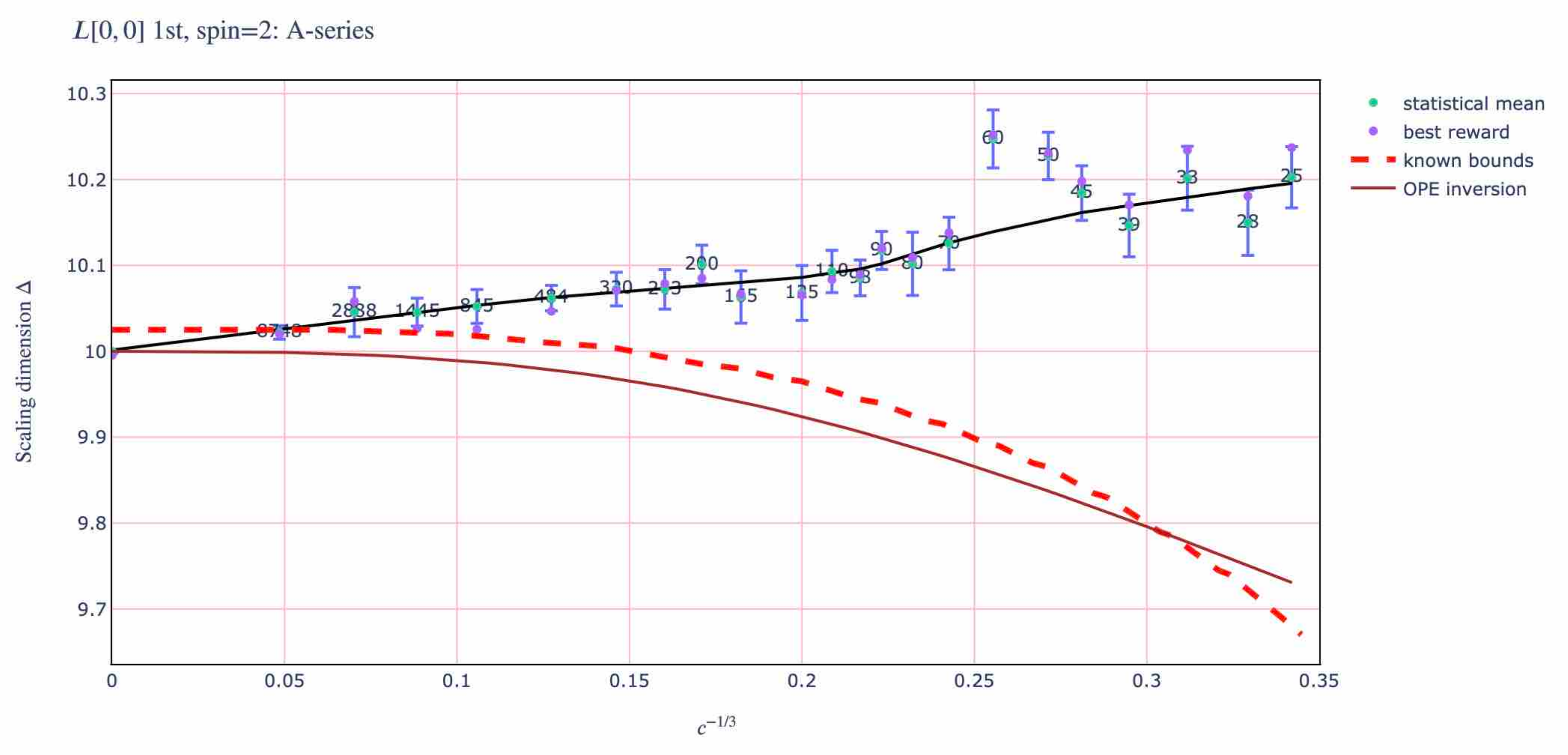}
\includegraphics[width=8.22cm, height=5cm]{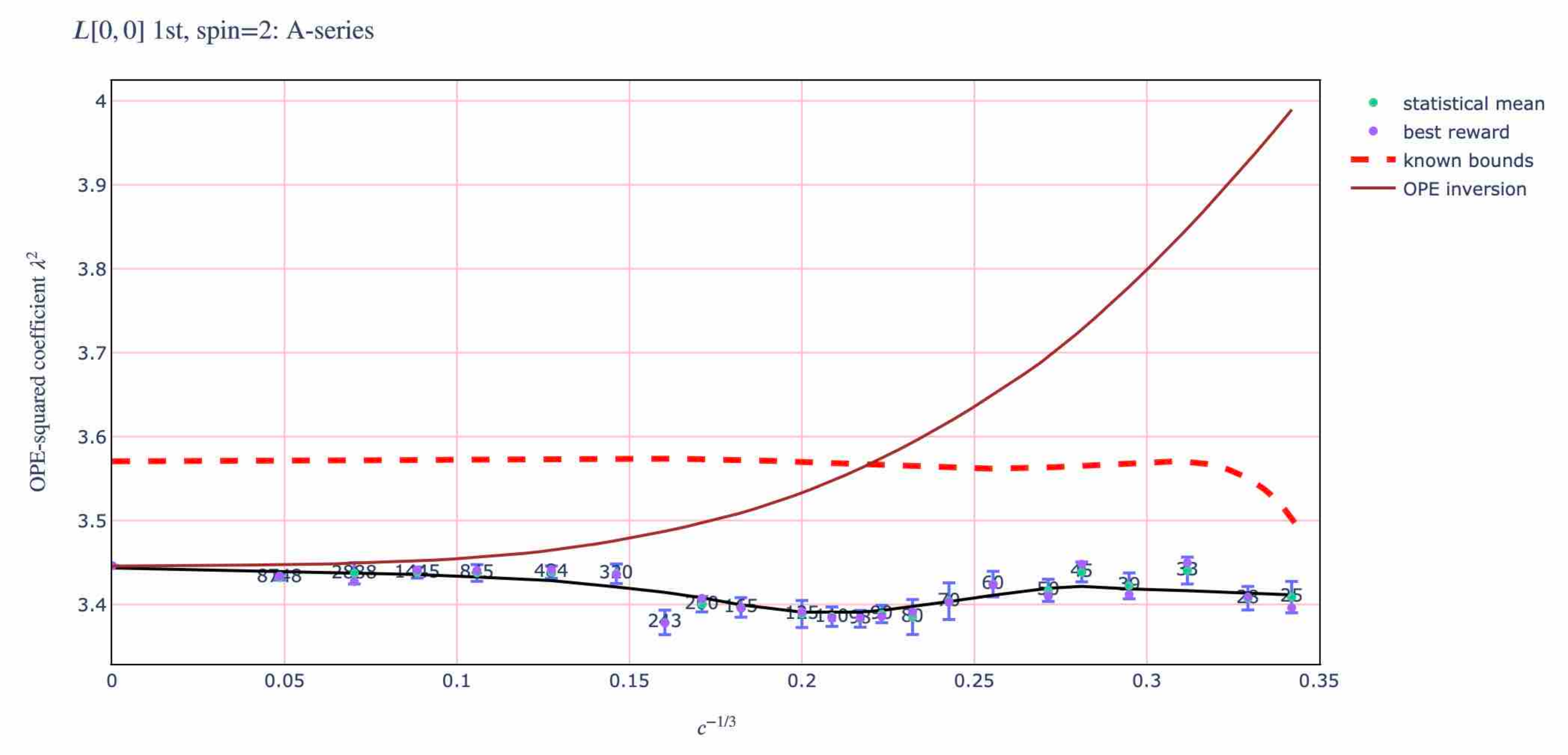} \\
\includegraphics[width=8.22cm, height=5cm]{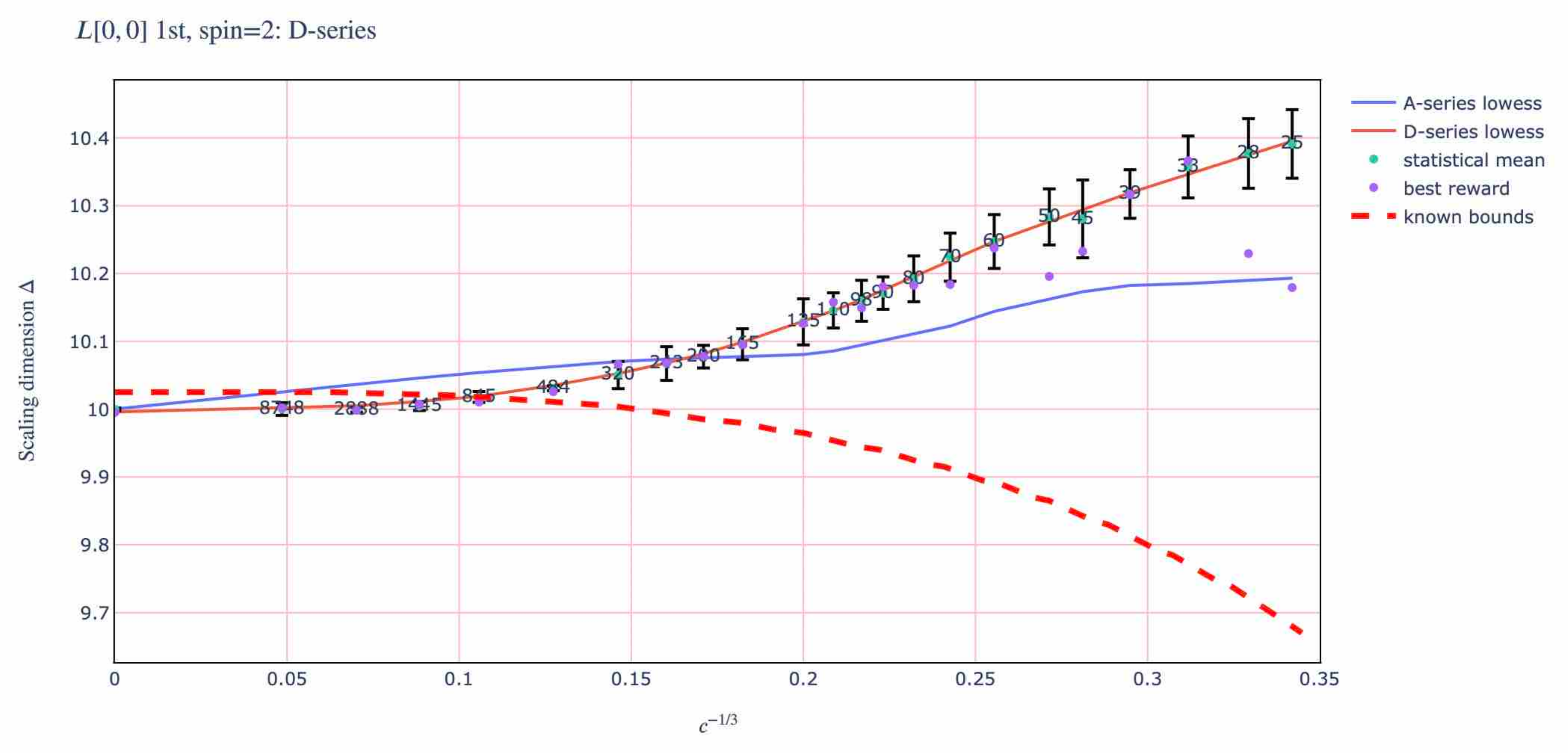}
\includegraphics[width=8.22cm, height=5cm]{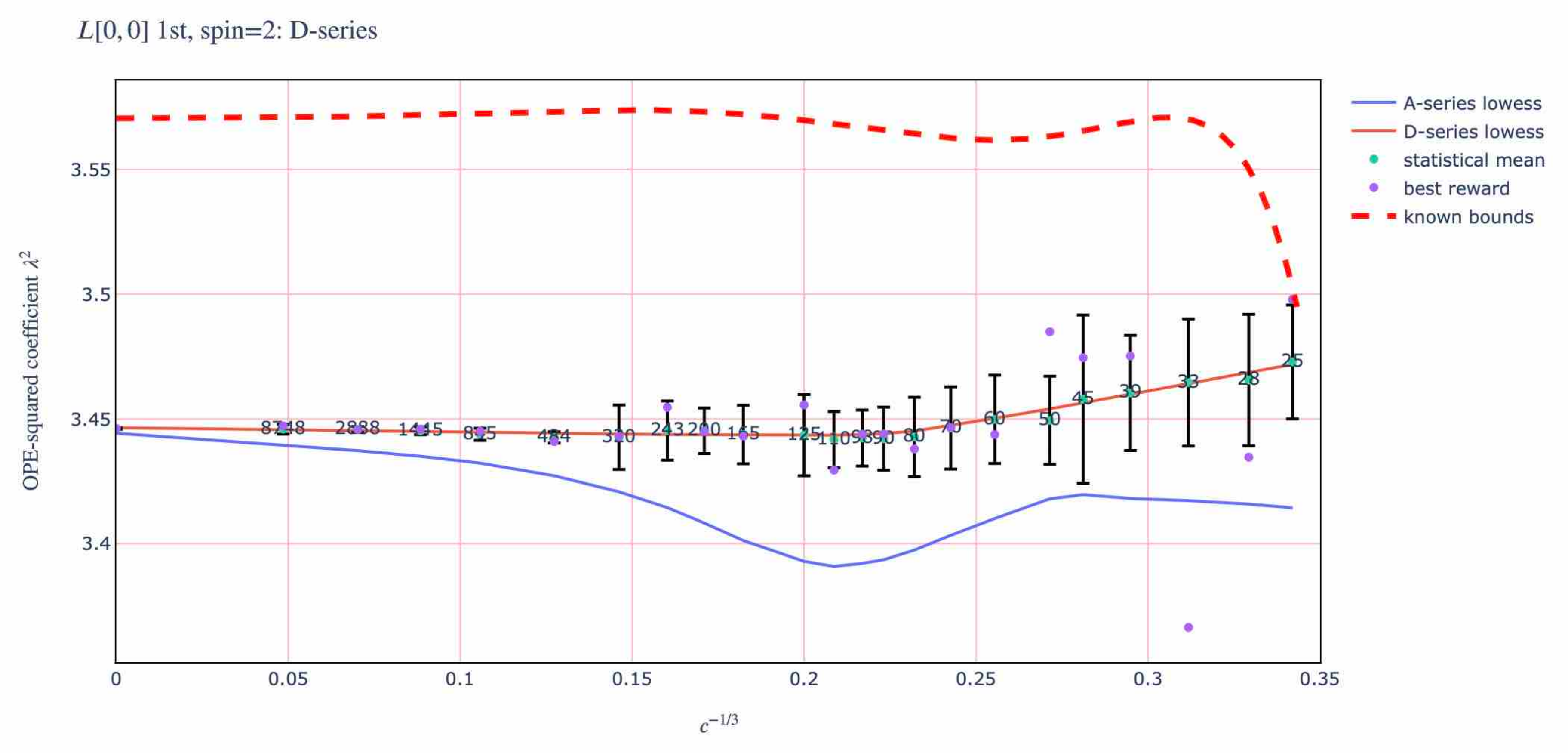}
\caption{Plots of the $A$- and $D$-series curves for the scaling dimension and OPE-squared coefficient of the leading non-protected spin-2 operators in the $\mathcal L[0,0]$ multiplet. The plots on the left depict the scaling-dimension results and the plots on the right the OPE-squared coefficient results. The brown curves on the top two plots exhibit the long-inverted curves in \cite{Lemos:2021azv}.}
\label{spin2_unprotected}
\end{figure}

Based on the above interpretation, the $A$- and $D$-series data of the leading spin-2 operators in the $\mathcal L[0,0]$ multiplet appear in Fig.\ \ref{spin2_unprotected}. The $A$-series curve for the scaling dimension (top-left plot) violates the known numerical bootstrap bounds for most values of $c$. It behaves significantly worse than the corresponding OPE inversion curve of Ref.\ \cite{Lemos:2021azv}, which also violates the numerical bootstrap bounds, but with a violation that occurs at much smaller values of $c$. For the OPE-squared coefficient (right-top plot) the situation is reversed. The (indicative) bootstrap bounds are respected throughout our $A$-series curve, but are violated significantly by the corresponding OPE inversion curve of Ref.\ \cite{Lemos:2021azv}. For the $D$-series scaling-dimension curve (bottom-left plot) we observe the same effect as in previous data: a significant violation occurs only for values of $c^{-1/3}$ above $\sim 0.11$. The OPE-squared curve (bottom-right plot) remains everywhere below the bound.

\begin{figure}[t]
\centering
\includegraphics[width=8.2cm, height=5cm]{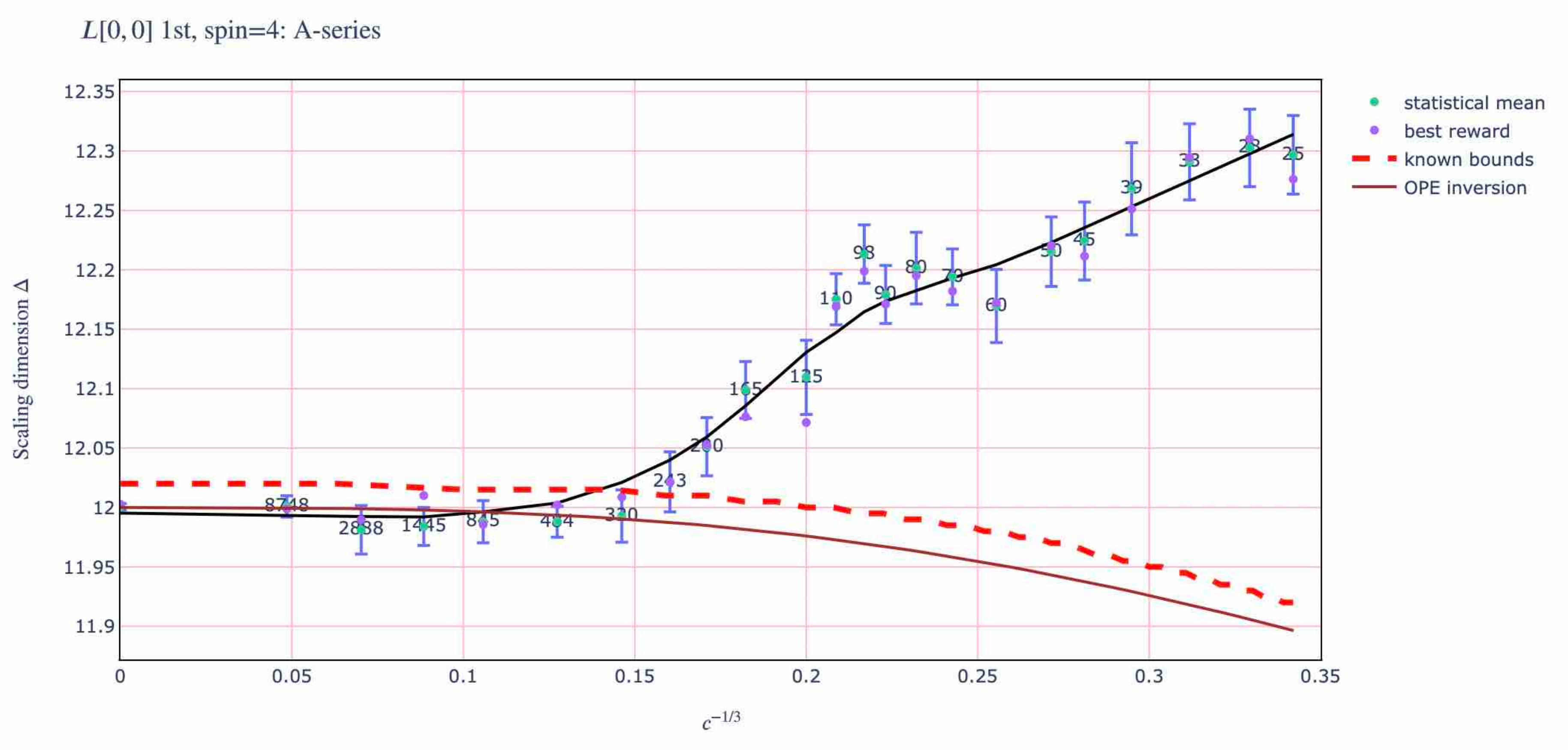}
\includegraphics[width=8.2cm, height=5cm]{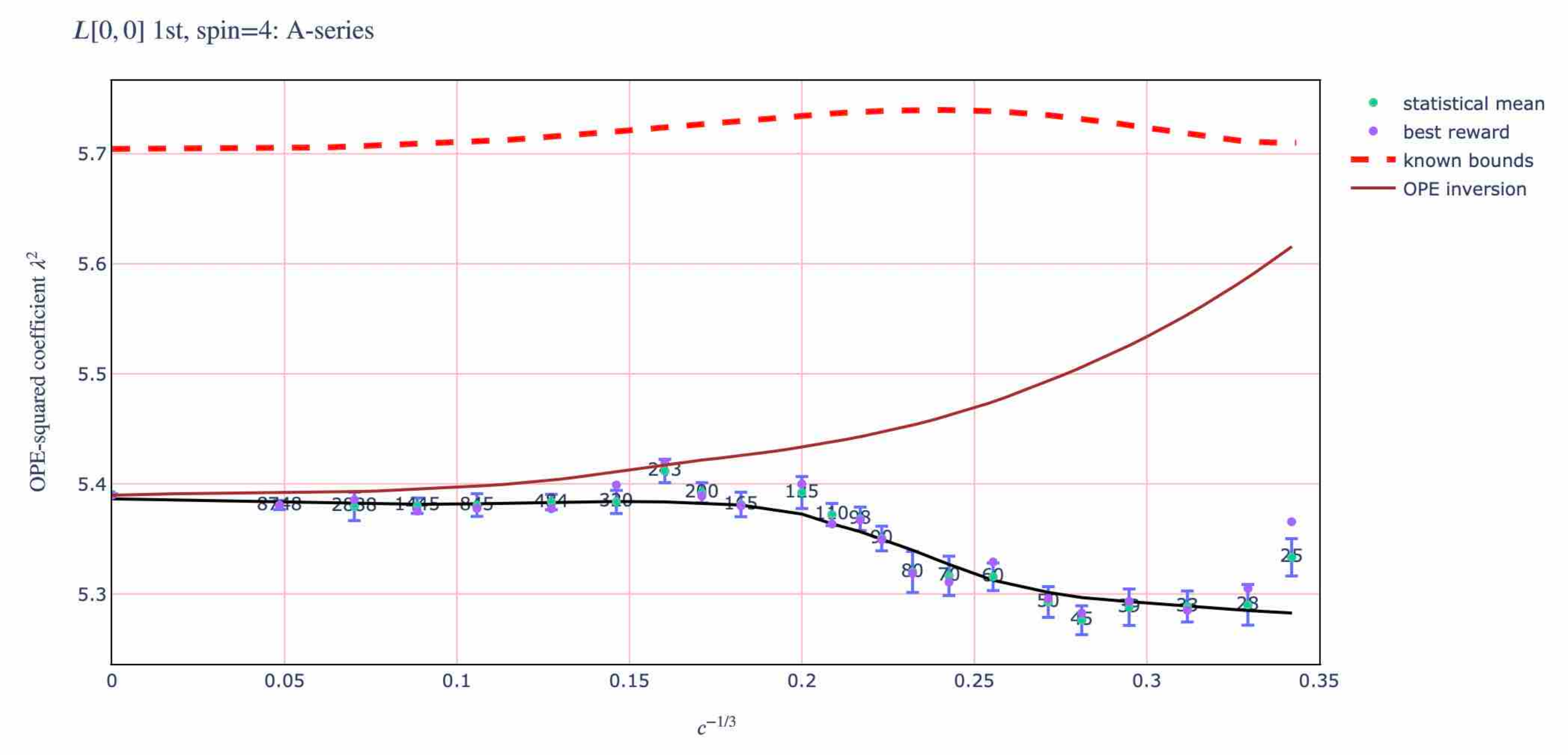} \\
\includegraphics[width=8.2cm, height=5cm]{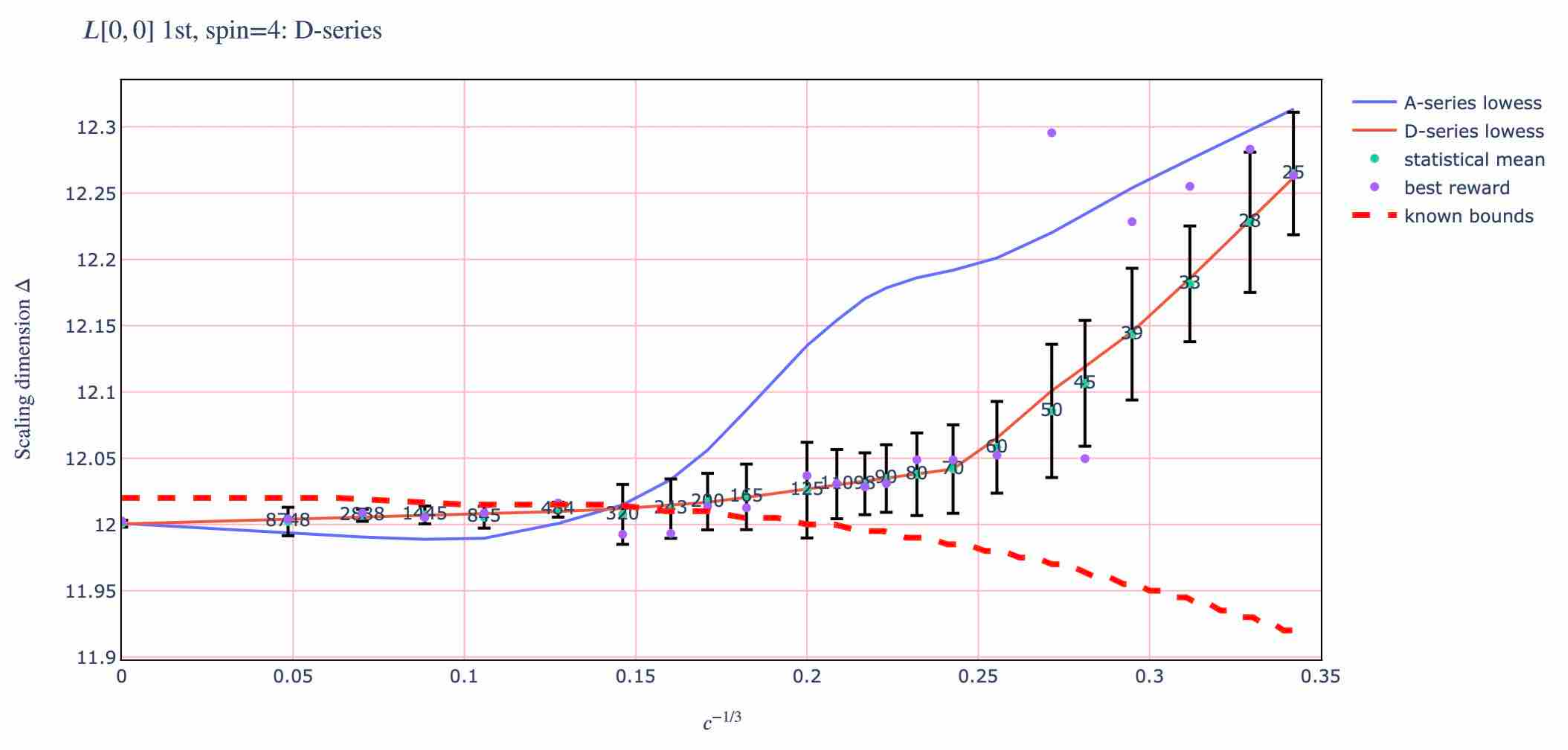}
\includegraphics[width=8.2cm, height=5cm]{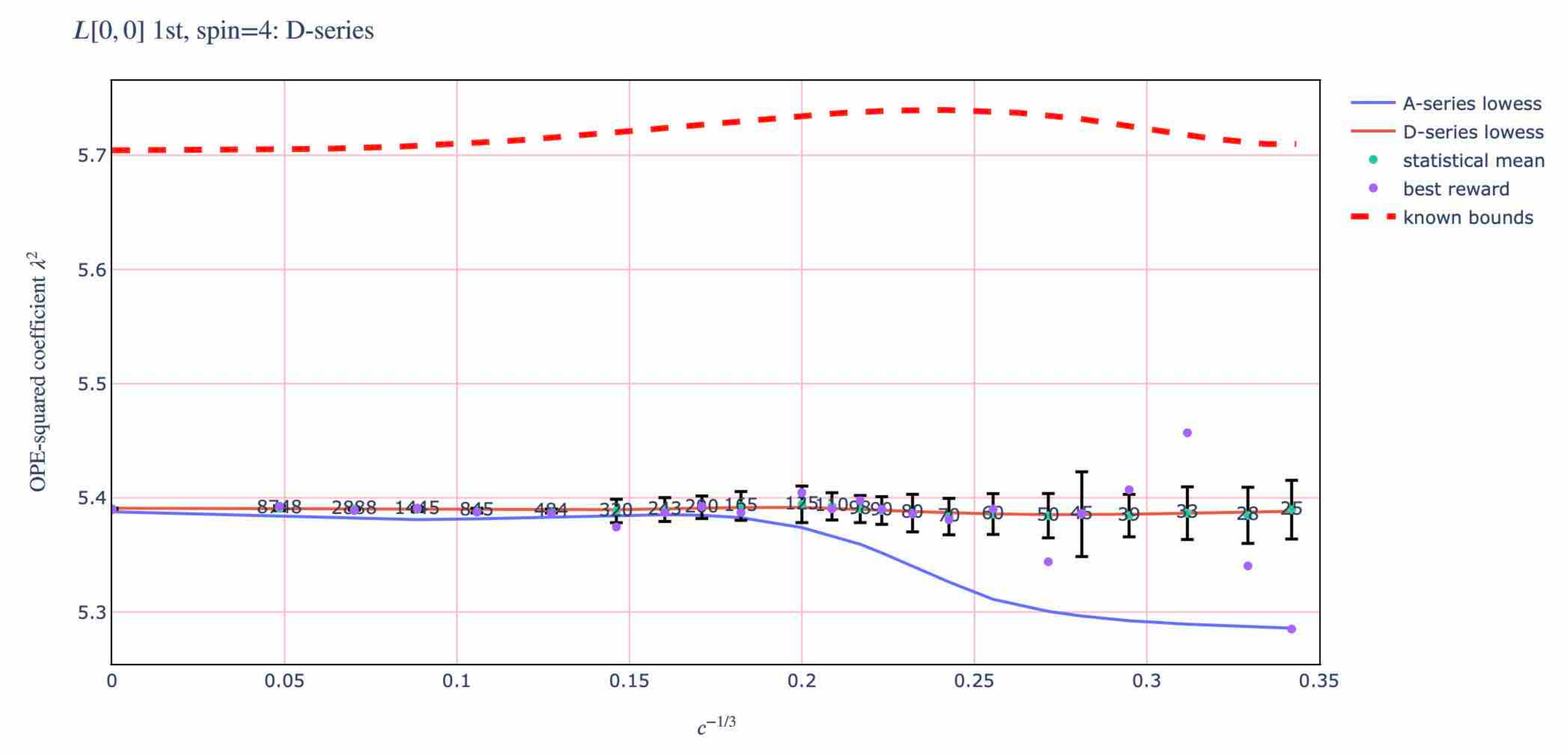}
\caption{Plots of the $A$- and $D$-series curves for the scaling dimension and OPE-squared coefficient of the leading non-protected spin-4 operators in the $\mathcal L[0,0]$ multiplet. The plots on the left depict the scaling-dimension results and the plots on the right the OPE-squared coefficient results. The brown curves on the top two plots exhibit the long-inverted curves in \cite{Lemos:2021azv}.}
\label{spin4_unprotected}
\end{figure}

The corresponding data of the leading spin-4 operators in the $\mathcal L[0,0]$ multiplet exhibit comparable features (see Fig.\ \ref{spin4_unprotected}). The $A$-series curve for the scaling dimension (left-top plot) violates the bound significantly for low enough values of $c$, but the $D$-series scaling dimensions do not, if the curve is terminated around $c^{-1/3}\sim 0.11$. In comparison, the corresponding OPE-inversion scaling-dimension curve in \cite{Lemos:2021azv} (see brown curve in the left-top plot) does not violate the bootstrap bounds. Finally, both of our OPE-squared curves are well within the (indicative) allowed regions at slightly lower values compared to the corresponding brown OPE-inversion curve (see brown curve in the right-top plot).

\begin{figure}[t]
\centering
\includegraphics[width=8.22cm, height=5cm]{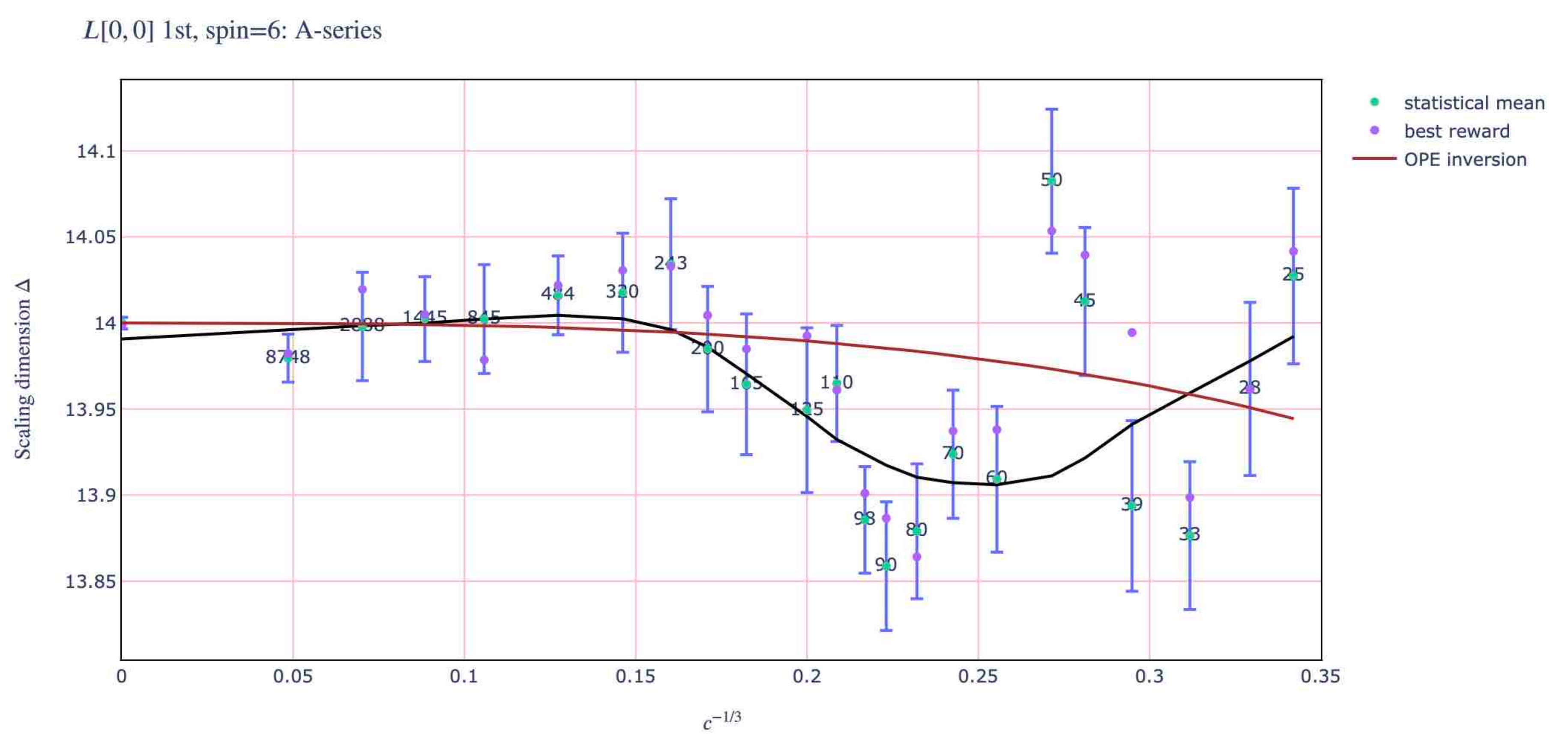}
\includegraphics[width=8.22cm, height=5cm]{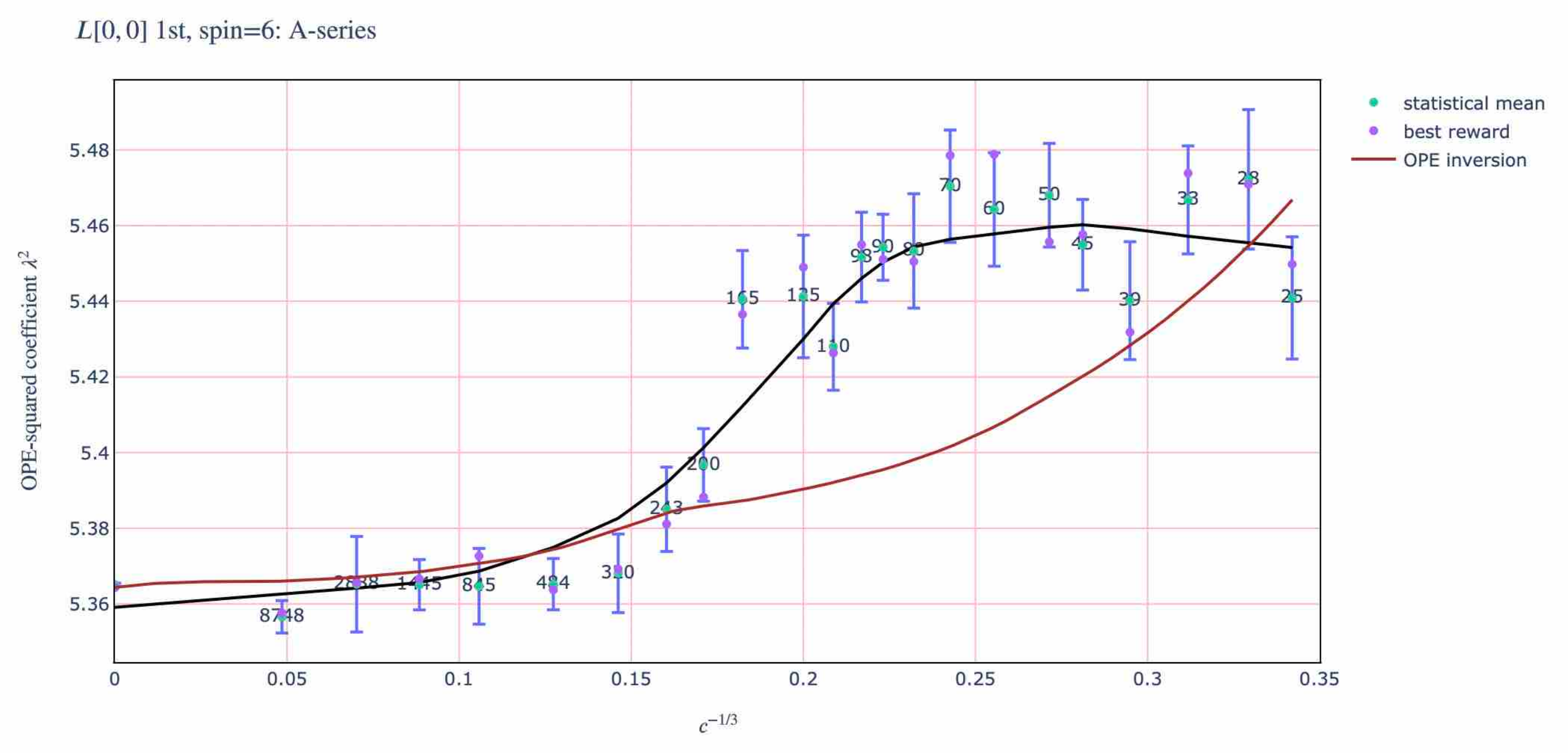} \\
\includegraphics[width=8.22cm, height=5cm]{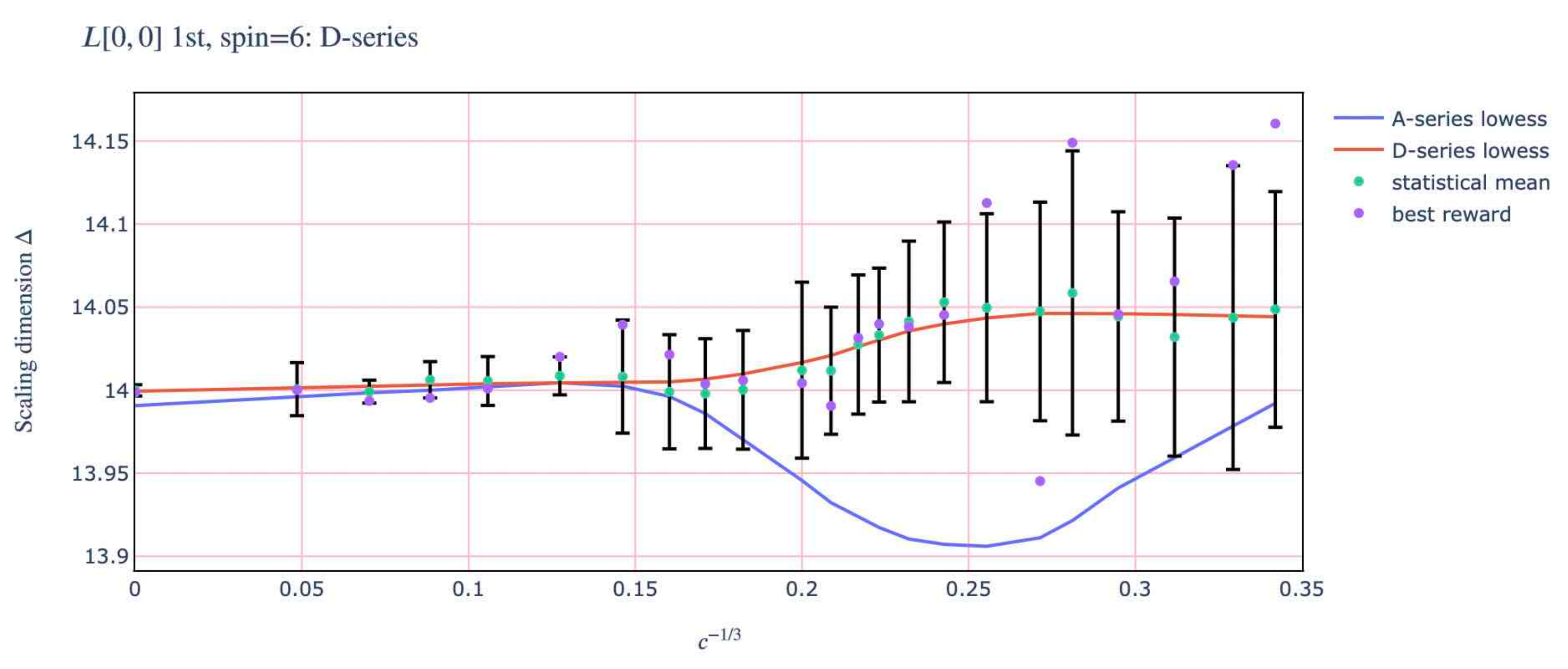}
\includegraphics[width=8.22cm, height=5cm]{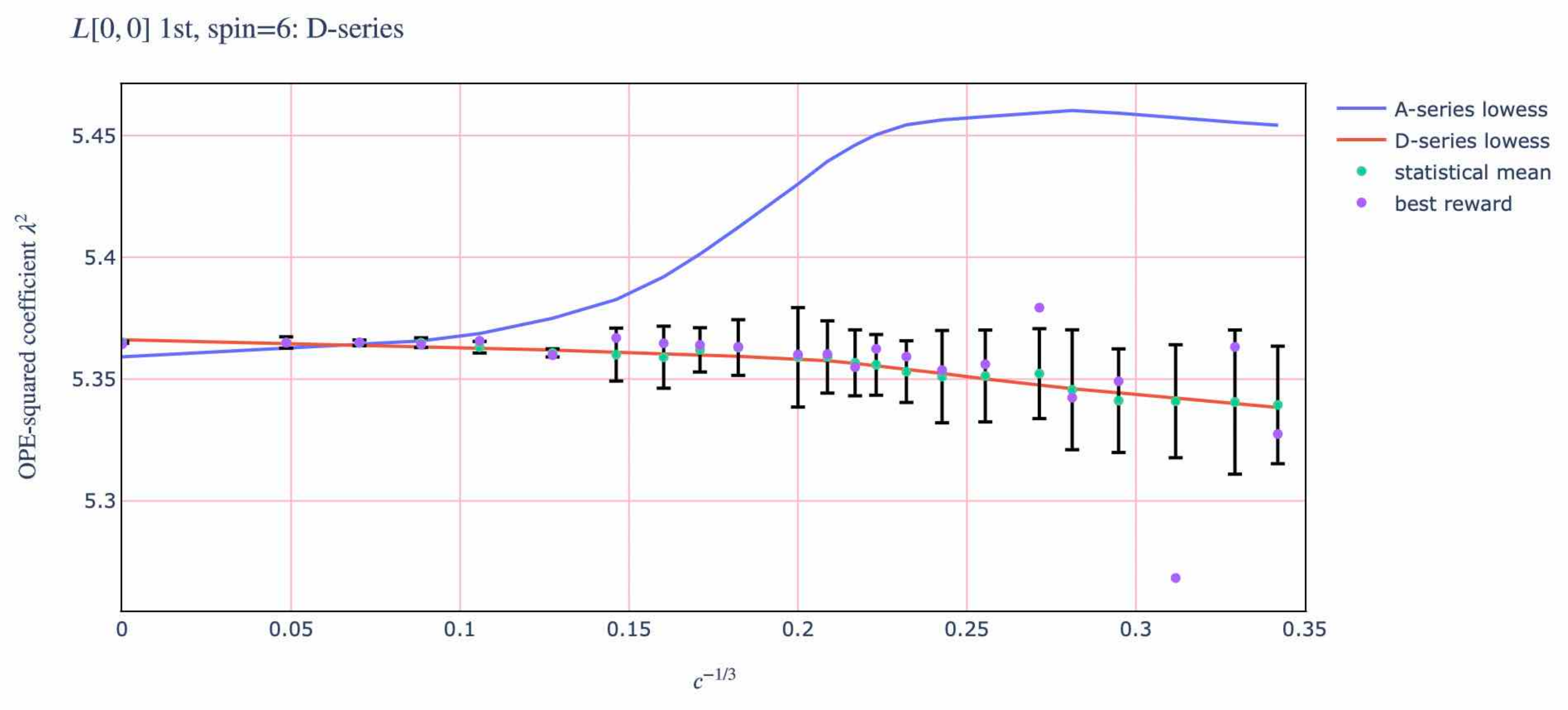}
\caption{Plots of the $A$- and $D$-series curves for the scaling dimension and OPE-squared coefficient of the leading non-protected spin-6 operators in the $\mathcal L[0,0]$ multiplet. The plots on the left depict the scaling-dimension results and the plots on the right the OPE-squared coefficient results. The brown curves on the top two plots exhibit the long-inverted curves in \cite{Lemos:2021azv}.}
\label{spin6_unprotected}
\end{figure}

For the leading spin-6 operators in the $\mathcal L[0,0]$ multiplet we are not aware of any published numerical bootstrap bounds. Our $A$- and $D$-series results appear in Fig.\ \ref{spin6_unprotected}. Compared with the corresponding curves derived using the OPE-inversion formula, our $A$-series data evolve in the same neighborhood of values (with visible differences in the shape of the curves). The scaling dimensions are in the neighborhood of 13.9 and 14 and the OPE-squared coefficients in the neighborhood 5.35 and 5.45. Once again, we observe that the error bars of the $D$-series curves exhibit a visible increase around the point $c^{-1/3}\sim 0.15$.

\subsubsection{$\mathcal B[0,2]$ for Spin $\ell=1,3,5$}

\begin{figure}[t!]
\centering
\includegraphics[width=8.22cm, height=5.3cm]{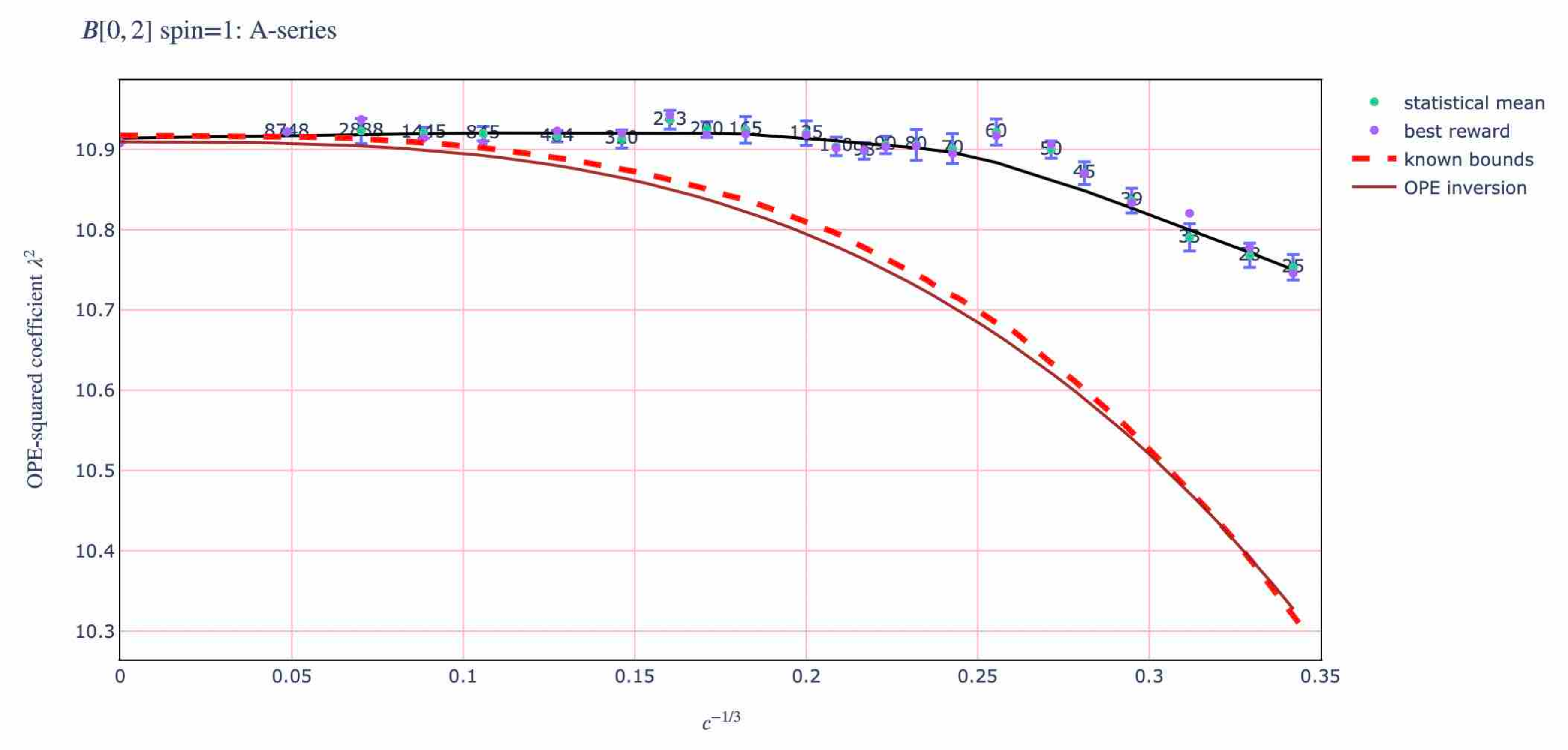}
\includegraphics[width=8.22cm, height=5.3cm]{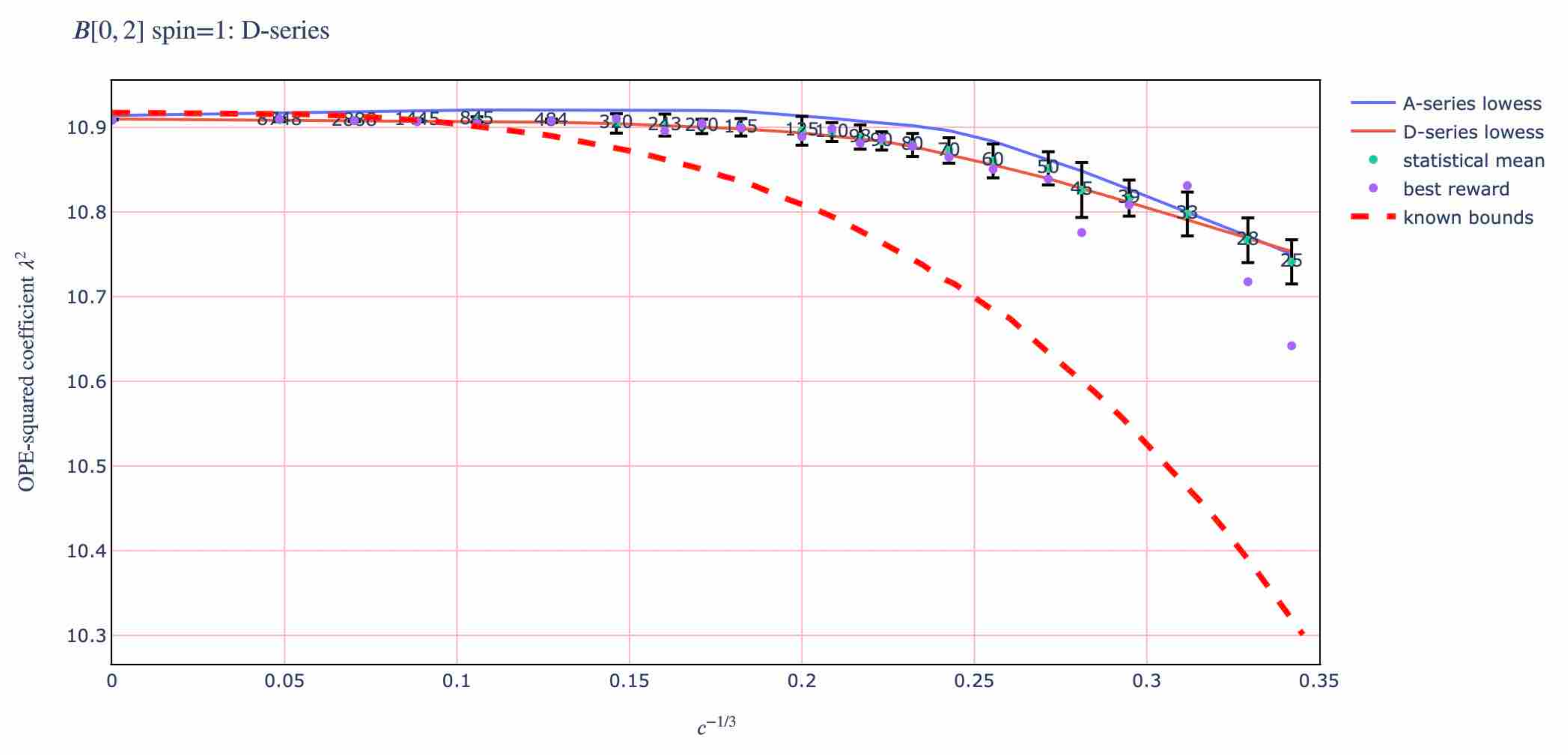}
\includegraphics[width=8.22cm, height=5.3cm]{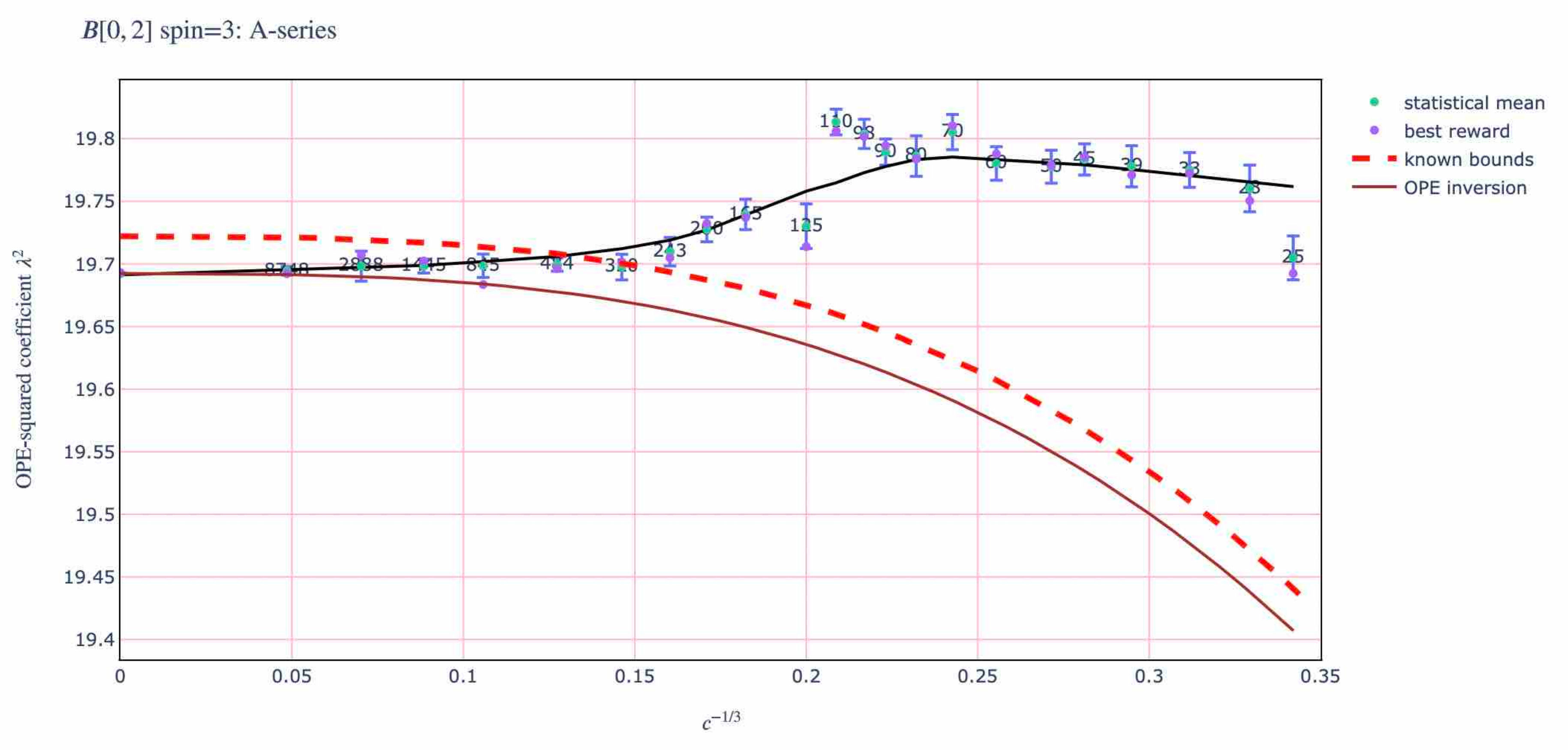}
\includegraphics[width=8.22cm, height=5.3cm]{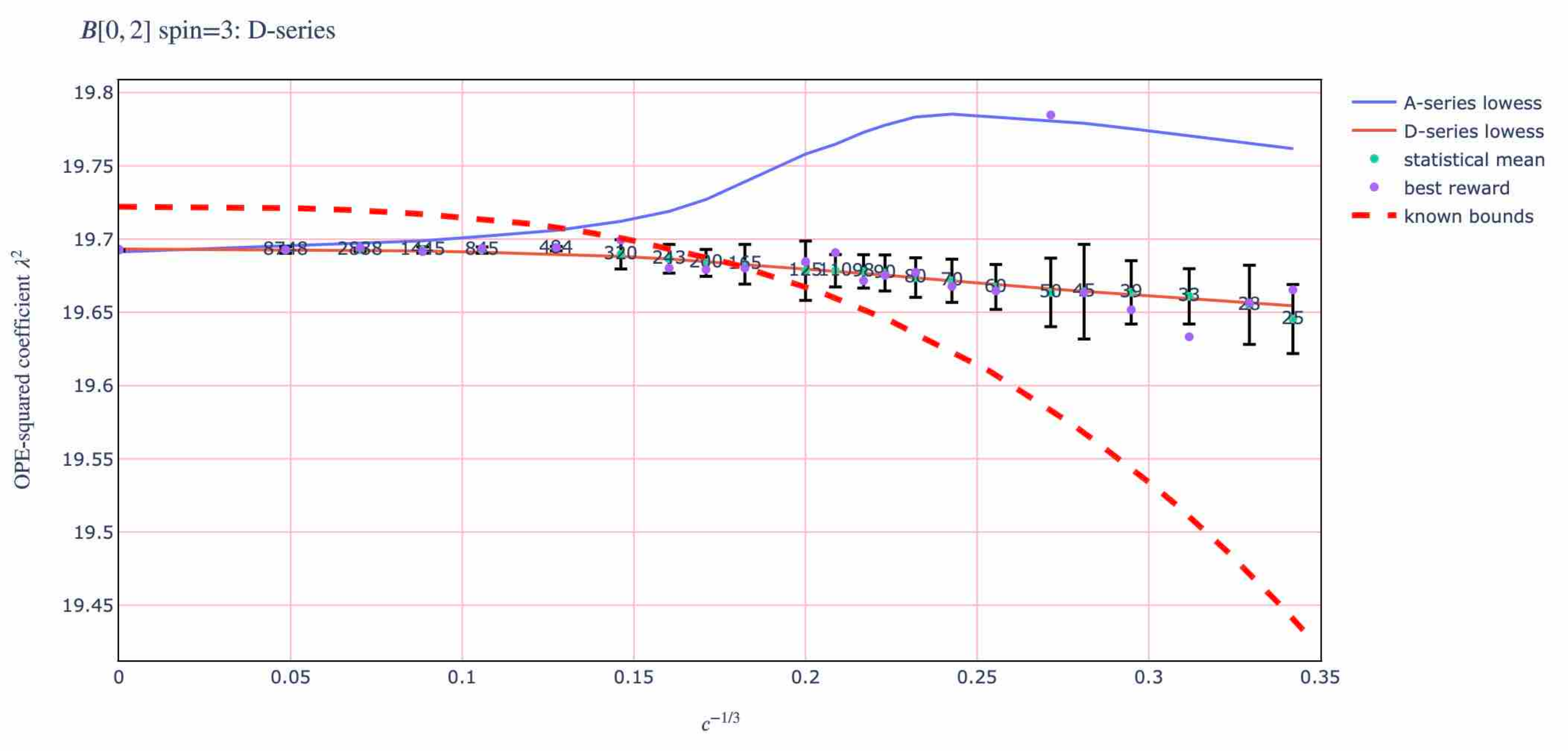}
\includegraphics[width=8.22cm, height=5.3cm]{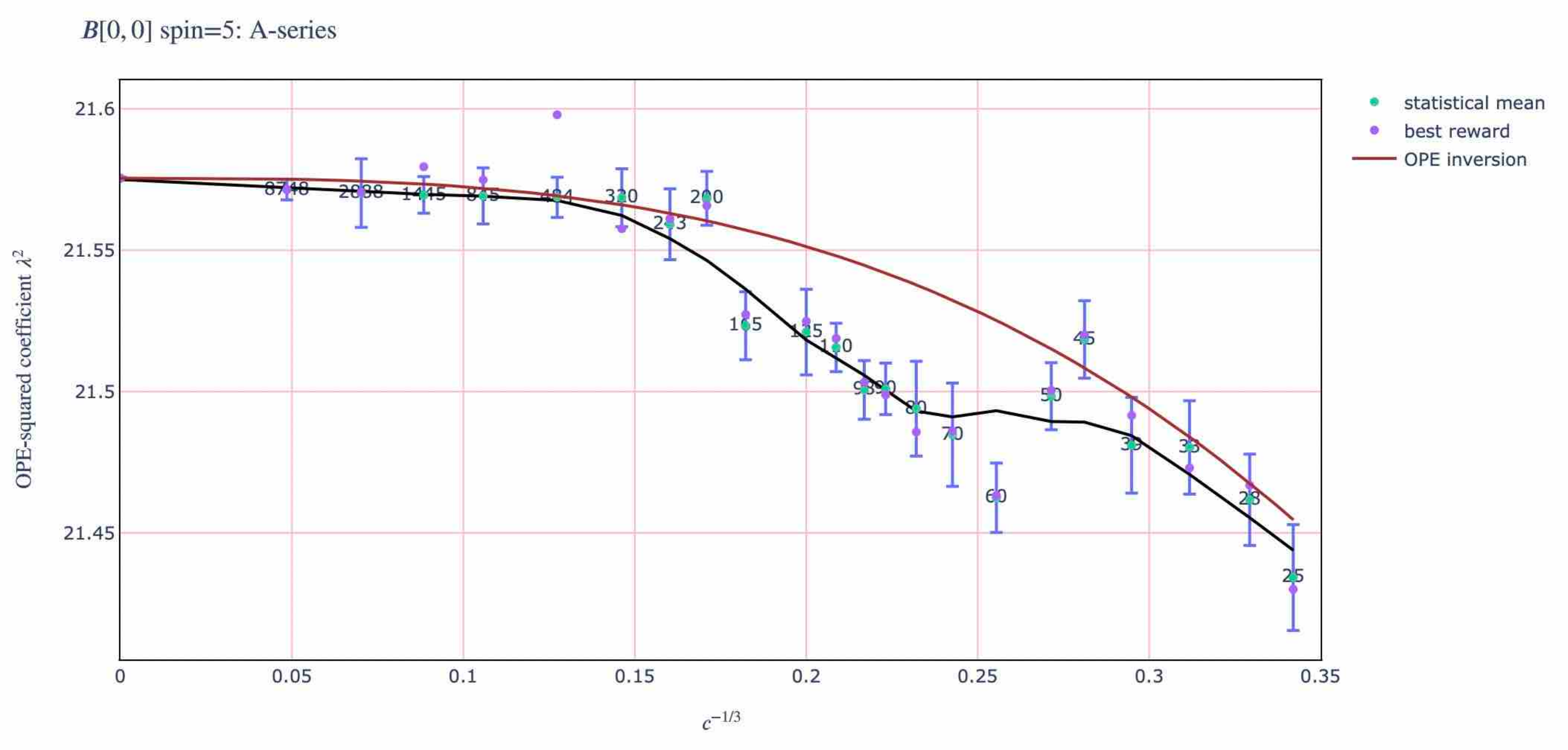}
\includegraphics[width=8.22cm, height=5.3cm]{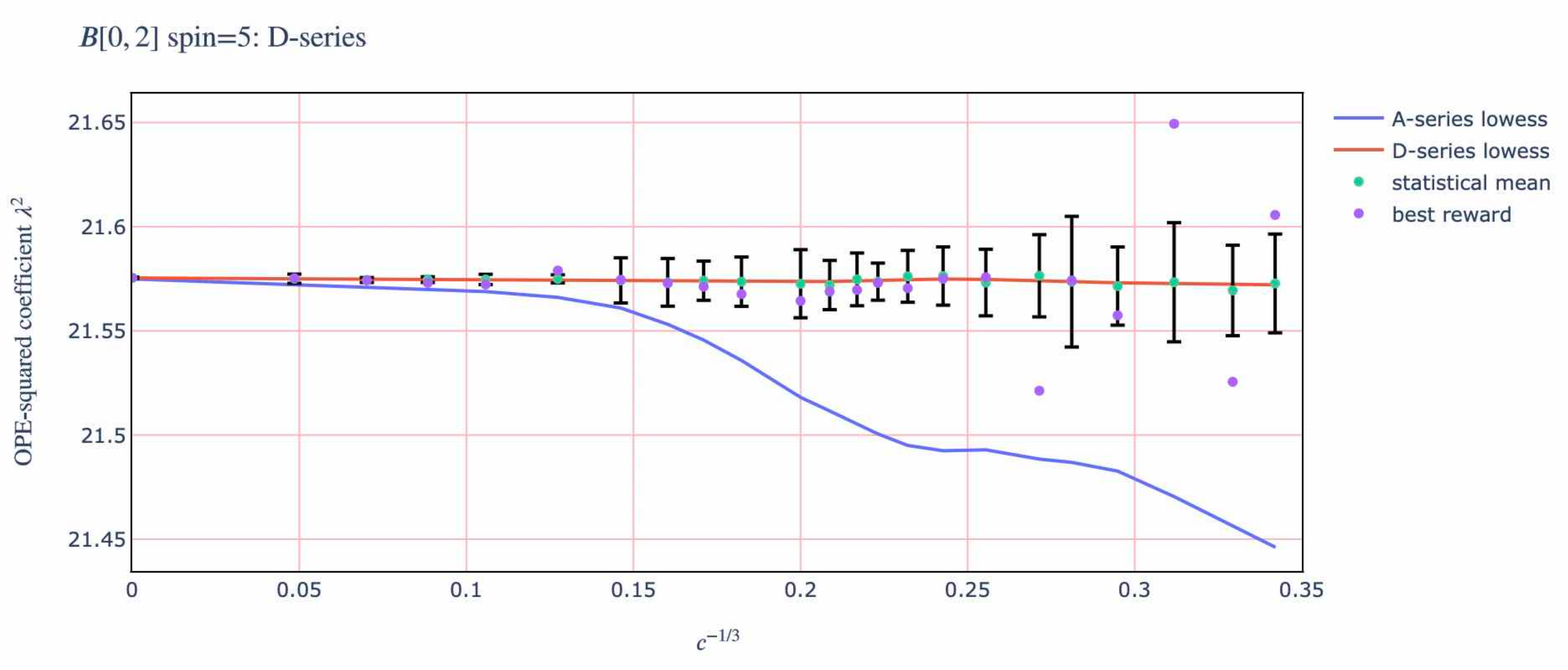}
\caption{Plots of the $A$- and $D$-series curves for the OPE-squared coefficients of the protected operators in the $\mathcal B[0,2]$ multiplets at spin $\ell = 1,3,5$. The brown curves on the left three plots exhibit the long-inverted curves in \cite{Lemos:2021azv}.}
\label{B135_protected}
\end{figure}

To complete the comparison with the data obtained in Ref.\ \cite{Lemos:2021azv}, we exhibit in Fig.\ \ref{B135_protected} the $A$ and $D$ versions of the curves for the OPE-squared coefficients of the $\mathcal B[0,2]$ multiplets at spin $\ell = 1, 3, 5$. The $A$-series curves appear on the left column while the $D$-series curves on the right column. Violations of the numerical bootstrap bounds are observed in the spin-1 and spin-3 cases for the $A$-series. For spin 5 our results are comparable with the results of \cite{Lemos:2021azv}, with the most significant deviation in the rough region of $c^{-1/3}$ between 0.2 and 0.25. Once again, the $D$-series curves confirm the pattern observed in previous data.  

\begin{figure}[t!]
\centering
\includegraphics[width=8.22cm, height=5cm]{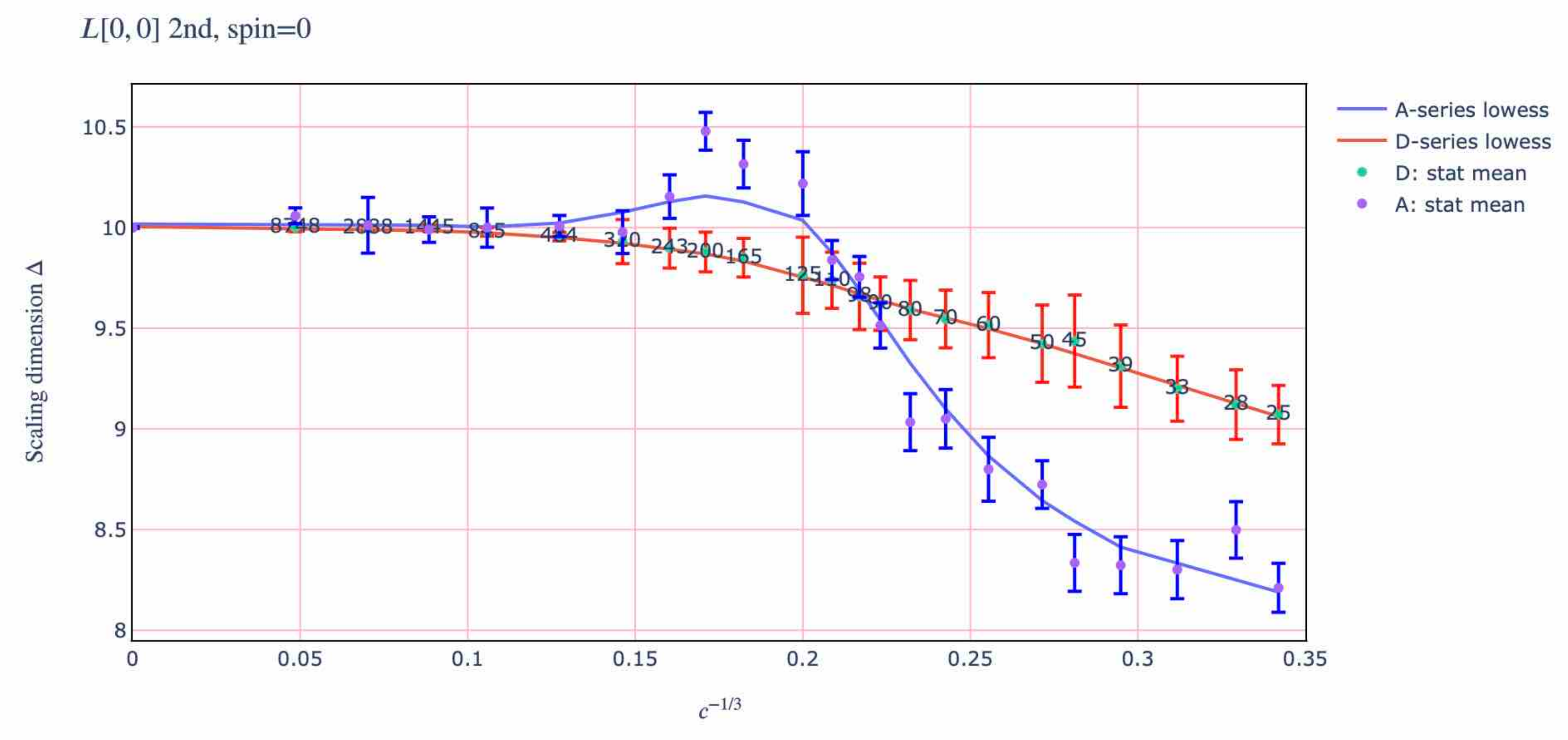}
\includegraphics[width=8.22cm, height=5cm]{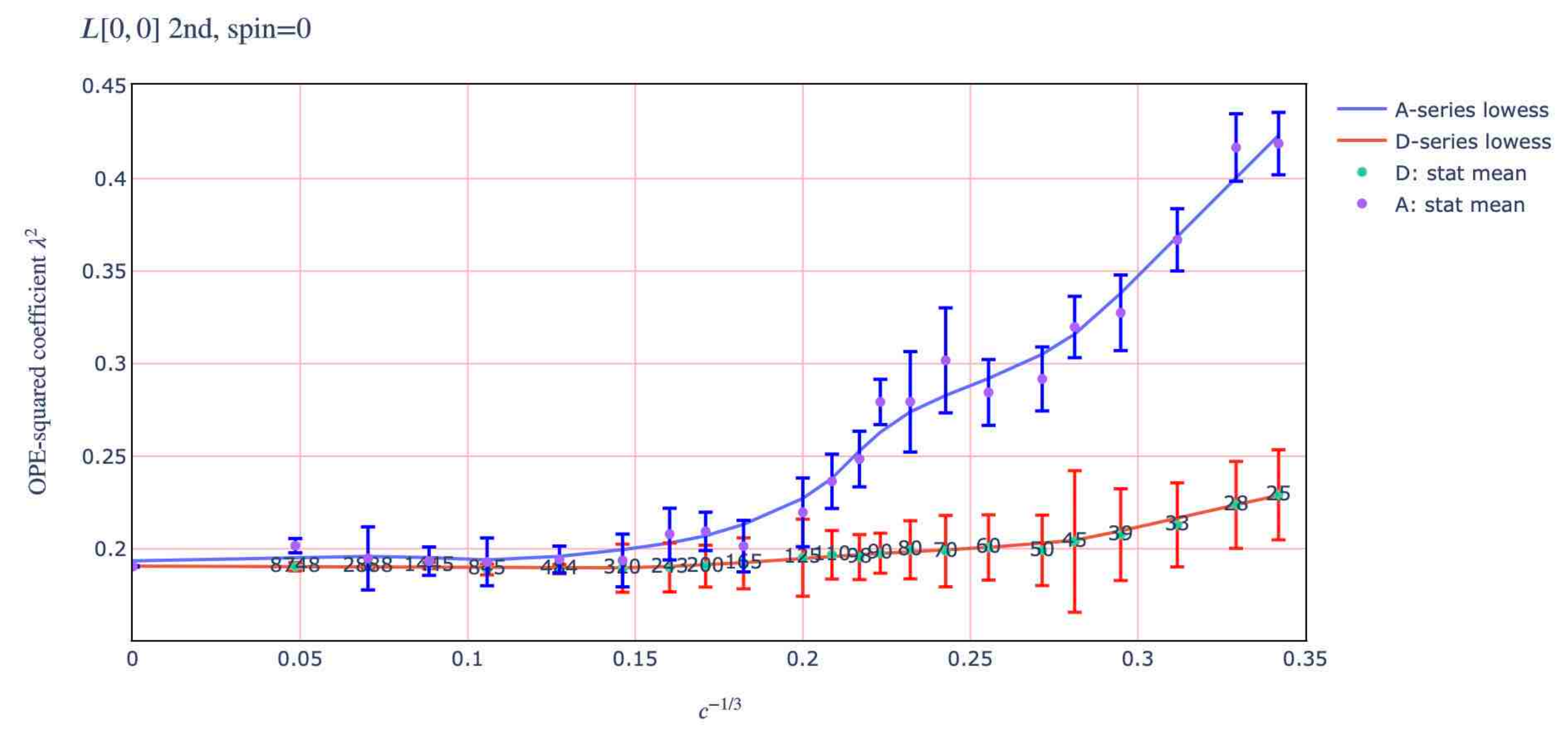}
\includegraphics[width=8.22cm, height=5cm]{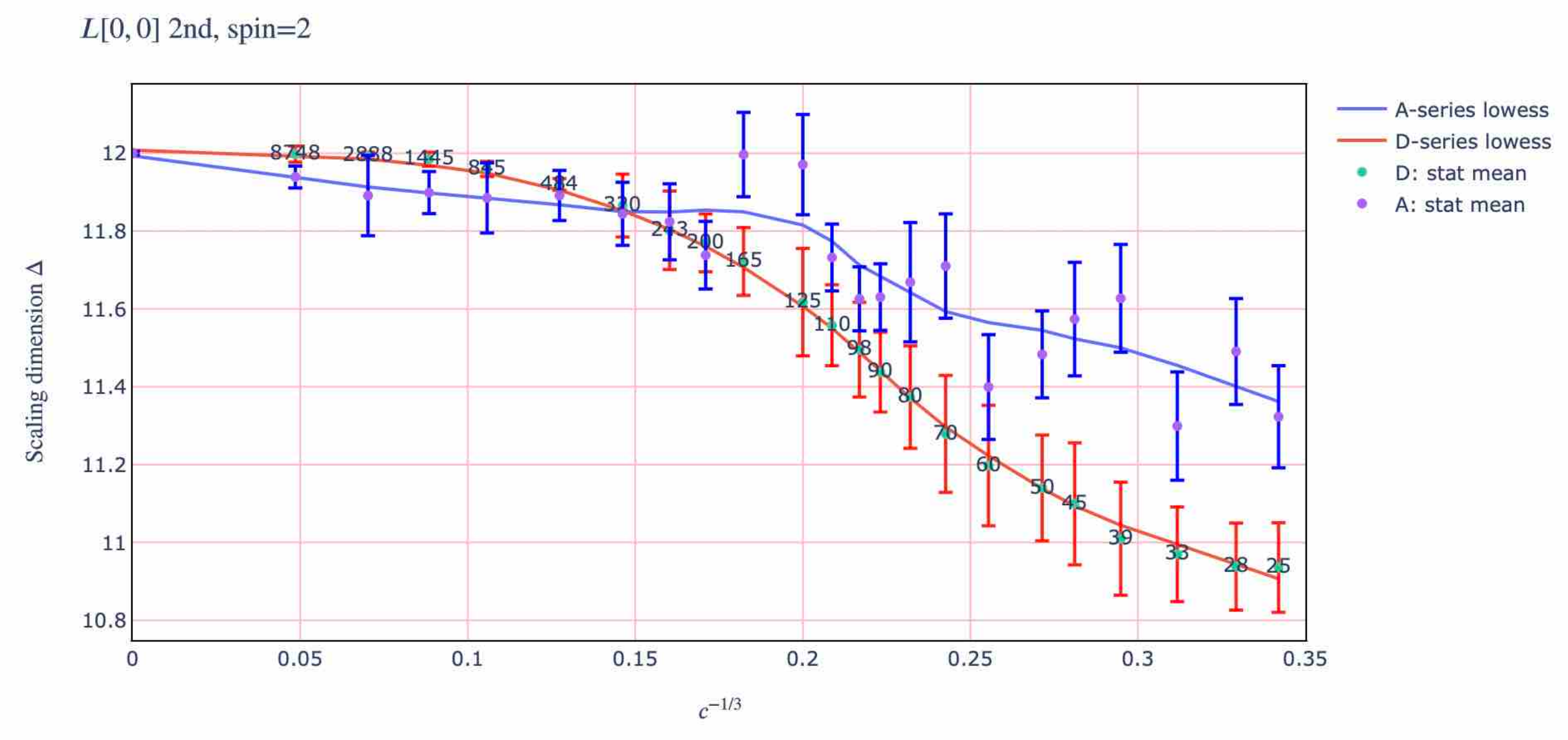}
\includegraphics[width=8.22cm, height=5cm]{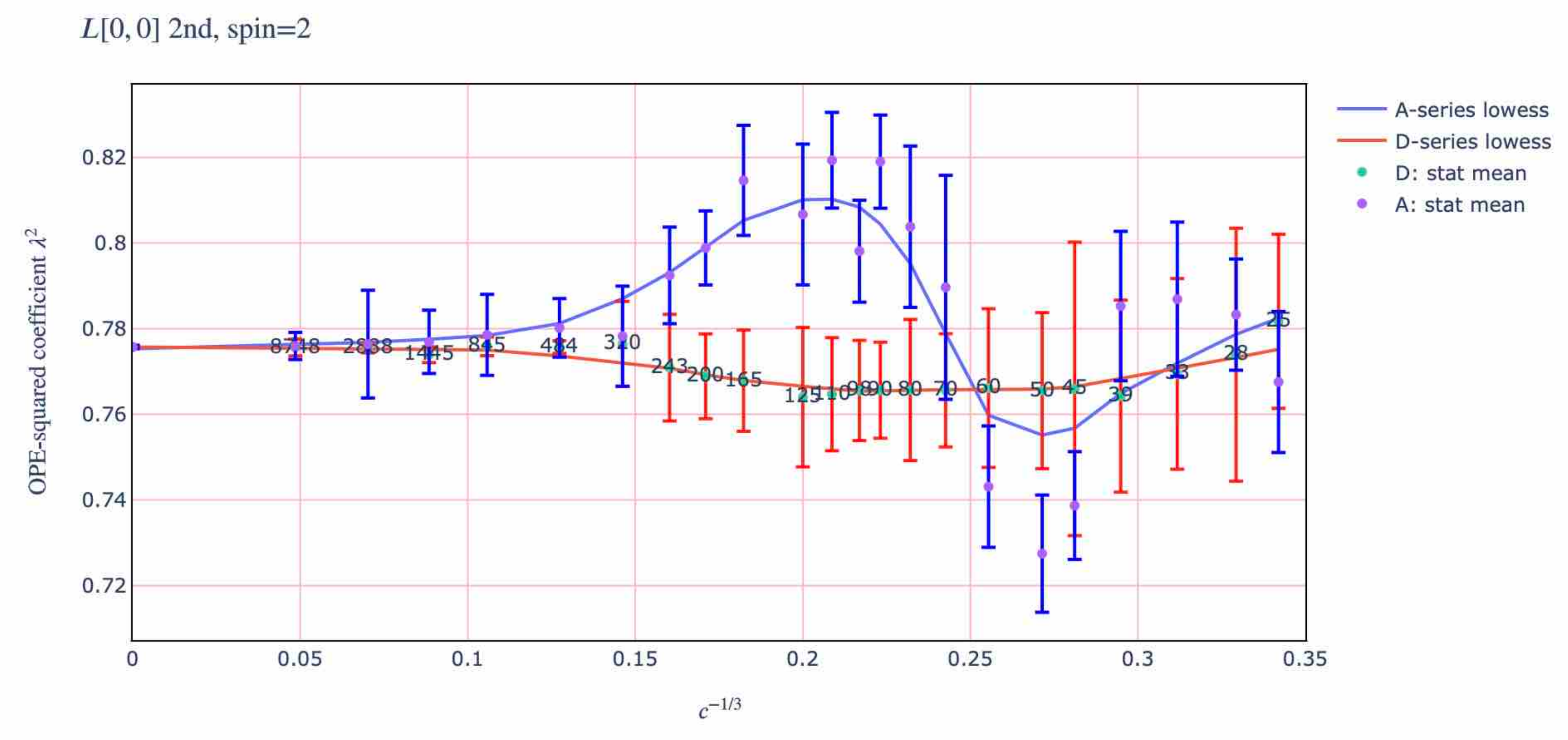}
\includegraphics[width=8.22cm, height=5cm]{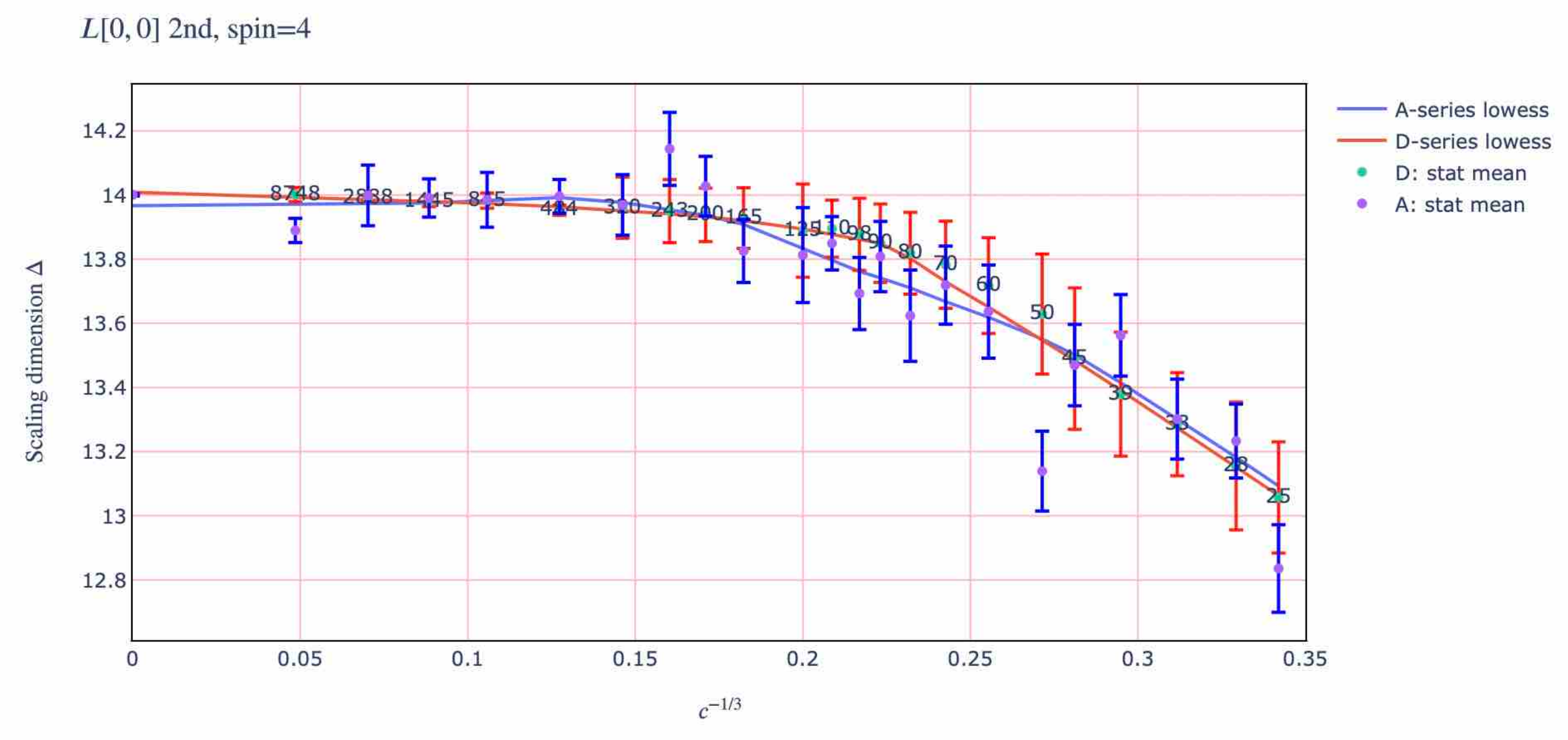}
\includegraphics[width=8.22cm, height=5cm]{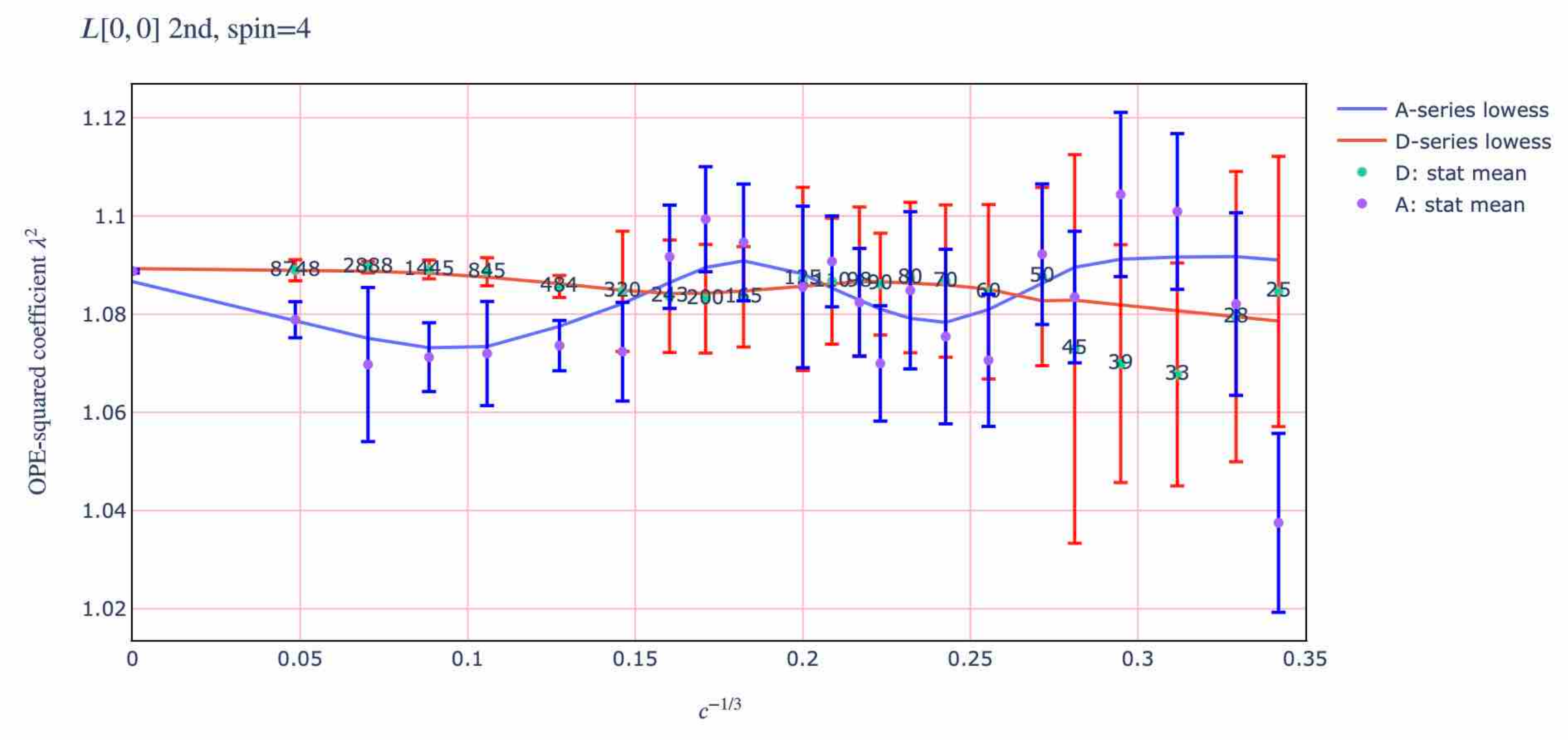}
\includegraphics[width=8.22cm, height=5cm]{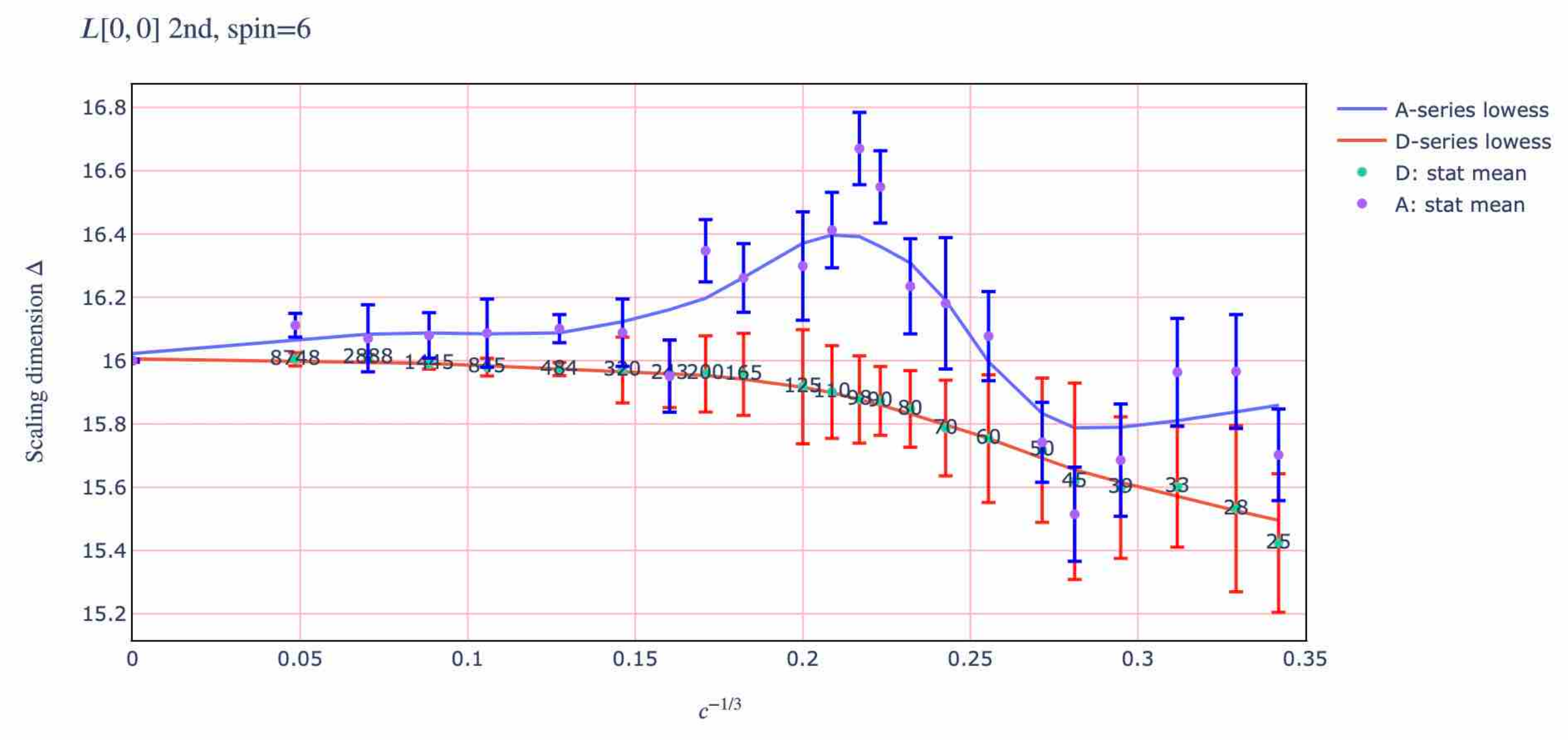}
\includegraphics[width=8.22cm, height=5cm]{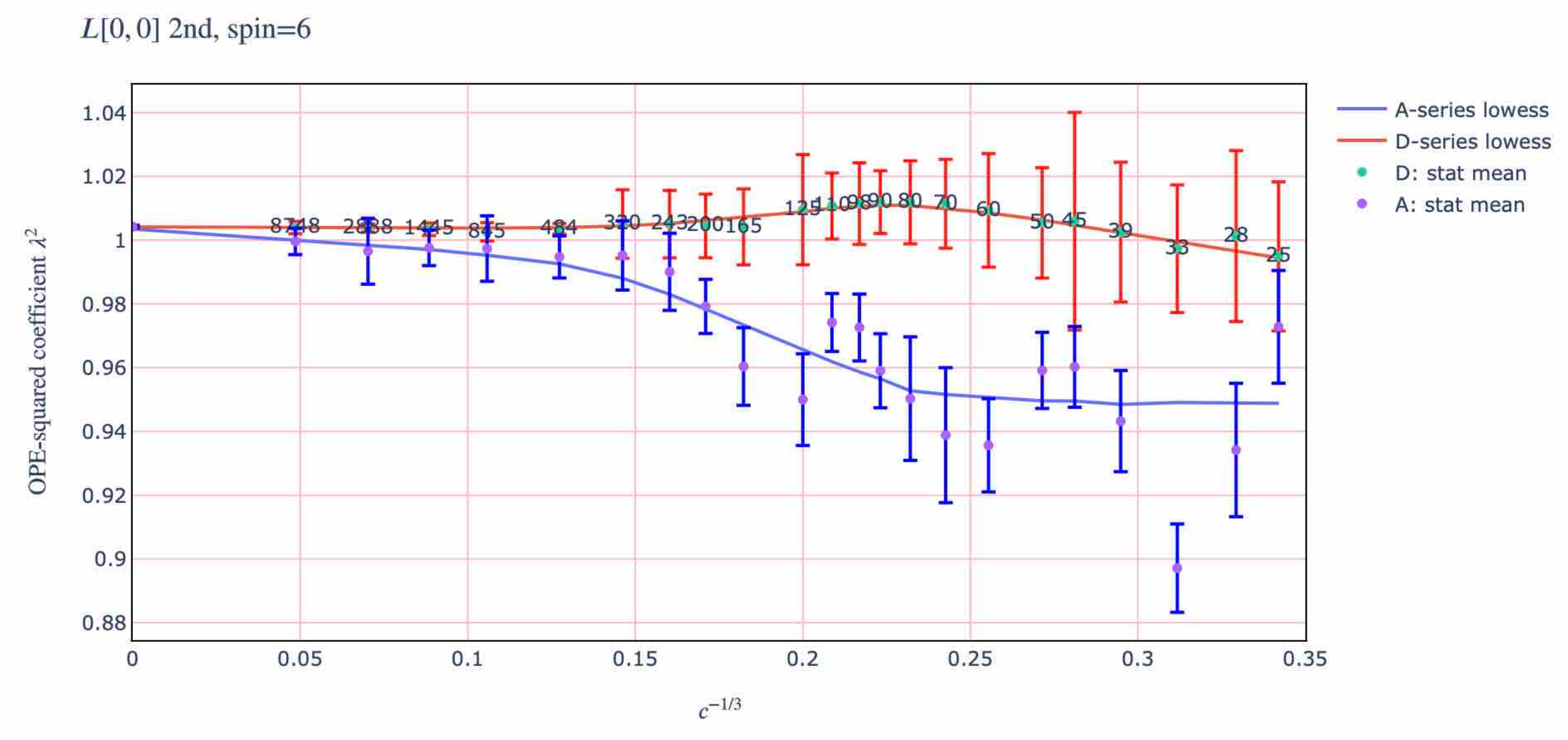}
\caption{Plots of the $A$- and $D$-series curves for the scaling dimensions and OPE-squared coefficients of the first subleading non-protected operators at spin $\ell=0, 2, 4, 6$.}
\label{subleading}
\end{figure}

\subsubsection{Summary of the Comparison with Known Bootstrap Results}

In summary, so far in this section we presented two versions of 12 CFT data. In 9 of them we could compare with known numerical bootstrap bounds. When terminated in the vicinity of $c=676$, the $D$-series curves did not exhibit any significant violations of these bounds. The $A$-series curves extended all the way down to $c=25$ and violated the bounds for sufficiently low values of $c$ in 4 cases: the spin-2 and spin-4 scaling dimensions of $\mathcal L[0,0]$ multiplets in Figs.\ \ref{spin2_unprotected}, \ref{spin4_unprotected} and the spin-1 and spin-3 OPE-squared coefficients of the $\mathcal B[0,2]$ multiplets in the corresponding plots of Fig.\ \ref{B135_protected}. These violations reflect the limitations of the specific truncation of 45 operators that we used throughout the computation. It would be very interesting to increase significantly the number of operators and determine the extent to which these violations persist. We have not explored this very important aspect of the problem in the present work, because we want to treat it separately elsewhere within a more focused investigation of systematic errors.

\subsection{More Results Beyond the Leading Regge Trajectory}

In total, the above application of the SAC algorithm has provided an $A$ and a $D$-series version of 80 CFT data. Most of these data are new predictions that have not appeared previously in the literature. We expect the quality of these predictions to be worse for operators that are closer to the upper bound (in scaling dimension) of the truncation. In our case, the truncation contained long operators up to spin 12 and short operators up to spin 17. In the previous subsections we observed that the results for the lowest-lying operators are consistent with known bounds, but some higher spin data in leading Regge trajectories exhibit violations of known bounds. We expect such potential violations for other higher-dimension data in leading and subleading Regge trajectories. Nevertheless, for future reference we present in Fig.\ \ref{subleading} data for the first subleading long operators at spin $\ell = 0, 2, 4, 6$. PDF files with all the plots presented in this paper, and more plots that were not included here, can be found at \href{https://github.com/vniarchos/BootSTOP/tree/main/Applications/6D/2022_08_results}{this GitHub folder}.

\section{Conclusions and Outlook}\label{outlook}

In this paper we employed the SAC algorithm as a stochastic optimizer to obtain 80 CFT data for protected and unprotected operators in interacting 6D (2,0) theories. These data appear in the truncated superconformal block expansion of the 4-point function of superconformal primaries in the energy-momentum multiplet. The computations were carried out using our BootSTOP package. Our best results were achieved via an adiabatic strategy, whereupon starting from the analytic supergravity values for the CFT data, the central charge was gradually changed to $c = 25$, the lowest expected value corresponding to a physical theory. We were able to identify two curves for each datum. There is evidence that suggests that each of these curves corresponds to the $A$- and $D$-series (2,0) SCFTs. On the whole, these results are competitive when compared to numerical bootstrap bounds \cite{Beem:2015aoa} and more recent approaches using the OPE inversion formula \cite{Lemos:2021azv}, even with a truncation of only 45 unknown operators (a number which is still low compared to our ultimate target for large-scale optimization searches). We expect that the accuracy of the reported results will be improved upon increasing the size of the truncation. The rate and limit of improvement are important open questions, which are part of our general on-going investigation. The addition of tens of extra operators can be explored immediately with the current version of BootSTOP. We encourage the interested reader to do so.

Although BootSTOP is currently using SAC exclusively to optimize the 6D (2,0) crossing equations, we intend to enlarge its functionality in the very near future. This will include more traditional stochastic-optimization algorithms, such as Metropolis--Hastings Monte--Carlo that was recently used in a similar context \cite{Laio:2022ayq}, and will allow it to serve as a benchmarking tool. Similarly, we will incorporate conformal blocks for a number of dimensions, so that BootSTOP can be used by the inquisitive reader to attack a number of interesting CFTs.

\ack{ \bigskip We would like to thank J.~Halverson, P.~Kravchuk,  F.~Ruehle, K.~Skenderis, A.~Stergiou and especially M.~Lemos for stimulating discussions. We would also like to thank the authors of \cite{Lemos:2021azv} for sharing the data contained in their plots. C.P. is supported in part by a Royal Society University Research Fellowship  URF\textbackslash R\textbackslash 180009 and the STFC Consolidated Grant ST/P000754/1. P.R. is funded through the Royal Society Research Fellows Enhancement Award RGF\textbackslash EA\textbackslash 181049. C.P. and V.N. would like to acknowledge the Mainz Institute for Theoretical Physics for hospitality during the workshop  ``A Deep-Learning Era of Particle Theory". V.N. would also like to acknowledge Queen Mary University of London for hospitality during the completion of this work. This research utilised Queen Mary's Apocrita HPC facility, supported by QMUL Research-IT \href{http://doi.org/10.5281/zenodo.438045}{http://doi.org/10.5281/zenodo.438045}.}

\begin{appendix}

\section{Example of a Wide Search from Scratch at $c=25$}
\label{blind}
 
In this appendix we showcase the results of a blind search without any additional, theory-dependent prior information, at $c=25$. The parameters used for this search were: $\displaystyle \textbf{guess\_sizes\_deltas} = 10$ for all scaling dimensions, $\displaystyle \textbf{guess\_sizes\_opes} = 20$ for all OPE-squared coefficients, while choosing  $\displaystyle \textbf{faff\_max} = 12000$, $\displaystyle{\bf pc\_ max}=5$ and $\displaystyle{\bf window\_ rate}=0.5$, to allow the SAC algorithm enough time to explore the search space. Each search started in guessing mode, the functionality of which was detailed in Sec.\ \ref{functionality}. We ran 10k agents on the computing cluster, with each run terminating after 24hrs. 

\begin{figure}[t!]
\centering
\includegraphics[width=8.22cm, height=6cm]{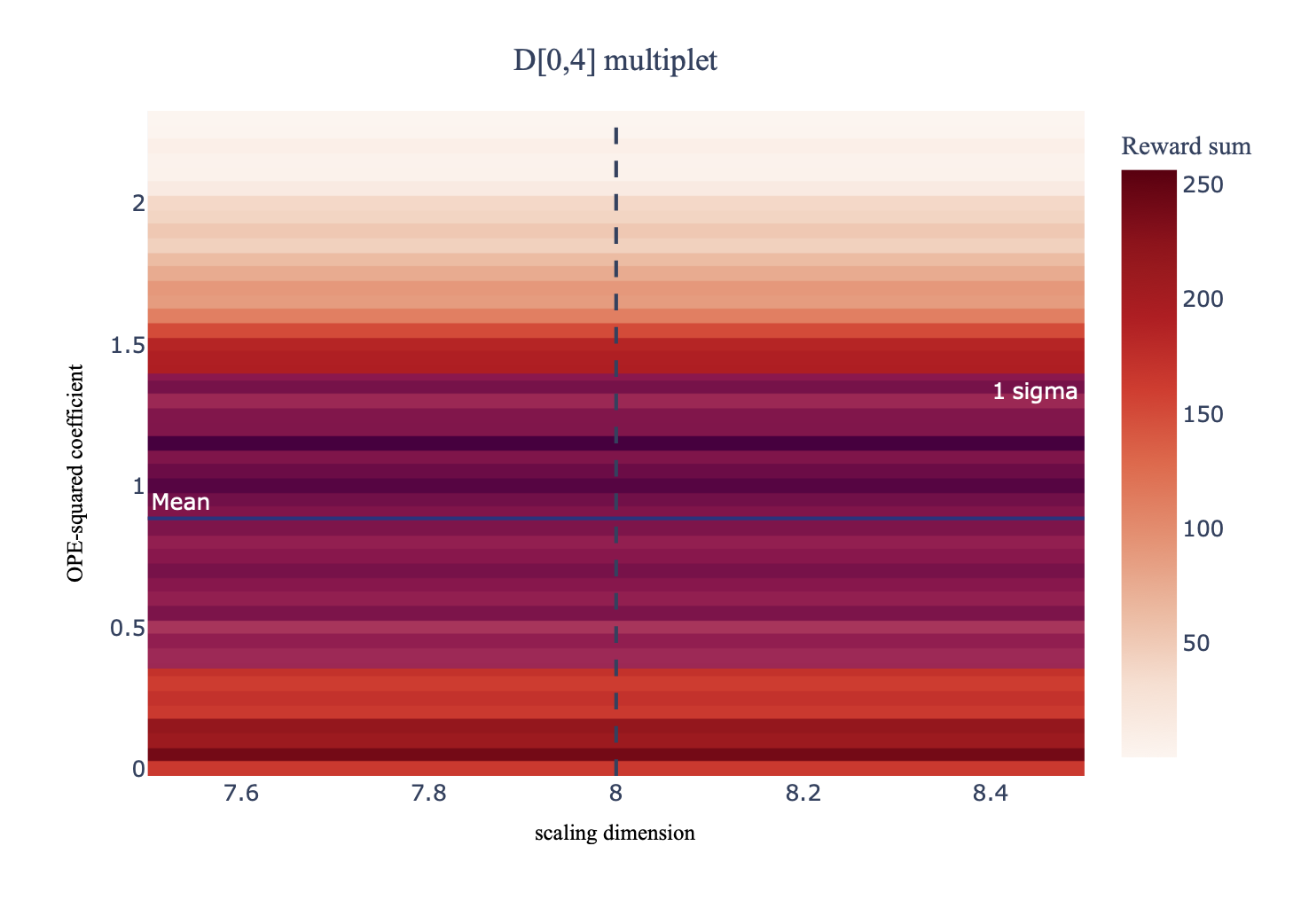}
\includegraphics[width=8.22cm, height=6cm]{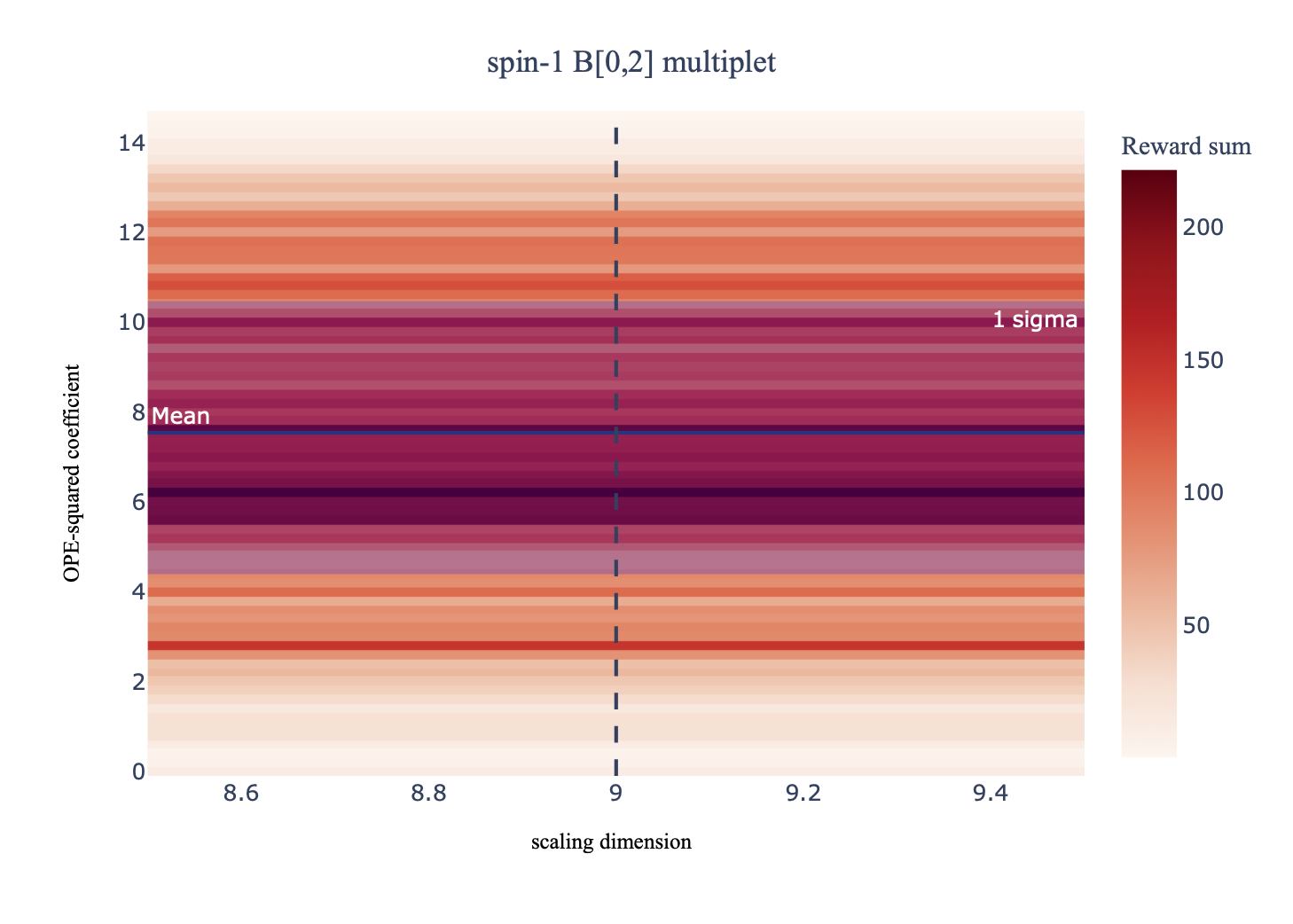}
\includegraphics[width=8.22cm, height=6cm]{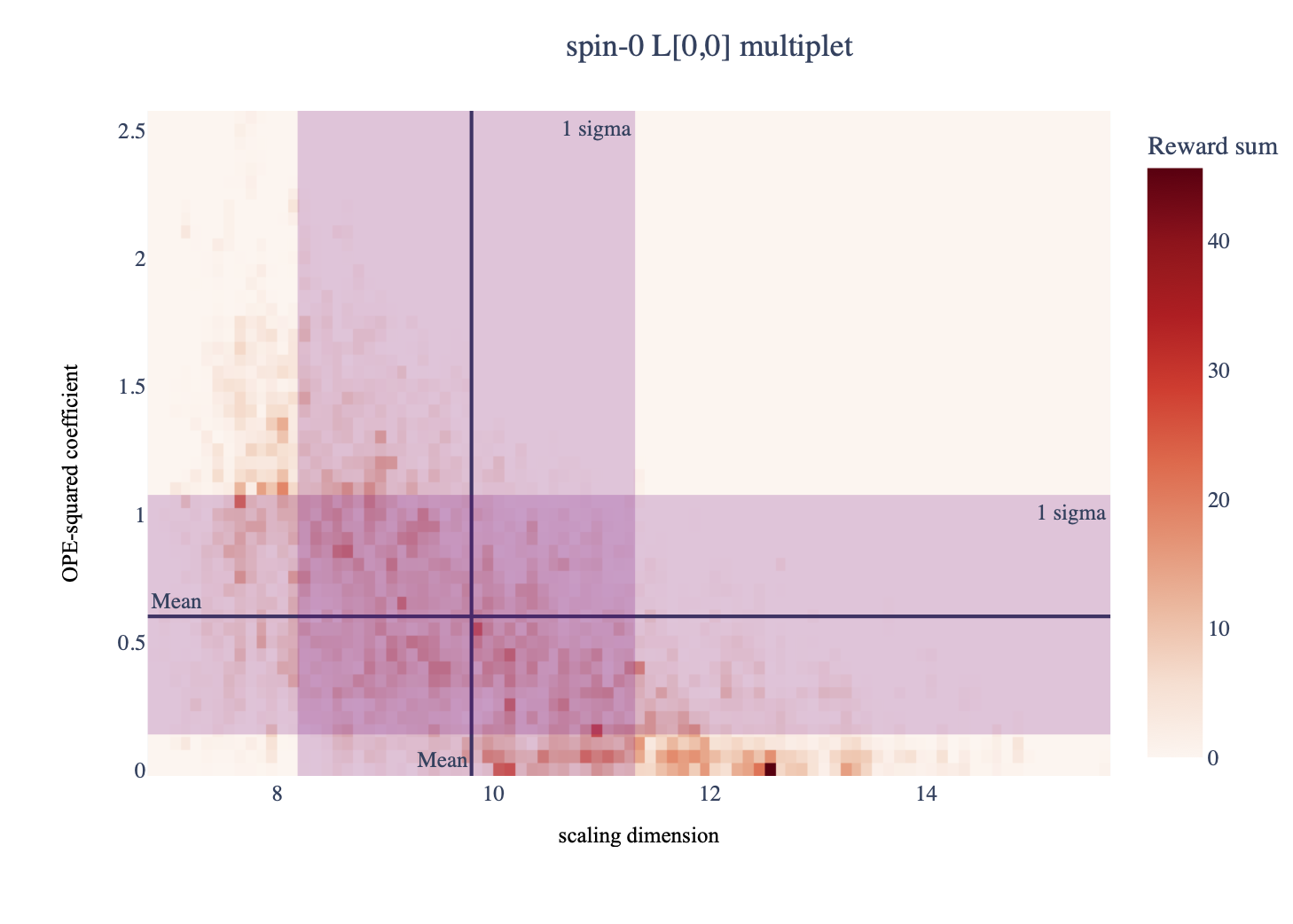}
\includegraphics[width=8.22cm, height=6cm]{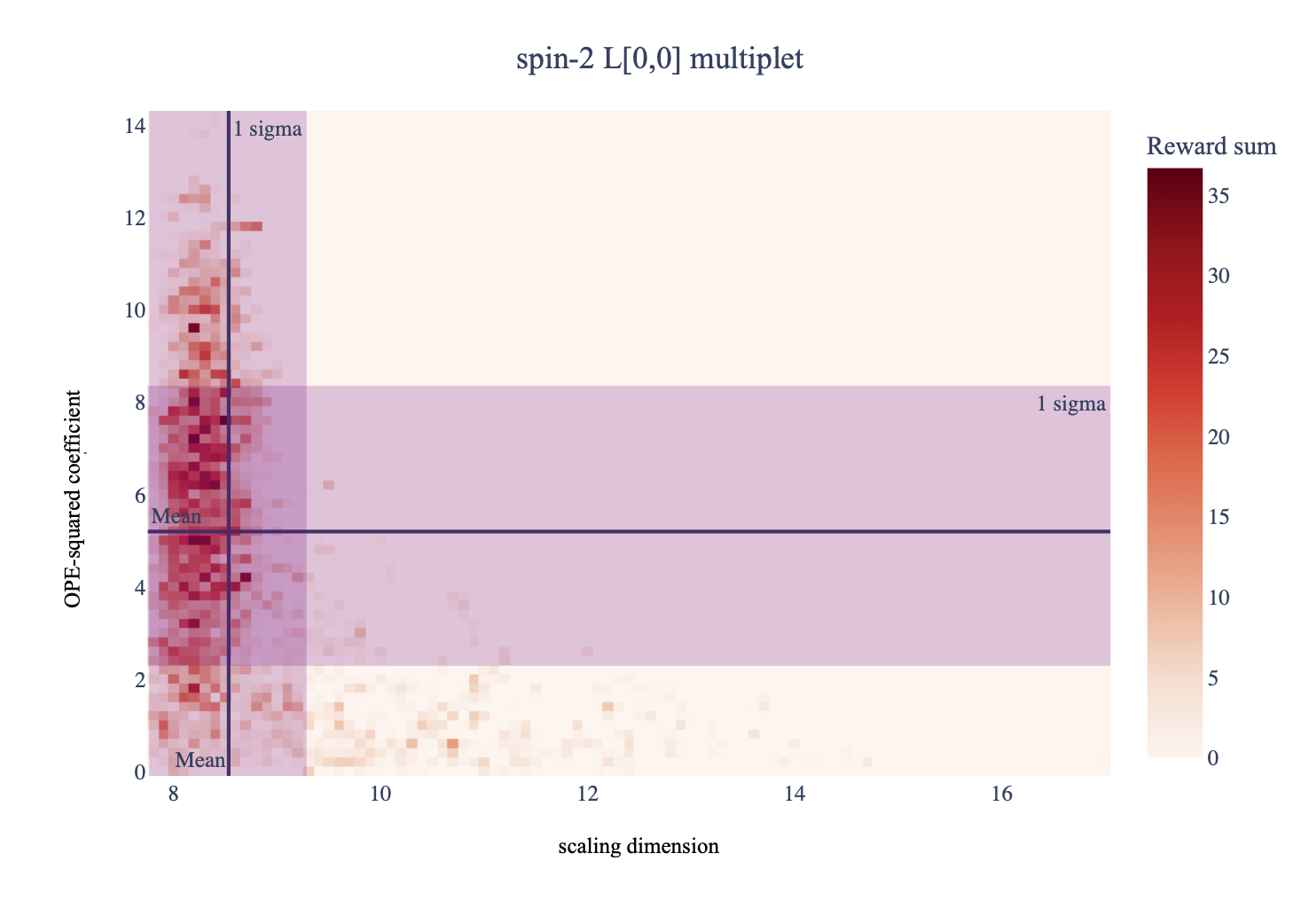}
\caption{Sample heat-map plots for the blind-search results of 10k agents for $\mathcal D[0,4]$, $\ell=1$ $\mathcal B[0,2]$ and $\ell = 0, 2$  $\mathcal L[0,0]$ multiplets in the leading Regge trajectory. In the two top plots the scaling dimension is fixed at the value denoted by the vertical line. We plot the heat map with uniform color intensity on the horizontal axis for visual effect. The shaded region, which is more clearly visible in the two bottom plots, represents the 1$\sigma$ standard deviation spread of the results.}
\label{heatmaps}
\end{figure}

A small sample of our results is collected in Fig.~\ref{heatmaps}, where we display the CFT data for the $\mathcal D[0,4]$ and spin-1 $\mathcal B[0,2]$ protected operators, as well as for the spin-0 and spin-2 $\mathcal L[0,0]$  unprotected operators in the leading Regge trajectory. The plots contain distributions of CFT-data values, for the top 100k reward improvements achieved by any agent among the 10k runs. Note that this is not the final, best reward for each run, so there could be imbalanced contributions from a small number of particularly lucky agents.\footnote{We experimented with many blind-search runs in different contexts and with different parameters. Another way to plot the results is with heat maps that record the 5 best rewards of each agent. Typically, these maps exhibit very similar qualitative features to the heat maps presented here.} For the rewards of data thus selected within a given bin, a corresponding weighted average is computed, where the weight was chosen as the square of the reward normalized by the overall best reward. The resulting heat map exhibits the formation of basins of attraction, with the highest concentrations of weighted-average reward denoted as deeper red on the plots. Although the basins of attraction exist in the full 80-dimensional search space, we find it natural to portray a two-dimensional projection involving the scaling dimension and OPE-squared coefficient for each operator. It is also useful to include the statistical mean and 1$\sigma$ standard deviation. 

Overall, the SAC algorithm proves quite effective in identifying basins of attraction. We observe that the values expected from bootstrap bounds, the OPE inversion formula or our adiabatic searches are in general within the blind-search basins. 

For example, for the OPE-squared coefficient of the $\mathcal D[0,4]$ multiplet, we observe that the basin in the top-left of plot of Fig.\ \ref{heatmaps} contains both the $A$-series value at $\lambda^2_{\mathcal D[0,4]} \sim 0.8$ and the $D$-series value at $\lambda^2_{\mathcal D[0,4]} \sim 1.4$ obtained in the adiabatic search of Sec.\ \ref{adiabatic}. 
For the spin-1 $\mathcal B[0,2]$ multiplet the conformal bootstrap bound gives a value $\lambda^2_{\mathcal B[0,2]_1} \sim 10.3$ and our adiabatic searches $\lambda^2_{\mathcal B[0,2]_1} \sim 10.75$. Both values are contained within the upper part of the region of interest identified by the basin in the top-right plot of Fig.\ \ref{heatmaps}.

The plots for the unprotected $\mathcal L[0,0]$ operators exhibit a similar clustering. The expected values for the scaling dimensions and OPE-squared coefficients for spin 0 are $\Delta_{\mathcal L[0,0]_0}\sim 7$ and $\lambda^2_{\mathcal L[0,0]_0}\sim 1.1$ from the $A$-series adiabatic search and the numerical bootstrap bounds, and $\Delta_{\mathcal L[0,0]_0}\sim 8.4$ and $\lambda^2_{\mathcal L[0,0]_0}\sim 0.8$ from the $D$-series adiabatic search. The $1\sigma$ regions of the bottom-left plot in Fig.\ \ref{heatmaps} favors the $D$-series result. 
The clustering is more pronounced in the spin-2 case. The expected values from the bootstrap bounds ($\Delta_{\mathcal L[0,0]_2}\sim 9.6$ and $\lambda^2_{\mathcal L[0,0]_2}\sim 3.5$) are within the $1\sigma$ region. The scaling dimension expected from the adiabatic searches is higher at $\Delta_{\mathcal L[0,0]_2}\sim 10.2$ and lies in the upper part of the distribution, slightly above the $1\sigma$ boundary. 

The rest of our results exhibit similar qualitative features.\footnote{Plots of the remaining CFT data can be made available upon request.} In these results, one can also observe that there are cases like the OPE-squared coefficient of the spin-5 ${\mathcal B}[0,2]$ multiplet, where although the original region of search in the guessing mode was between 0 and 20, the subsequent search in the non-guessing mode dynamically evolved to explore values above 20, where one anticipates to find the expected result. 

The blind-search results should be understood as the first step in a longer process. One should perform additional iterations around the initial statistical averages obtained here, with the search windows adjusted, for example, to the $1\sigma$ ranges for each unknown. This typically improves the reward, makes the basins of attraction immediately more pronounced and further reduces the statistical spread. It can also be accompanied by a corresponding tuning of other parameters. Unfortunately, as we also stressed in Sec.\ \ref{stochastic}, due to the generically complicated nature of high-dimensional search spaces there is no guarantee that the global minimum will be reached in this fashion, without an additional delicate and time-consuming fine tuning of parameters. Moreover, there can be situations where there are multiple minima corresponding to physical theories that this approach will not be able to distinguish. For these reasons, we find the adiabatic method of Sec.\ \ref{adiabatic} far more promising as a tool towards reaching the final more accurate answer. Blind searches with large windows will be less accurate in general, but can still be useful at providing further intuition about the problem at hand.

\end{appendix}


\bibliography{machine_learning}

\begin{thebibliography}{10}
\ifx\href\asklfhas\newcommand{\href}[2]{#2}\fi
\ifx\arxivref\asklfhas\newcommand{\arxivref}[2]{\href{http://arxiv.org/abs/#1}{#2}}\fi
\ifx\doiref\asklfhas\newcommand{\doiref}[2]{\href{http://dx.doi.org/#1}{#2}}\fi
\parskip 0pt
\normalsize

\bibitem{Kantor:2021kbx}
G.~K\'antor, V.~Niarchos \& C.~Papageorgakis,
\textit{``{Solving Conformal Field Theories with Artificial Intelligence}''},
\doiref{10.1103/PhysRevLett.128.041601}{Phys.~Rev.~Lett. \textbf{128}, 041601
  (2022)\ignorespaces}\ignorespaces,
\normalsize{\texttt{\arxivref{2108.08859}{arXiv:2108.08859}}}\ignorespaces
\bibitem{Kantor:2021jpz}
G.~K\'antor, V.~Niarchos \& C.~Papageorgakis,
\textit{``{Conformal bootstrap with reinforcement learning}''},
\doiref{10.1103/PhysRevD.105.025018}{Phys.~Rev.~D \textbf{105}, 025018
  (2022)\ignorespaces}\ignorespaces,
\normalsize{\texttt{\arxivref{2108.09330}{arXiv:2108.09330}}}\ignorespaces
\bibitem{Gliozzi:2013ysa}
F.~Gliozzi,
\textit{``{More constraining conformal bootstrap}''},
\doiref{10.1103/PhysRevLett.111.161602}{Phys.~Rev.~Lett. \textbf{111}, 161602
  (2013)\ignorespaces}\ignorespaces,
\normalsize{\texttt{\arxivref{1307.3111}{arXiv:1307.3111}}}\ignorespaces
\bibitem{Gliozzi:2014jsa}
F.~Gliozzi \& A.~Rago,
\textit{``{Critical exponents of the 3d Ising and related models from Conformal
  Bootstrap}''},
\doiref{10.1007/JHEP10(2014)042}{JHEP \textbf{1410}, 042
  (2014)\ignorespaces}\ignorespaces,
\normalsize{\texttt{\arxivref{1403.6003}{arXiv:1403.6003}}}\ignorespaces
\bibitem{Gliozzi:2015qsa}
F.~Gliozzi, P.~Liendo, M.~Meineri \& A.~Rago,
\textit{``{Boundary and Interface CFTs from the Conformal Bootstrap}''},
\doiref{10.1007/JHEP05(2015)036}{JHEP \textbf{1505}, 036
  (2015)\ignorespaces}\ignorespaces,
\normalsize{\texttt{\arxivref{1502.07217}{arXiv:1502.07217}}}\ignorespaces
\bibitem{Gliozzi:2016cmg}
F.~Gliozzi,
\textit{``{Truncatable bootstrap equations in algebraic form and critical
  surface exponents}''},
\doiref{10.1007/JHEP10(2016)037}{JHEP \textbf{1610}, 037
  (2016)\ignorespaces}\ignorespaces,
\normalsize{\texttt{\arxivref{1605.04175}{arXiv:1605.04175}}}\ignorespaces
\bibitem{Li:2017ukc}
W.~Li,
\textit{``{New method for the conformal bootstrap with OPE truncations}''},
\normalsize{\texttt{\arxivref{1711.09075}{arXiv:1711.09075}}}\ignorespaces
\bibitem{DBLP:journals/corr/abs-1801-01290}
T.~Haarnoja, A.~Zhou, P.~Abbeel \& S.~Levine,
\textit{``Soft Actor-Critic: Off-Policy Maximum Entropy Deep Reinforcement
  Learning with a Stochastic Actor''},
CoRR \textbf{abs/1801.01290},  (2018)\ignorespaces\ignorespaces,
\normalsize{\texttt{\arxivref{1801.01290}{arXiv:1801.01290}}}\ignorespaces,
\href{http://arxiv.org/abs/1801.01290}{\texttt{http://arxiv.org/abs/1801.01290}}
\bibitem{Beem:2014kka}
C.~Beem, L.~Rastelli \& B.~C. van~Rees,
\textit{``{$ \mathcal{W} $ symmetry in six dimensions}''},
\doiref{10.1007/JHEP05(2015)017}{JHEP \textbf{1505}, 017
  (2015)\ignorespaces}\ignorespaces,
\normalsize{\texttt{\arxivref{1404.1079}{arXiv:1404.1079}}}\ignorespaces
\bibitem{Beem:2015aoa}
C.~Beem, M.~Lemos, L.~Rastelli \& B.~C. van~Rees,
\textit{``{The (2, 0) superconformal bootstrap}''},
\doiref{10.1103/PhysRevD.93.025016}{Phys.~Rev.~D \textbf{93}, 025016
  (2016)\ignorespaces}\ignorespaces,
\normalsize{\texttt{\arxivref{1507.05637}{arXiv:1507.05637}}}\ignorespaces
\bibitem{Henriksson:2022gpa}
J.~Henriksson, S.~R. Kousvos \& M.~Reehorst,
\textit{``{Spectrum continuity and level repulsion: the Ising CFT from
  infinitesimal to finite $\boldsymbol\varepsilon$}''},
\normalsize{\texttt{\arxivref{2207.10118}{arXiv:2207.10118}}}\ignorespaces
\bibitem{Reehorst:2021ykw}
M.~Reehorst, S.~Rychkov, D.~Simmons-Duffin, B.~Sirois, N.~Su \& B.~Van~Rees,
\textit{``{Navigator Function for the Conformal Bootstrap}''},
\normalsize{\texttt{\arxivref{2104.09518}{arXiv:2104.09518}}}\ignorespaces
\bibitem{El-Showk:2012vjm}
S.~El-Showk \& M.~F. Paulos,
\textit{``{Bootstrapping Conformal Field Theories with the Extremal Functional
  Method}''},
\doiref{10.1103/PhysRevLett.111.241601}{Phys.~Rev.~Lett. \textbf{111}, 241601
  (2013)\ignorespaces}\ignorespaces,
\normalsize{\texttt{\arxivref{1211.2810}{arXiv:1211.2810}}}\ignorespaces
\bibitem{El-Showk:2016mxr}
S.~El-Showk \& M.~F. Paulos,
\textit{``{Extremal bootstrapping: go with the flow}''},
\doiref{10.1007/JHEP03(2018)148}{JHEP \textbf{1803}, 148
  (2018)\ignorespaces}\ignorespaces,
\normalsize{\texttt{\arxivref{1605.08087}{arXiv:1605.08087}}}\ignorespaces
\bibitem{Lemos:2021azv}
M.~Lemos, B.~C. van~Rees \& X.~Zhao,
\textit{``{Regge trajectories for the (2, 0) theories}''},
\doiref{10.1007/JHEP01(2022)022}{JHEP \textbf{2201}, 022
  (2022)\ignorespaces}\ignorespaces,
\normalsize{\texttt{\arxivref{2105.13361}{arXiv:2105.13361}}}\ignorespaces
\bibitem{Caron-Huot:2017vep}
S.~Caron-Huot,
\textit{``{Analyticity in Spin in Conformal Theories}''},
\doiref{10.1007/JHEP09(2017)078}{JHEP \textbf{1709}, 078
  (2017)\ignorespaces}\ignorespaces,
\normalsize{\texttt{\arxivref{1703.00278}{arXiv:1703.00278}}}\ignorespaces
\bibitem{Minwalla:1997ka}
S.~Minwalla,
\textit{``{Restrictions imposed by superconformal invariance on quantum field
  theories}''},
\doiref{10.4310/ATMP.1998.v2.n4.a4}{Adv.~Theor.~Math.~Phys. \textbf{2}, 783
  (1998)\ignorespaces}\ignorespaces,
\normalsize{\texttt{\arxivref{hep-th/9712074}{hep-th/9712074}}}\ignorespaces
\bibitem{Dobrev:2002dt}
V.~K. Dobrev,
\textit{``{Positive energy unitary irreducible representations of D = 6
  conformal supersymmetry}''},
\doiref{10.1088/0305-4470/35/33/308}{J.~Phys.~A \textbf{35}, 7079
  (2002)\ignorespaces}\ignorespaces,
\normalsize{\texttt{\arxivref{hep-th/0201076}{hep-th/0201076}}}\ignorespaces
\bibitem{Buican:2016hpb}
M.~Buican, J.~Hayling \& C.~Papageorgakis,
\textit{``{Aspects of Superconformal Multiplets in D\ensuremath{>}4}''},
\doiref{10.1007/JHEP11(2016)091}{JHEP \textbf{1611}, 091
  (2016)\ignorespaces}\ignorespaces,
\normalsize{\texttt{\arxivref{1606.00810}{arXiv:1606.00810}}}\ignorespaces
\bibitem{Cordova:2016emh}
C.~Cordova, T.~T. Dumitrescu \& K.~Intriligator,
\textit{``{Multiplets of Superconformal Symmetry in Diverse Dimensions}''},
\doiref{10.1007/JHEP03(2019)163}{JHEP \textbf{1903}, 163
  (2019)\ignorespaces}\ignorespaces,
\normalsize{\texttt{\arxivref{1612.00809}{arXiv:1612.00809}}}\ignorespaces
\bibitem{Heslop:2004du}
P.~J. Heslop,
\textit{``{Aspects of superconformal field theories in six dimensions}''},
\doiref{10.1088/1126-6708/2004/07/056}{JHEP \textbf{0407}, 056
  (2004)\ignorespaces}\ignorespaces,
\normalsize{\texttt{\arxivref{hep-th/0405245}{hep-th/0405245}}}\ignorespaces
\bibitem{Eden:2001wg}
B.~Eden, S.~Ferrara \& E.~Sokatchev,
\textit{``{(2,0) superconformal OPEs in D = 6, selection rules and
  nonrenormalization theorems}''},
\doiref{10.1088/1126-6708/2001/11/020}{JHEP \textbf{0111}, 020
  (2001)\ignorespaces}\ignorespaces,
\normalsize{\texttt{\arxivref{hep-th/0107084}{hep-th/0107084}}}\ignorespaces
\bibitem{Dolan:2003hv}
F.~A. Dolan \& H.~Osborn,
\textit{``{Conformal partial waves and the operator product expansion}''},
\doiref{10.1016/j.nuclphysb.2003.11.016}{Nucl.~Phys.~B \textbf{678}, 491
  (2004)\ignorespaces}\ignorespaces,
\normalsize{\texttt{\arxivref{hep-th/0309180}{hep-th/0309180}}}\ignorespaces
\bibitem{Dolan:2011dv}
F.~A. Dolan \& H.~Osborn,
\textit{``{Conformal Partial Waves: Further Mathematical Results}''},
\normalsize{\texttt{\arxivref{1108.6194}{arXiv:1108.6194}}}\ignorespaces
\bibitem{Kantor:thesis}
G.~Kantor,
\textit{``{Exact and Numerical Methods in (Super)Conformal Field Theories}''},
\href{https://qmro.qmul.ac.uk/xmlui/handle/123456789/79600}{\texttt{https://qmro.qmul.ac.uk/xmlui/handle/123456789/79600}}
\bibitem{CastedoEcheverri:2016fxt}
A.~Castedo~Echeverri, B.~von~Harling \& M.~Serone,
\textit{``{The Effective Bootstrap}''},
\doiref{10.1007/JHEP09(2016)097}{JHEP \textbf{1609}, 097
  (2016)\ignorespaces}\ignorespaces,
\normalsize{\texttt{\arxivref{1606.02771}{arXiv:1606.02771}}}\ignorespaces
\bibitem{NIPS2014_17e23e50}
Y.~N. Dauphin, R.~Pascanu, C.~Gulcehre, K.~Cho, S.~Ganguli \& Y.~Bengio,
\textit{``Identifying and attacking the saddle point problem in
  high-dimensional non-convex optimization''},
in \textit{``Advances in Neural Information Processing Systems''},
ed: Z.~Ghahramani, M.~Welling, C.~Cortes, N.~Lawrence \& K.~Weinberger,
Curran Associates, Inc. (2014)\ignorespaces\bibitem{king_thomas_2017_438045}
T.~King, S.~Butcher \& L.~Zalewski,
\textit{``{Apocrita - High Performance Computing Cluster for Queen Mary
  University of London}''},
\href{https://doi.org/10.5281/zenodo.438045}{\texttt{https://doi.org/10.5281/zenodo.438045}}
\bibitem{Alday:2020tgi}
L.~F. Alday, S.~M. Chester \& H.~Raj,
\textit{``{6d (2,0) and M-theory at 1-loop}''},
\doiref{10.1007/JHEP01(2021)133}{JHEP \textbf{2101}, 133
  (2021)\ignorespaces}\ignorespaces,
\normalsize{\texttt{\arxivref{2005.07175}{arXiv:2005.07175}}}\ignorespaces
\bibitem{Alday:2022ldo}
L.~F. Alday \& S.~M. Chester,
\textit{``{Pure anti-de Sitter supergravity and the conformal bootstrap}''},
\normalsize{\texttt{\arxivref{2207.05085}{arXiv:2207.05085}}}\ignorespaces
\bibitem{Laio:2022ayq}
A.~Laio, U.~L. Valenzuela \& M.~Serone,
\textit{``{Monte~Carlo approach to the conformal bootstrap}''},
\doiref{10.1103/PhysRevD.106.025019}{Phys.~Rev.~D \textbf{106}, 025019
  (2022)\ignorespaces}\ignorespaces,
\normalsize{\texttt{\arxivref{2206.05193}{arXiv:2206.05193}}}\ignorespaces
\end{thebibliography}
\end{document}